\documentclass[authoryear, preprint, 12pt]{elsarticle}

\usepackage{graphicx}
\usepackage{caption}
\usepackage{subcaption}
\usepackage{amssymb}
\usepackage{amsthm}
\usepackage{amsmath}
\usepackage{float}
\usepackage{bigfoot}
\usepackage{pdflscape}
\usepackage{adjustbox}
\usepackage{longtable}

\usepackage[utf8]{inputenc}

\usepackage{lineno}
\usepackage{lipsum}

\usepackage{natbib}

\usepackage{tabularx}
\newcolumntype{L}[1]{>{\raggedright\arraybackslash}p{#1}}
\newcolumntype{C}[1]{>{\centering\arraybackslash}p{#1}}
\newcolumntype{R}[1]{>{\raggedleft\arraybackslash}p{#1}}
\usepackage{multirow}

\usepackage[a4paper, margin=1in]{geometry}
\usepackage{setspace}
\usepackage{geometry}

\usepackage{xr-hyper}
\usepackage{hyperref}

\hypersetup{
	colorlinks   = true,    
	urlcolor     = blue,    
	linkcolor    = blue,    
	citecolor    = blue     
}

\usepackage[nogroupskip,acronym,nonumberlist,nopostdot]{glossaries}
\makeglossaries
\newacronym{FD}{FD}{fundamental diagram}
\newacronym{FDs}{FDs}{fundamental diagrams}
\newacronym{AFC}{AFC}{automated fare collection}
\newacronym{AVL}{AVL}{automatic vehicular location}
\newacronym{MTR}{MTR}{Mass Transit Railway}
\newacronym{TSC}{TSC}{Transport Strategy Centre}
\newacronym{NPIV}{NPIV}{non-parametric instrumental variables}
\newacronym{NP}{NP}{non-parametric}
\newacronym{IV}{IV}{instrumental variables}
\newacronym{OLS}{OLS}{ordinary least squares}
\newacronym{POLS}{POLS}{pooled ordinary least squares}
\newacronym{2SLS}{2SLS}{two-stage least squares}
\newacronym{OD}{OD}{origin-destination}
\newacronym{DGP}{DGP}{data generating process}
\newacronym{DPM}{DPM}{Dirichlet process mixture}
\newacronym{DP}{DP}{Dirichlet process}
\newacronym{CA}{CA}{Cellular Automata}
\newacronym{MCMC}{MCMC}{Markov chain Monte Carlo}

\newcommand*\sref[1]{%
    Appendix Section \ref{#1}}

\newcommand*\stref[1]{%
    Appendix Table \ref{#1}}
    
\newcommand*\seref[1]{%
    Appendix equation \ref{#1}}

\begin{document}

\begin{frontmatter}

\title{Optimal congestion control strategies for near-capacity urban metros: informing intervention via fundamental diagrams} 
\author{Anupriya$^1$\corref{cor1}}
\author{Daniel J. Graham$^1$}
\author{Prateek Bansal$^{2}$}
\author{Daniel H\"{o}rcher$^1$}
\author{Richard Anderson$^1$}
\cortext[cor1]{Corresponding author.  Email address: anupriya15@imperial.ac.uk}
\address{$^1$Transport Strategy Centre, Department of Civil and Environmental Engineering, \\ Imperial College London, UK \\
$^2$Department of Civil and Environmental Engineering, National University of Singapore, Singapore
\vspace{-1cm}}

\begin{abstract}
\noindent Congestion; operational delays due to a vicious circle of passenger-congestion and train-queuing; is an escalating problem for metro systems because it has negative consequences from passenger discomfort to eventual mode-shifts. Congestion arises due to large volumes of passenger boardings and alightings at bottleneck stations, which may lead to increased stopping times at stations and consequent queuing of trains upstream, further reducing line throughput and implying even greater accumulation of passengers at stations. Alleviating congestion requires control strategies such as regulating the inflow of passengers entering bottleneck stations. The availability of large-scale smartcard and train movement data from day-to-day operations facilitates development of models that can inform such strategies in a data-driven way. In this paper, we propose to model station-level passenger-congestion via empirical passenger boarding-alightings and train flow relationships, henceforth, \acrfull{FDs}. We emphasise that estimating \acrshort{FDs} using station-level data is empirically challenging due to confounding biases arising from the interdependence of operations at different stations, which obscures the true sources of congestion in the network. We thus adopt a causal statistical modelling approach to produce \acrshort{FDs} that are robust to confounding and as such suitable to properly inform control strategies. The closest antecedent to the proposed model is the \acrshort{FD} for road traffic networks, which informs traffic management strategies, for instance, via locating the optimum operation point. Our analysis of data from the Mass Transit Railway, Hong Kong indicates the existence of concave \acrshort{FDs} at identified bottleneck stations, and an associated critical level of boarding-alightings above which congestion sets-in unless there is an intervention.

\end{abstract}
 
\end{frontmatter}

\bigskip

\noindent \textbf{Declarations of interest}

\noindent None.

\bigskip

\noindent \textbf{Funding}

\noindent This research did not receive any specific grant from funding agencies in the public, commercial, or not-for-profit sectors.

\newpage

\section{Introduction}

Efficient urban rail transit systems or metro systems are a key element of sustainable transportation in most cities worldwide. London's rail-based transit services, for instance, carried approximately 6.0 million daily trips in 2019, comprising more than 22\% and 35\% of the overall mode share and the sustainable mode share of daily trips, respectively \citep{TfL2020}. Metro services are attractive to passengers because of their high travel speeds, large passenger-carrying capacities and reliability.  Metro systems often operate at service frequencies approaching their capacity (that is,  maximum allowable frequency) to transport large volumes of passengers, especially during peak periods. Such operations render the system sensitive to service disruptions from technical failures and incidents, extensively analysed in studies such as \citep{Silva2015}. However, as peak-hour ridership continues to grow unprecedentedly, metro operations are also disrupted by congestion and frequent scheduling delays on a day-to-day basis, that have substantial socioeconomic implications \citep{Tirachini2013, Seo2017}. For instance, the London Underground reported 504 congestion-related delays of two minutes or more in 2018 \citep{LA2019}, and passengers lost almost 400,000 hours due to these delays \citep{Ind2017}. Moreover, such disruptions can potentially lead to a shift in demand towards alternative and potentially unsustainable travel modes, which may also precipitate a financial shortfall for metro agencies. 

Congestion in metros can be classified into two main categories: (1) passenger-congestion due to longer boarding and alighting times, and (2) train-congestion due to queuing and reduction in train velocity. High numbers of passenger boardings and alightings may lead to substantial increases in stopping or dwell times of trains at stations, which gives rise to passenger-congestion at stations \citep{Seo2017}. As transit systems are operating at  high, often near-capacity service  frequencies, increased  and irregular dwell  times  of  trains  may  eventually disrupt  service  frequencies due  to  queuing  of  trains. This queuing phenomenon is referred to as train-congestion or knock-on-delays \citep{Carey1994}. Since the headway of train arrivals at stations increase as a result of train-congestion, passenger-congestion at stations intensifies due to further accumulation of passengers on the platform \citep{Seo2017, Keiji2015, Daganzo2009}. Thus, passenger-congestion and train-congestion develop into a vicious cycle, and passenger-congestion is generally the root cause of this phenomenon \citep{Seo2017,Daganzo2009,Zhang2019}. The stations where passenger-congestion arises can be characterised as active bottlenecks in the transit network. Based on network configuration and operational attributes, congestion may spread from these bottleneck stations to other parts of the network, resulting in larger overall delays and degradation in system-wide performance. Reduction in recurrent congestion requires control strategies, such as controlling the inflow of passengers entering bottleneck stations.  The overarching aim of this study is to develop an appropriate data-driven model of passenger-congestion; the nucleus of congestion-related delays in metro networks; that can inform optimal control strategies.

\subsection{Overview of the analysis}

Congestion in a given facility can be described using a model representing the variation in performance of the facility over its intensity of use \citep{Small2007,Daganzo1997}. We propose to model passenger-congestion via \textit{novel station-level causal relationships} between the total number of passenger boardings and alightings per train and train flows, signifying usage and performance, respectively. Based on the station-level congestion phenomenon introduced above, we expect our model to capture the following causal process: Before the outset of passenger-congestion, we hypothesise that the relationship between train-flow and passenger boarding-alightings is bi-directional (that is, both variables are simultaneously determined) or could even be independent of each other. However, once passenger-congestion sets in, that is, when station-level operations reach a critical level, passenger boarding-alightings becomes the primary causal variable because it increases the dwell-times of trains at stations and consequently, reduces the train flow through the station. 

Our objective is to determine the underlying station-level curve that represents equilibrium interactions between passenger boarding-alightings and train flow in both uncongested and congested conditions. We refer to these curves as the station-level \acrfull{FDs}, and aim to estimate them using \acrfull{AFC} and \acrfull{AVL} data recorded by metro systems. However, the bi-directionality and change of causality in different segments of the relationship makes the task at hand empirically challenging because they are potential sources of confounding biases in the estimation. Another major empirical challenge emerges from the interdependence of operations at different stations in the network, particularly those along the same line. This interdependence may cause station-level observations to be confounded by operations at other stations in the network, in particular, obscuring the points in the network that are the true source of congestion.  Moreover, the functional form of the \acrshort{FD} is not known a priori and may vary from one station to another. To adjust for such statistical biases in obtaining robust estimates of the station-level \acrshort{FD} using data on station-level metro operations, we adopt a causal statistical modelling approach. Specifically, we adopt a Bayesian \acrfull{NPIV} approach, proposed by \cite{Wisenfarth2014}, that adjusts for such confounding biases and does not require parametric assumptions on the functional form of the fundamental relationship.  We thus aim to deliver reproducible \acrshort{FDs} that are robust to potential confounding biases and as such suitable to properly inform congestion control strategies. The proposed model is analogous to the traffic \acrshort{FD} for a road section, where estimation of the \acrshort{FD} can inform optimal traffic control strategies, for instance, finding the optimum operation point for the section. 

The estimated \acrshort{FD} can be viewed as an inherent property of the station, representing a reproducible supply side station-level relationship between passenger boarding-alightings and train flow. We argue that the characteristics determining the particular form of the observed station-level \acrshort{FD} comes from station-level characteristics (for instance, station design), although some also comes from the \emph{quality} of usage (for instance, percentage of slow and fast moving passengers). What does not enter in the determination of the underlying curve is the quantity of usage (that is, train flow or passenger boarding-alightings). Thus, this property exists even in the absence of passengers and a particular amount of equilibrium interaction simply determines a point on this curve. 

One of the primary objectives of this analysis is to identify potential bottlenecks in the metro network and inform the critical or optimal operation point at such stations. As discussed earlier,  passenger-congestion arises particularly at bottleneck stations due to excessive passenger boarding-alightings. Thus for such stations, we expect the estimated \acrshort{FDs} to comprise a congested part beyond a \textit{critical passenger boarding-alighting} level above which train flow or throughput of the station reduces. For all other stations, we do not expect the congested limb. Our analysis of relevant data from \acrfull{MTR} in Hong Kong suggests the existence of a concave-shaped \acrshort{FD} at each bottleneck station, with a well-defined optimal boarding-alighting level where throughput of the station is maximum. We emphasise that these estimates of critical passenger boarding-alightings would be instrumental for metro operators in developing data-driven control strategies to avoid congestion delays.

Prior to the empirical analysis, we present a brief simulation study of a synthetic metro system to provide an intuitive depiction of the vicious circle of passenger-train-congestion and the importance of estimating optimal passenger boarding-alightings in developing station-level control strategies. While the main focus of this study remains the station-level empirical analysis, the objective of the simulation study is to illustrate the implications of our empirical analysis in practice.

\subsection{Relevance}

\begin{figure}[h!]
  \centering
  \includegraphics[width= 0.5\linewidth]{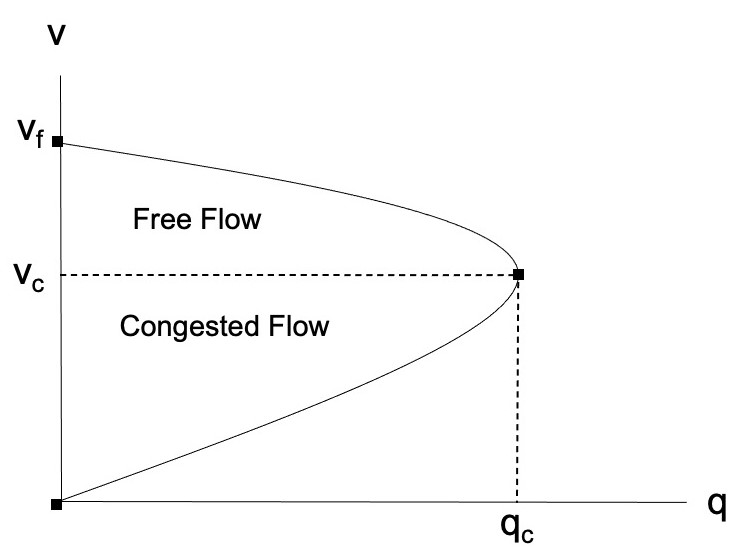}
  \caption{The fundamental diagram of traffic flow.}
  \label{fig:traffic_FD}
\end{figure}

To understand the relevance of the estimated \acrshort{FDs} and optimum passenger boarding-alightings from a conceptual perspective, we briefly discuss the analogous concept of \acrshort{FDs} from traffic flow theory. Estimating the \acrshort{FD} of traffic flow for a road section (see Figure \ref{fig:traffic_FD}) and understanding the optimal operation point for the section has been the main focus in the modelling of traffic flow \citep{Loder2019,Geroliminis2008,Small2007,Daganzo1997}. According to Figure \ref{fig:traffic_FD}, the increase of flow on a road section, that is, the average number of vehicles passing the section per unit time, causes average speed to fall below the free flow speed ($v_f$), until maximum or optimal flow ($q_c$) is eventually achieved on the link at critical speed $v_c$. Note that the relationship between speed and flow is double valued at this point (that is, it bends at $v_c$), distinguishing free-flow conditions from the phenomena known as congestion, in which both speed and flow decrease simultaneously due to dysfunctional driving behaviour arising from instabilities in congested flow. This can eventually result in gridlock with both speed and flow approaching zero. Traffic control and management strategies such as traffic signals and congestion pricing aim to regulate the demand such that the congested region of the curve dominates traffic conditions less frequently \citep{Amirgholy2017,Papageorgiou2003}. The existence of a well-defined \acrshort{FD} and the corresponding estimates of the optimal operation point over which congestion sets in and flow through the section falls have been crucial inputs to such strategies. 

Similar to the \acrshort{FD} for a road section, we illustrate the existence of well-defined station-level relationships between passenger boarding-alightings and train flow in metro networks. Furthermore, we find that there exists a unique critical level of passenger boarding-alightings at each bottleneck station over which the throughput of the station decreases. These estimates could be crucial inputs in the design of passenger inflow control strategies that are similar to vehicular control strategies in road traffic.

\section{Simulation of passenger-congestion and delays}

As a precursor to our empirical study, we reproduce the passenger-train-congestion phenomenon by simulating a typical high-frequency metro operation. The purpose of this simulation is to demonstrate the efficacy of station-level passenger inflow control measures in avoiding congestion-related delays. While the details of the train operation model \citep[adopted from][]{Yan2012} are attached in \sref{S:II}, we discuss the model parameters and results of the simulation exercise in this section. The model parameters relate to an Asian metro system with near capacity operations. We merge the train operation model with a train dwell time model proposed by \cite{Zhang2019} and \cite{Seo2017} to develop our simulation model. 

\subsection{Model parameters}

We consider a metro line of length $L=4000$ metres with three stations, namely Station 1, Station 2 and Station 3. These stations are located at positions 1000 metres, 2000 metres and 3000 metres respectively. The system is simulated for $T_s=3600$ seconds. Based on the characteristics of the Asian metro system, we assume that the velocity $v_{n,t}$ of a train in our system varies between 0 and 20 m/s, its acceleration is $a=1$ m/s\textsuperscript{2}, and its deceleration is $b=1$ m/s\textsuperscript{2} \citep{Yan2012}. We consider that the safety margin is $SM=50$ metres. We assume that the dwell time of trains at Station 1 $T_{d_1}$ and Station 2 $T_{d_2}$ is 30 seconds. We consider Station 3 as an active bottleneck station along the simulated metro line where the dwell time $T_{d_3}$ increases with increasing boarding-alightings per train $N_p$  (that is, the total boarding and alighting movements). Consistent with \cite{Zhang2019} and \cite{Seo2017}, we assume that $N_p = A_p  H$, where $A_p$ represents the passenger boarding-alighting rate and $H$ is the time headway of successive trains at Station 3. Following \cite{Zhang2019} and \cite{Keiji2015}, we adopt the following dwell time model for Station 3:

  \begin{equation*}
    T_{d_3} =
    \begin{cases}
      40\ \text{seconds}, & \text{if}\ (A_pH \leq N_o) \\
      40 + \gamma (A_p H - N_o)\ \text{seconds}, & \text{if}\ (A_p H > N_o)
    \end{cases}
  \end{equation*}

\noindent that is, $T_{d_3}$ remains constant until a critical passenger number $N_o$ is reached, following which it starts increasing. $\gamma$ represents the growth rate of dwell time with the increase in number of boarding-alightings. We consider $\gamma$ to be equal to 0.1. Moreover, we assume that the passenger boarding-alighting rate $A_p$ increases gradually from a value of 0 passenger per second to a maximum value of 10 passengers per second, with an increment of 0.005 passenger per second at each time step.

Furthermore, we assume that the interval $D$ of trains entering into metro system decreases by 5 seconds at every 2 minutes interval, starting from a value of 120 seconds until it attains a value of 60 seconds. This specification implies that with increasing boarding-alightings, the operator increases train departure frequency up to the maximum value or the capacity value beyond which further increase in frequency is not possible.

\subsection{Results of simulation}

\begin{figure}[h]
    \centering
        \begin{subfigure}{0.35\textwidth}
            \centering
            \includegraphics[width=1\textwidth]{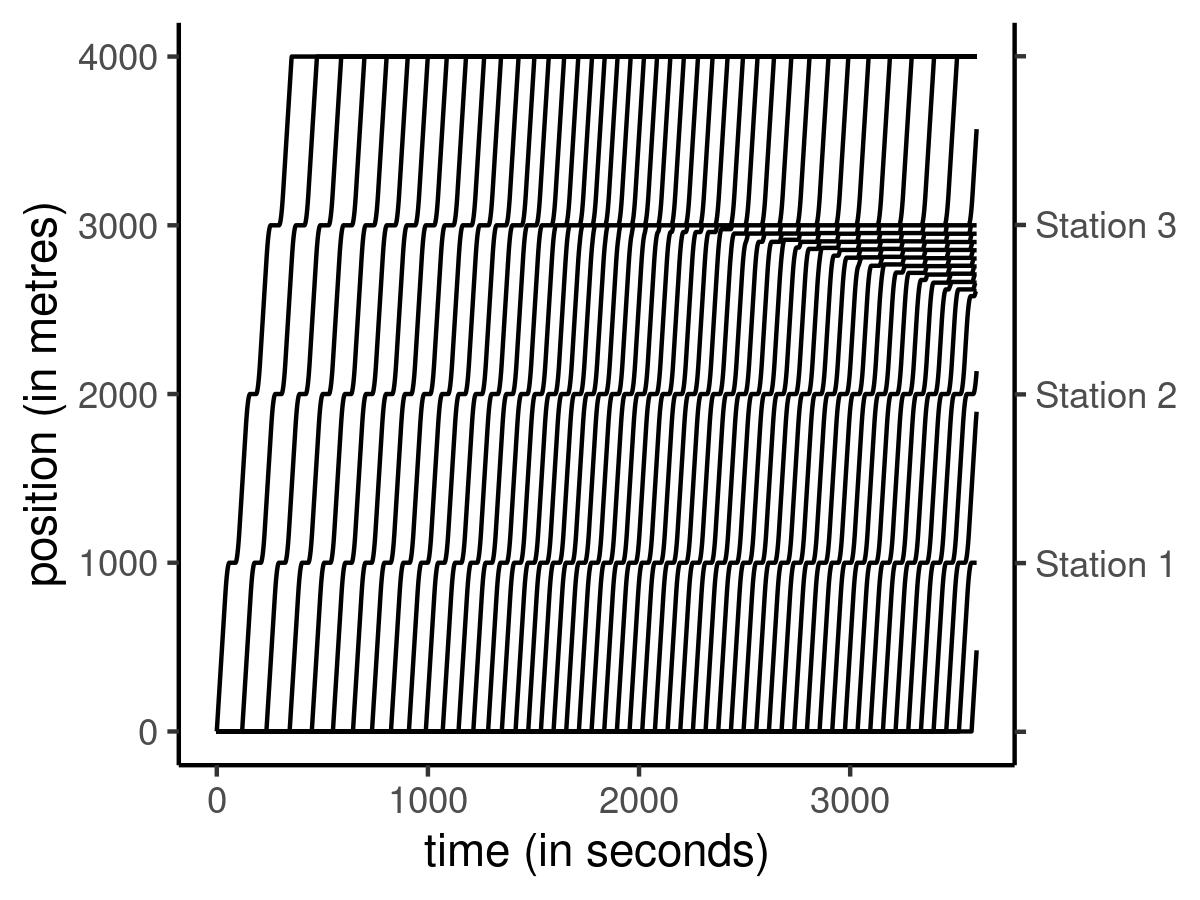}
            \caption{The time-space diagram representing train trajectories.}
            \label{fig:Sim_NoControl}
        \end{subfigure}%
        \begin{subfigure}{0.35\textwidth}
            \centering
            \includegraphics[width=1\textwidth]{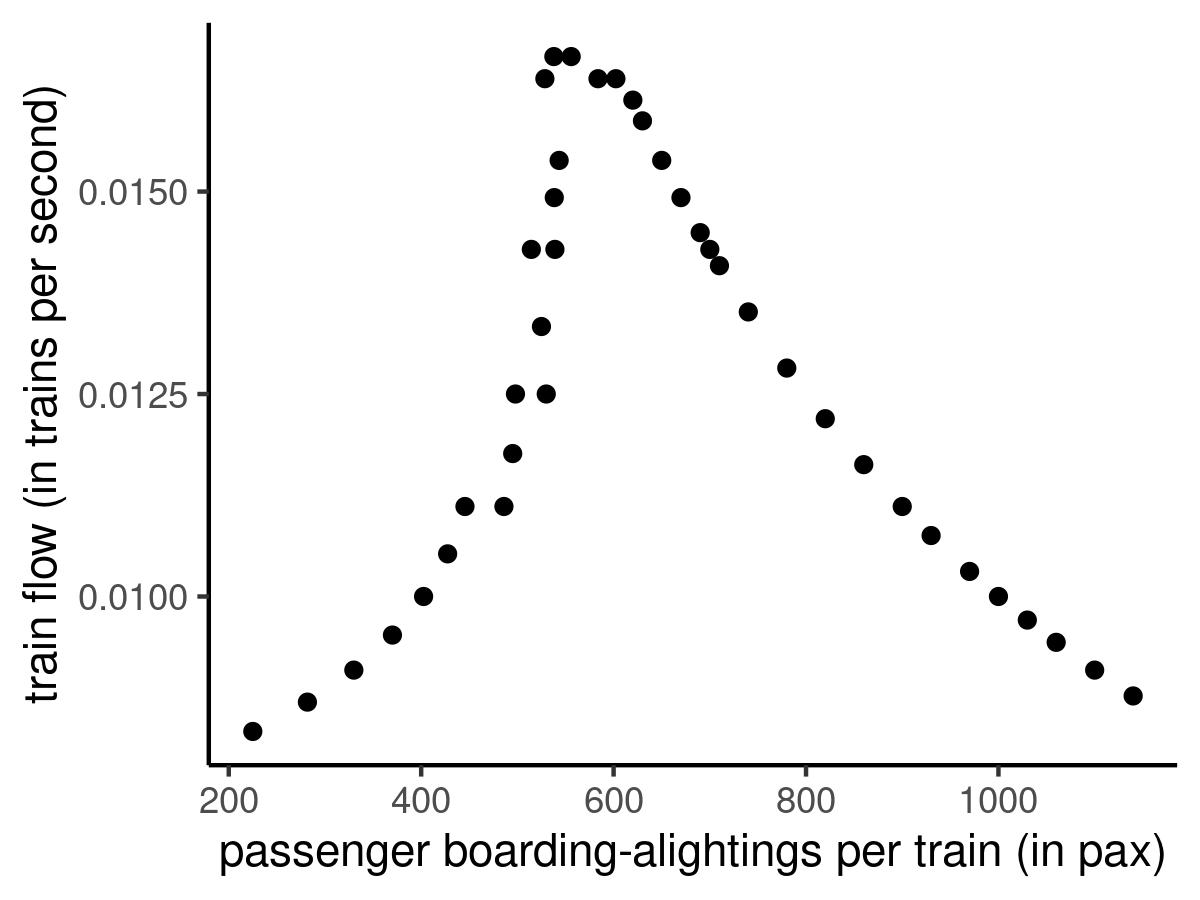}
            \caption{The fundamental diagram at Station 3.}
            \label{fig:QP}
        \end{subfigure}
    \caption{Train operations under no control scenario.}
\end{figure}

Figure \ref{fig:Sim_NoControl} shows a time-space diagram representing the train trajectories during the simulation period. From this figure, we note that longer station dwell time of trains at Station 3 eventually leads to queuing of trains or train-congestion in upstream of Station 3, starting at about $t=2100$ seconds. The queuing-related delays increase the time headway of arrivals of trains at Station 3, which leads to  increase in passenger boarding-alightings because $N_p=A_pH$. As a consequence, station dwell time at Station 3 increases further and queuing of trains in the upstream increases significantly. Based on the spacing of trajectories downstream of Station 3, we note that the throughput of the line first increases as a result of increase in train departure frequency. However, with increasing passenger-congestion and delays, this throughput decreases substantially. The number of trains that pass through the system is 39 in an hour.

Figure \ref{fig:QP} plots the equilibrium interactions between passenger boarding-alightings and train flow at Station 3, where train flow is obtained by taking the inverse of time headway of trains arriving at Station 3. We note from this figure that the underlying curve comprises of two parts: a non-congested segment (the upward sloping part) and a congested segment (the downward sloping part). The non-congested segment represents equilibrium states that result from interactions between increasing train flow (due to increase in train departure frequency) and increasing passenger boarding-alighting numbers (due to increase in demand). However, beyond a certain critical value of passenger boarding-alightings, train flow decreases with increasing passenger boarding-alightings due to high levels of passenger-train-congestion, leading to equilibrium states in the congested limb of the curve. The overall relationship is concave in shape. The maximum observed train flow is 0.0167 train per second or equivalently, 60 trains per hour, and the corresponding optimum level of passenger boarding-alightings per train is around 580 passengers per hour. Using $N_p=A_pH$, the optimum boarding-alighting rate turns out to be 9.65 passengers per second.

\begin{figure}[h]
    \centering
        \begin{subfigure}{0.35\textwidth}
            \centering
            \includegraphics[width=1\textwidth]{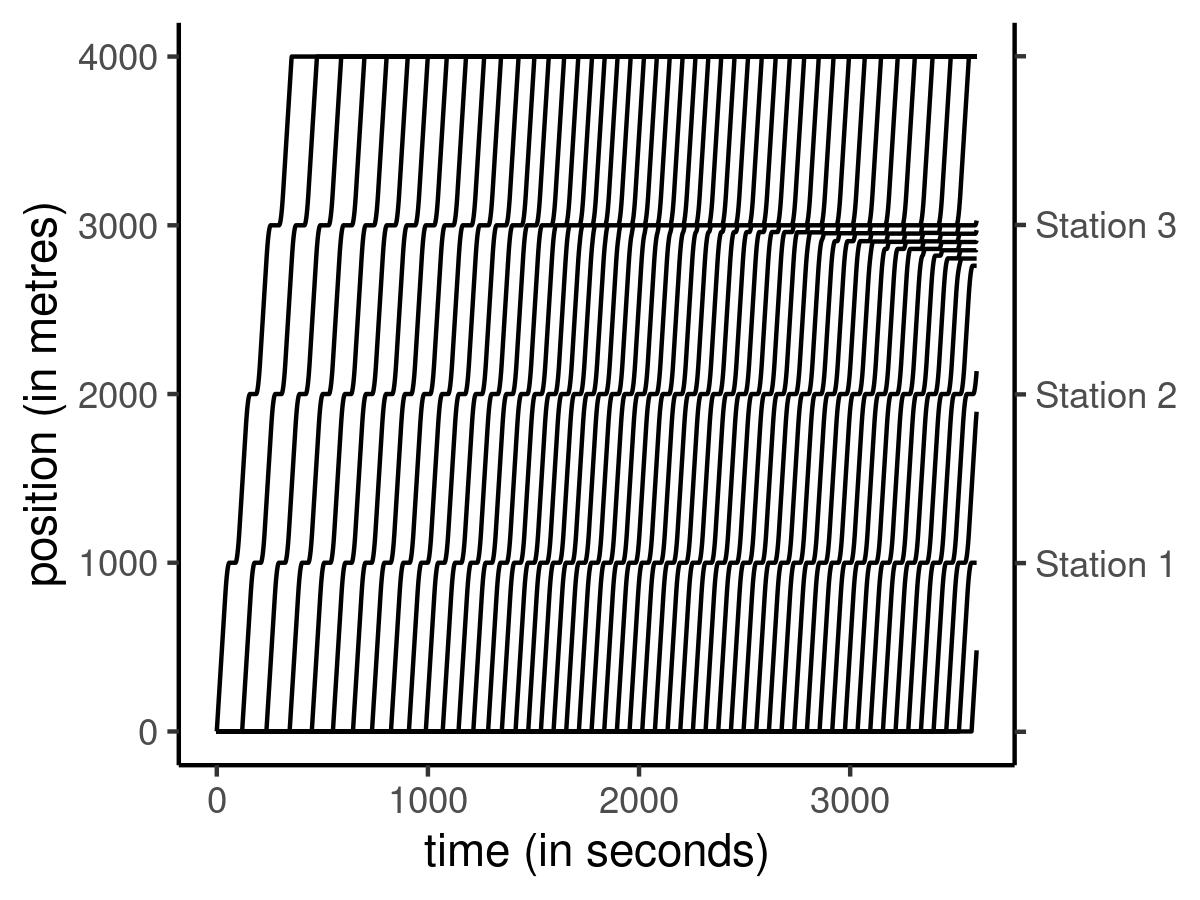}
            \caption{When maximum boarding-alighting rate at the bottleneck station is restricted at 9.75 passengers per second.}
            \label{fig:Sim_Control_a}
        \end{subfigure}%
        \begin{subfigure}{0.35\textwidth}
            \centering
            \includegraphics[width=1\textwidth]{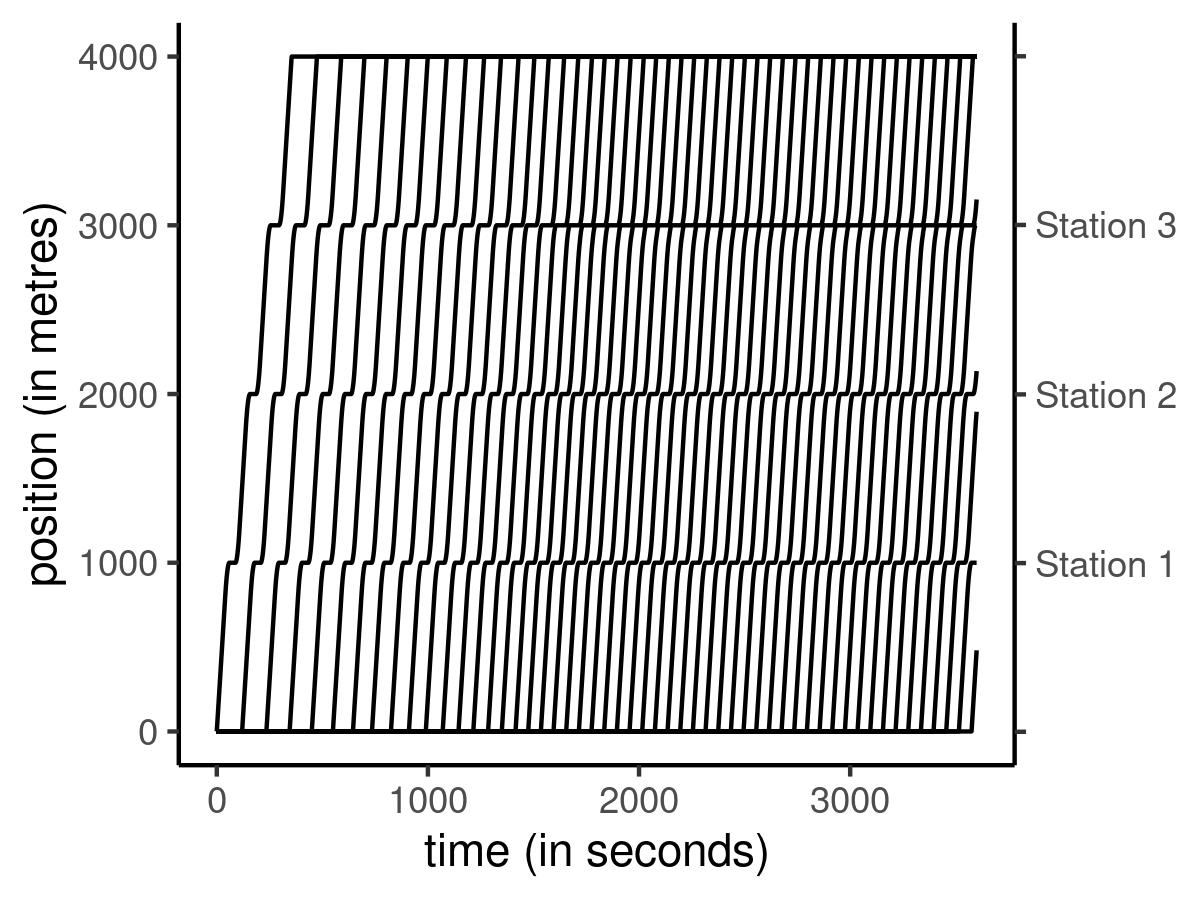}
            \caption{When maximum boarding-alighting rate at the bottleneck station is restricted at the optimal level of 9.65 passengers per second.}
            \label{fig:Sim_Control_b}
        \end{subfigure}
    \caption{The time-space diagrams representing train operations under passenger inflow control scenarios.}
    \label{fig:Sim_Control}
\end{figure}

We now consider two passenger inflow control scenarios: (i) when the boarding-alighting rate is restricted at 9.75 passengers per second (that is, slightly higher than the optimum rate), and (ii) when the boarding-alighting rate is restricted at the estimated optimal value of 9.65 passengers per second using station-level control strategies. Figure \ref{fig:Sim_Control} shows the time-space diagram for both scenarios. Although we observe lower train-congestion as compared to the \textit{no control} scenario (compare Figures \ref{fig:Sim_NoControl} and \ref{fig:Sim_Control_a}), train queuing and decrease in system throughput as compared to optimal are still substantial. The number of trains that pass through the system in the first scenario is 43 per hour. Interestingly, the queues are entirely eliminated, and the hourly system throughput increases to 47 trains in the optimal scenario (see Figure \ref{fig:Sim_Control_b}). 

The simulation study thus illustrates that maintaining the boarding-alighting per train at bottleneck stations below the critical value using station-level controls can be help avoid congestion-related delays and improve service reliability. In \sref{S:III}, we present a brief extension to this simulation study where we compare station-level inflow control strategies with headway-based control strategies that are often recommended in the literature on congestion management (see \sref{S:I} for a review). In the rest of this paper, we show how automated fare collection and train movement data can be used to empirically identify potential bottleneck stations in a metro network. Further, we discuss how the associated optimum passenger boarding-alightings can be estimated.

\section{Model and Data}

This section is divided into two subsections. The first subsection discusses the model specification and explains potential confounding bias in estimation of the station-level \acrshort{FDs}. This subsection also introduces the Bayesian \acrshort{NPIV} method briefly. This subsection is supplemented by \sref{S:IV} where we describe the Bayesian \acrshort{NPIV} method in the context of this study. In the second subsection, we describe the data and the relevant variables used in this analysis.

\subsection{Model Specification}

As discussed in the Introduction, we aim to estimate a \textit{causal relationship} between passenger boarding-alightings per train and train flows at each station. In other words, the objective of the empirical study is to estimate an equivalent of Figure \ref{fig:QP} (presented in the simulation exercise in the previous section).

We consider that the average train flow $q^s_{it}$ (that is, inverse of headway) at a station $s$ in the ten-minute interval $i$ on a particular day $t$ is a function of the average number of boarding and alighting movements per train $n^s_{it}$ occurring at the station in that interval: 

\begin{equation}
\label{eq:spec}
q^s_{it} = S(n^s_{it}) + \omega_{it} + \xi_{it} 
\end{equation}

\noindent where $\omega_{it}$ represents the unobserved interactions of the operations at station $s$  with other stations in the metro network or any unobserved station-specific operational attributes such as existing control measures adopted by station staff, $\xi_{it}$ is a idiosyncratic error term representing all random shocks to the dependent variable, and the unknown functional relationship of $n^s_{it}$ with $q^s_{it}$ is denoted by $S(.)$. Based on Figure \ref{fig:Sim_NoControl}(b), we can expect $S(.)$ to roughly concave and comprising of steps in the rising part of the curve that represent various regimes of the planned train frequencies. Adopting a parametric specification such as a quadratic function may be too restrictive to capture such non-linearities in the estimated relationship, and in any case, would make a strong a-priori assumption on functional form. Therefore, a non-parametric specification of $S(.)$ should be preferred.

To estimate equation \ref{eq:spec}, we adopt a \acrfull{NPIV} regression, which not only enables non-parametric specification of $S(.)$ but also addresses any potential confounding biases from bi-directionality or reverse causality (refer to the Introduction section for a detailed discussion) or from omitted variables represented by $\omega_{it}$. We discuss the implications of one such source of confounding bias on the estimated $S(.)$. Metro operators often adopt station-specific control measures to restrict passenger movements during peak hours so that the planned dwell times and headway of trains can be maintained. For instance, Transport for London often closes entrances/exits at various stations during peak hours to regulate passenger demand \citep{TfL2018}. Considering that $\omega_{it}$ represents these control measures, we expect a negative correlation between $\omega_{it}$ and $n^s_{it}$ and a positive correlation between $\omega_{it}$ and $q^s_{it}$. Moreover, $\omega_{it}$ also captures other unobserved effects such as increase in dwell times due to slow moving passengers. Note that this effect is expected to have a negative correlation with both $n^s_{it}$ and $q^s_{it}$. The unavailability of a measure for $\omega_{it}$ may lead to a confounding bias in the estimates of $S(.)$, commonly known as omitted variable bias. In particular,if $S(.)$ is a linear function, ordinary least squares estimation may underestimate or overestimate $S(.)$ (depending on the correlation of unobserved factors with the explanatory variable and the dependent variable) in the absence of a suitable measure or proxy for $\omega_{it}$. Therefore, it is imperative to adopt an appropriate statistical method that could adjust for such confounding biases. 

Classical (frequentist) \acrshort{NPIV} regression approaches are popular in theoretical econometrics \citep[such as,][]{Newey2003,Horowitz2011,Newey2013,Chetverikov2017}, but they are challenging to apply in practice due to two main reasons. First, tuning parameters to monitor the flexibility of $S(.)$ are often required to be specified by the analyst. Second, standard errors are generally computed using bootstrap, making these methods computationally prohibitive for large datasets. Therefore, we adopt a scalable \textit{Bayesian} \acrshort{NPIV} approach, proposed by \cite{Wisenfarth2014}, that can produce a consistent estimate of non-parametric $S(.)$, even if the analyst does not observe $\omega_{it}$. This Bayesian method addresses both challenges of the frequentist estimation because it \textit{learns} tuning parameters related to $S(.)$ during estimation and uncertainty in parameters estimates is inherently captured by credible intervals (analogous to classical confidence intervals). In addition, it also enables nonparametric specification of the unobserved error component $\xi_{it}$, precluding the need for making additional assumptions. In \sref{S:IV}, we describe the Bayesian \acrshort{NPIV} method in the context of this study. 

\subsection{Data and Relevant Variables}

We use the \acrfull{AFC} data or data from entry/exit gates at the stations and \acrfull{AVL} or train movement data provided by Hong Kong \acrfull{MTR}, the urban and suburban rail operator of Hong Kong and a member of the Community of Metros facilitated by the \acrfull{TSC} at Imperial College London. The \acrshort{MTR} dataset is practical for the present analysis because it is a closed system. All stations in the \acrshort{MTR} network are fenced, and thus, the AFC data contain information about all transactions at both the origin and destination stations. The data contain a record for millions of entry/exit transactions corresponding to trips occurring in the \acrshort{MTR} network over the period from January 1, 2019 to March 31, 2019. The \acrshort{AVL} data recovered from the signalling system contains a precise record of departure and arrival times of trains at each station in the \acrshort{MTR} network for the above mentioned period. We assign passengers to trains by matching \acrshort{AFC} data with \acrshort{AVL} data using the methodology detailed in \cite{Bansal2020} (an extension to \cite{Horcher2017}). 

\begin{figure}[tp]
  \centering
  \includegraphics[width= 0.7 \linewidth]{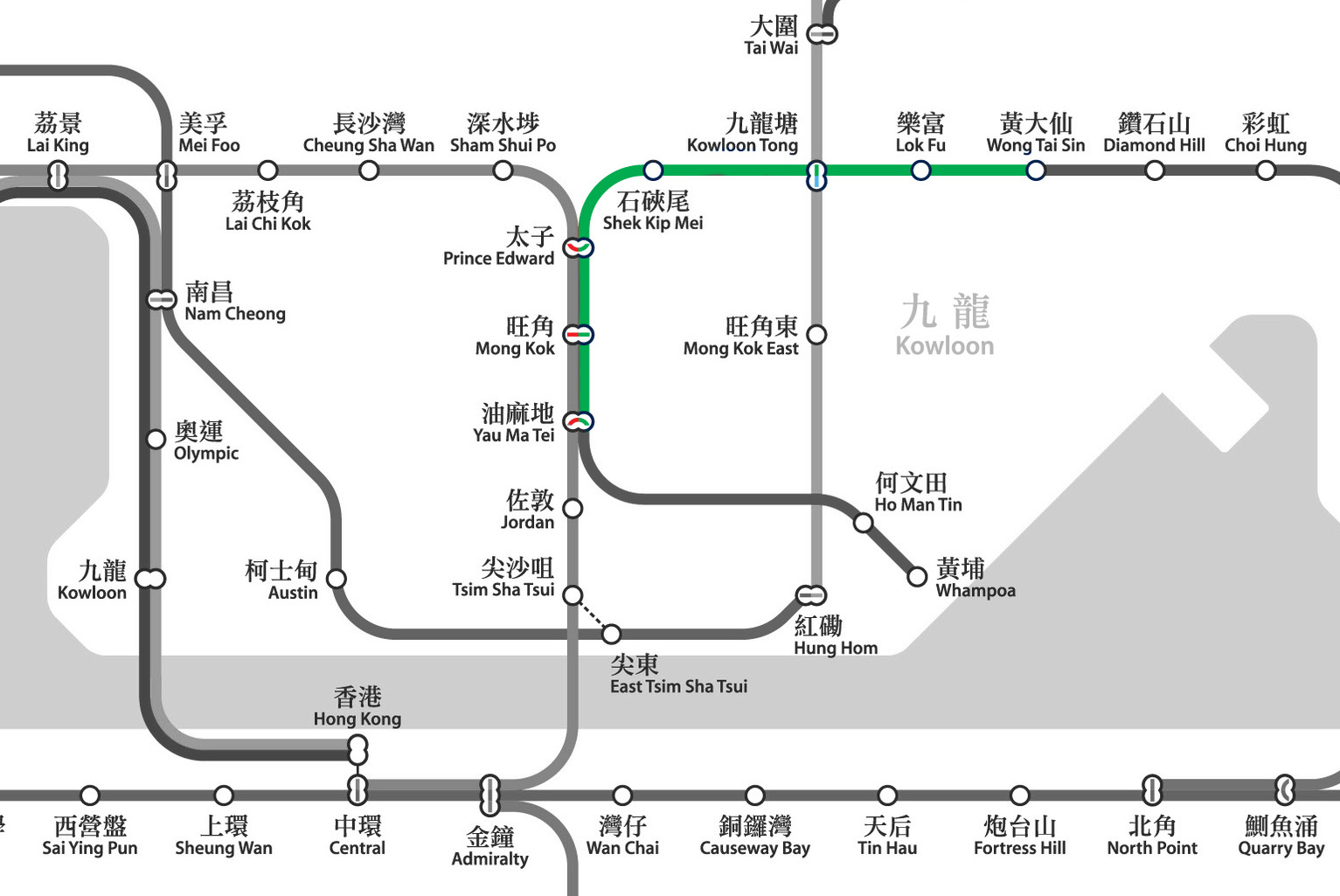}
  \caption{A part of the \acrshort{MTR} network where the line that we study is highlighted in green.}
  \label{fig:map}
\end{figure}

In this analysis, we focus on a group of stations on the Kwun Tong Line (green line) that are located in the central business district of Hong Kong. These stations are highlighted in Figure \ref{fig:map}. 

We analyse the trains moving in both downward and upward directions, that is, towards Whampao and Choi Hung stations respectively, on all work days. From the results of passenger-train assignment, we calculate train flow and average number of alightings and boardings per train or passenger boarding-alightings per train for different consecutive ten-minute time intervals between 5:30 hours to 11:30 hours and 15:30 hours to 21:30 hours for each station. The chosen hours comprise morning and afternoon peak hours and peak shoulder hours. Note that train flow in a ten-minute interval for a station is obtained by taking the mean of the inverse of time headway of trains arriving at that station within that interval.

\stref{tab:sumstats} presents the summary statistics for ten-minute train flows and average number of boardings and alightings at each station. We also provide the observed scatter plots of train flow versus passenger boarding-alightings per train for the considered stations in Figures \ref{fig:Res_Full_D} and \ref{fig:Res_Full_U}. The observed scatter in these diagrams can be attributed to differences in operational conditions and characteristics of demand, among other factors. We aim to uncover a reproducible relationship that may exist between passenger boarding-alightings per train and train flow by adjusting for such unobserved characteristics and noise in the data.

\section{Results}

This section is divided into two subsections. In the first subsection, we compare results from our \acrshort{IV}-based estimator with those from a non-\acrshort{IV} estimator. The non-\acrshort{IV} estimator is a counterpart of the Bayesian \acrshort{NPIV}, which does not address confounding bias (that is, $z=n; \epsilon_1 = 0; h: \textrm{identity function}$ in \seref{eq:NPIV1}). We conclude this section by describing Bayesian \acrshort{NPIV} results in detail and discussing how we identify the optimal boarding-alighting at the bottleneck stations. 

\subsection{Comparison of \acrshort{IV}-based and non-\acrshort{IV}-based estimators}

We compare the estimates of $S(.)$ in \seref{eq:NPIV1} (second-stage), which we obtain using the Bayesian \acrshort{NPIV}, and its non-\acrshort{IV}-based counterpart, for all analysed stations for train flows in both downward and upward direction as shown in Figures \ref{fig:Res_Full_D} and \ref{fig:Res_Full_U}. 

For all stations, the non-\acrshort{IV} estimator estimates approximately concave-shaped curves with a statistically significant downward-sloping part where train flow decreases with increase in passenger boarding-alightings. However, barring a few stations that we identify as bottlenecks (see the next subsection for details), the downward-sloping part either disappears or becomes statistically insignificant with use of the \acrshort{IV} estimator. As discussed in Model and Data Section, the downward-sloping part in the non-\acrshort{IV} estimates may not result from operations within the station but could be indicative of operational influences from downstream stations. For instance, for downward direction of train flow, the non-\acrshort{IV} estimator yields a curve with a statistically significant downward-sloping branch for Ship Mei Kei Station, although at very low levels of passenger boarding-alightings, that is, $\sim 200$. We argue that this downward-sloping segment could have resulted from congestion at Prince Edward Station located just downstream of Ship Mei Kei station, which may have caused queuing of trains upstream of Prince Edward Station. Upon use of the \acrshort{IV}-estimator, we do not obtain the downward-sloping segment in the estimated curve for Ship Mei Kei station. Therefore, we conclude that the results from the non-\acrshort{IV}-based estimator may be non-representative of the operations within any station itself as the estimator fails to control for unobserved interactions between various stations in the metro network, thus revealing ambiguous sources of congestion in the network. The confounding bias may be more severe for more congested metro operations such as the London underground where queuing of trains occurs is more frequent during peak hours. The advantages of adopting \acrshort{IV}-based estimators would be even more apparent in such cases.

\subsection{Bottlenecks and station-level optimal passenger boarding-alightings}

\begin{figure}[tp]
    \centering
    \begin{subfigure}{0.32\textwidth}
        \includegraphics[width=\linewidth]{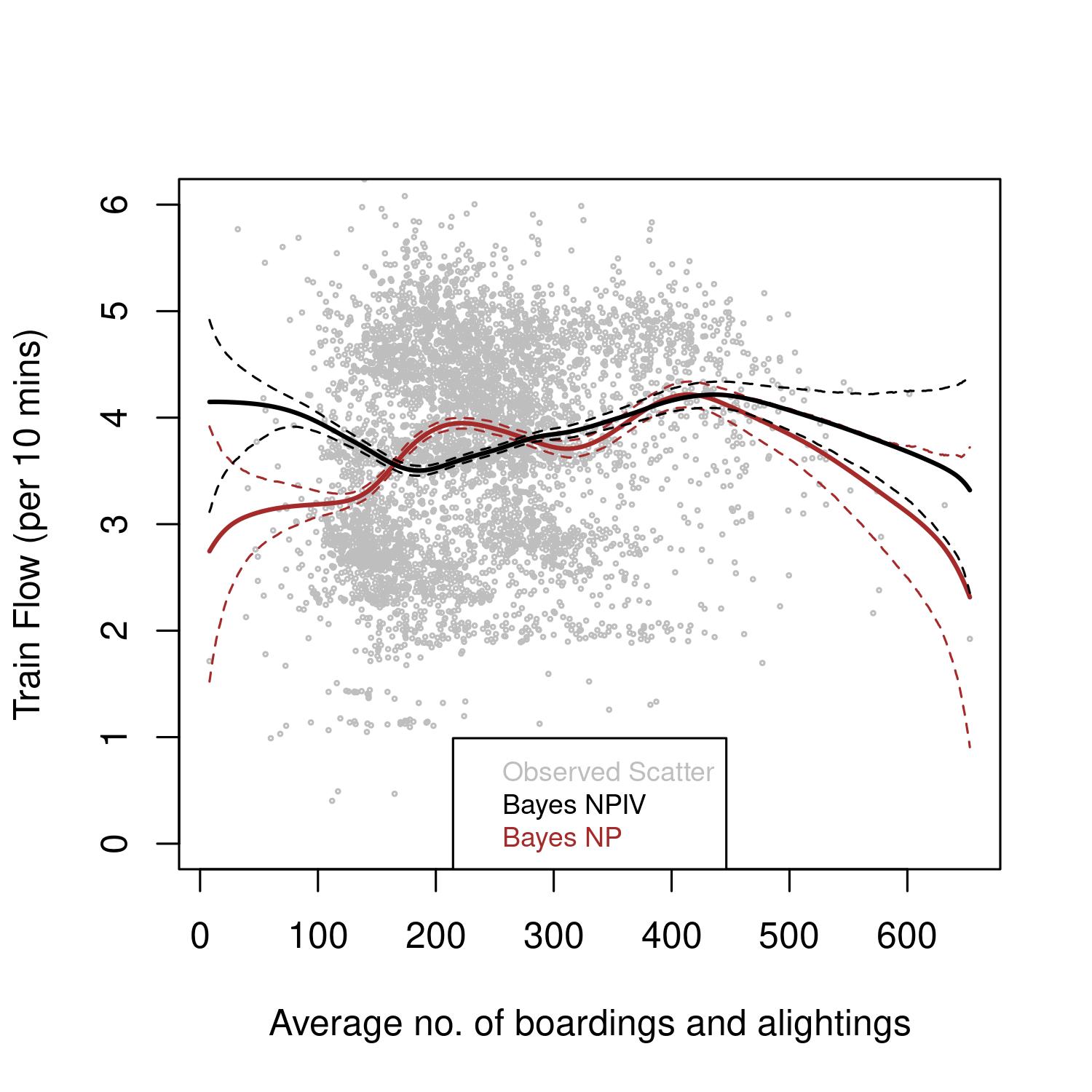}
        \subcaption{Wong Tai Sin Station}
    \end{subfigure}
\hfill
    \begin{subfigure}{0.32\textwidth}
        \includegraphics[width=\linewidth]{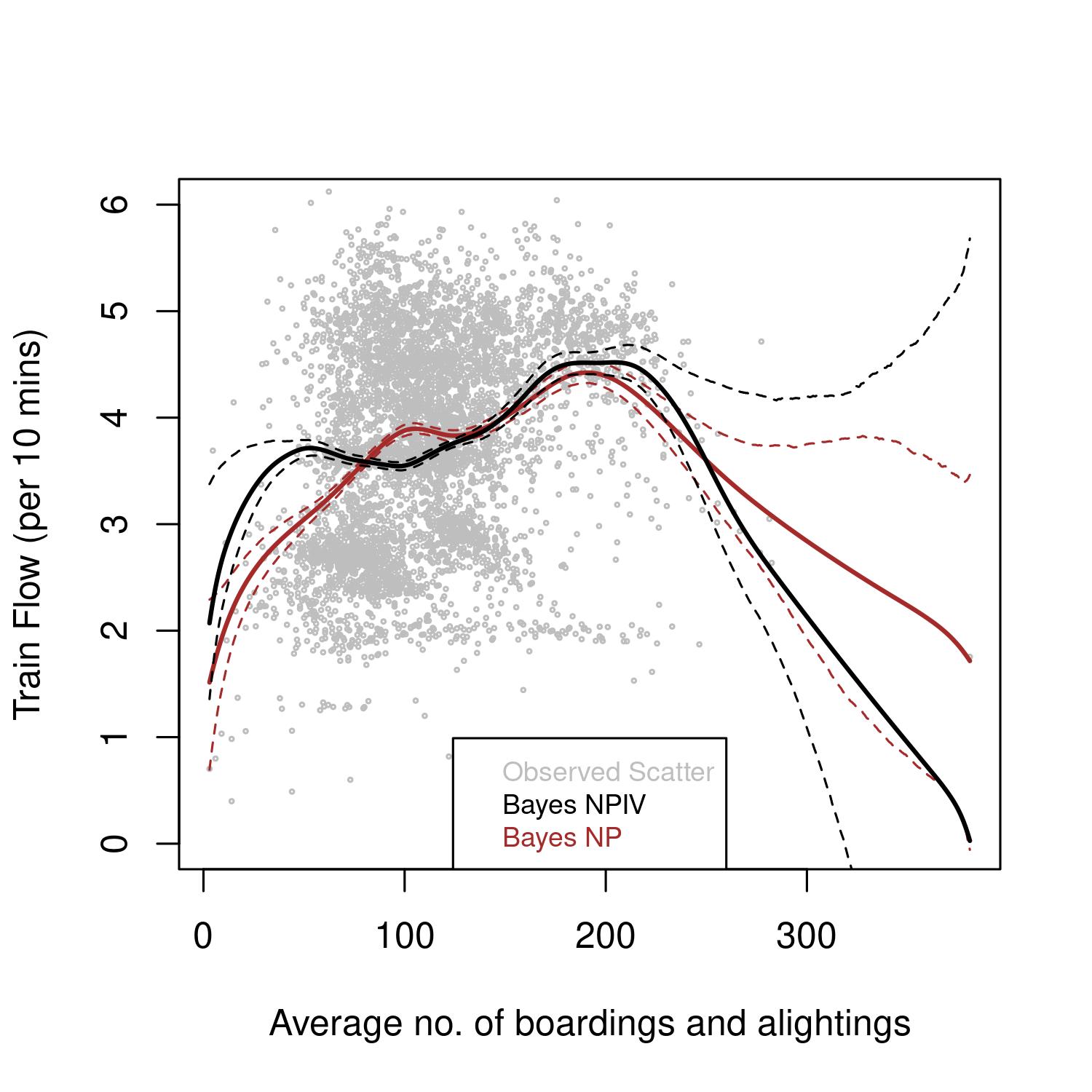}
        \subcaption{Lok Fu Station}
    \end{subfigure}
\hfill
    \begin{subfigure}{0.32\textwidth}
        \includegraphics[width=\linewidth]{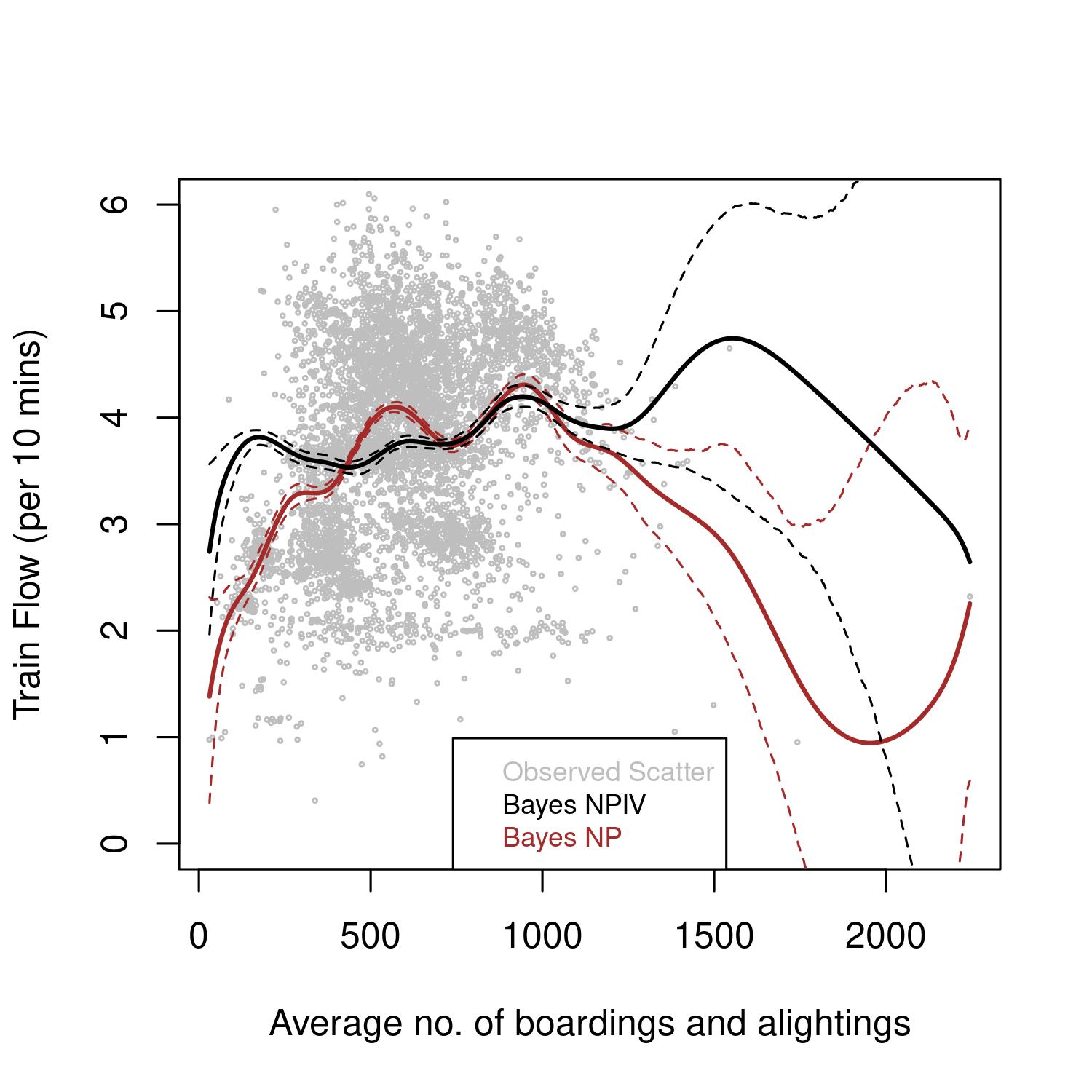}
        \subcaption{Kowloon Tong Station}
    \end{subfigure}
\hfill
    \begin{subfigure}{0.32\textwidth}
        \includegraphics[width=\linewidth]{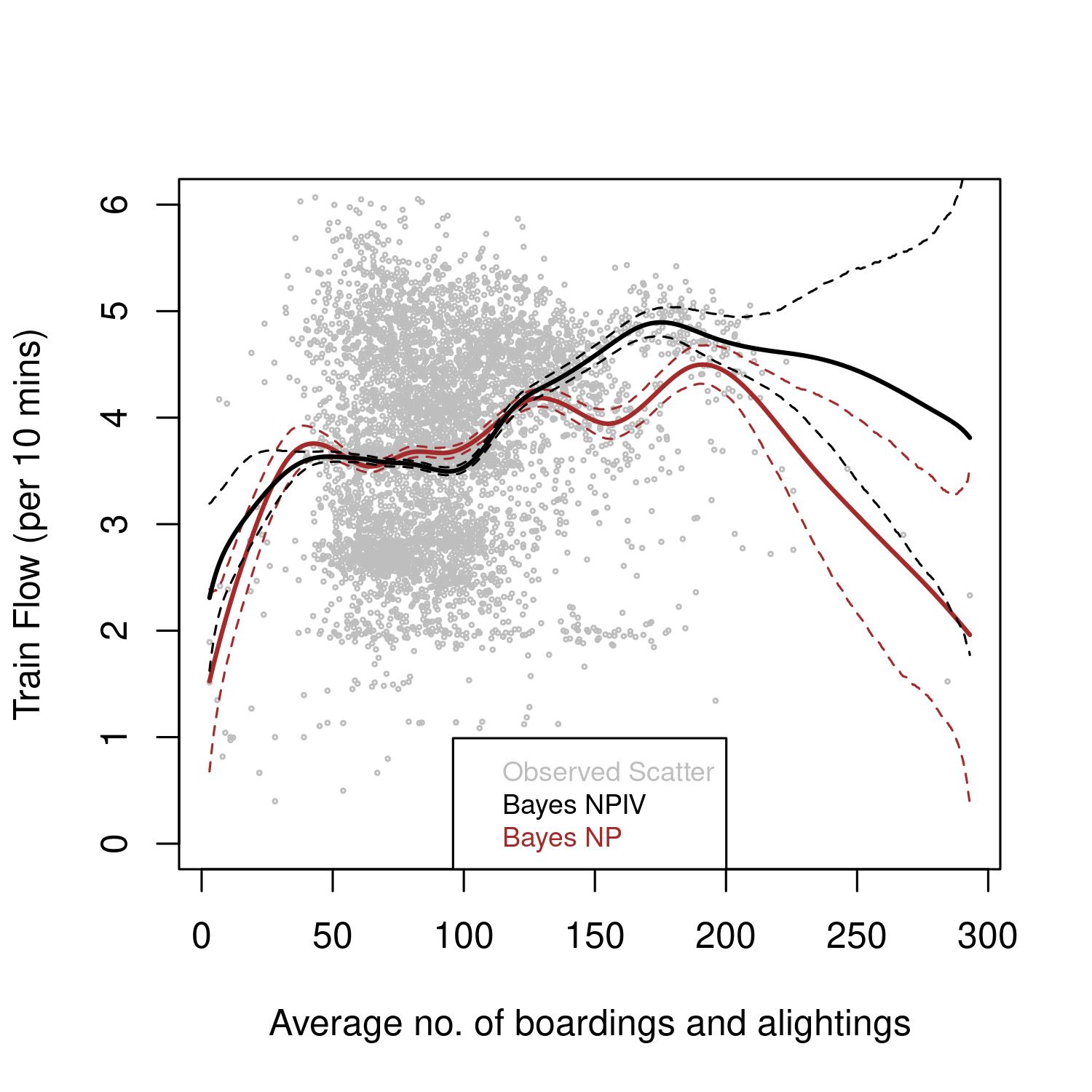}
        \subcaption{Shek Kip Mei Station}
    \end{subfigure}
\hfill
    \begin{subfigure}{0.32\textwidth}
        \includegraphics[width=\linewidth]{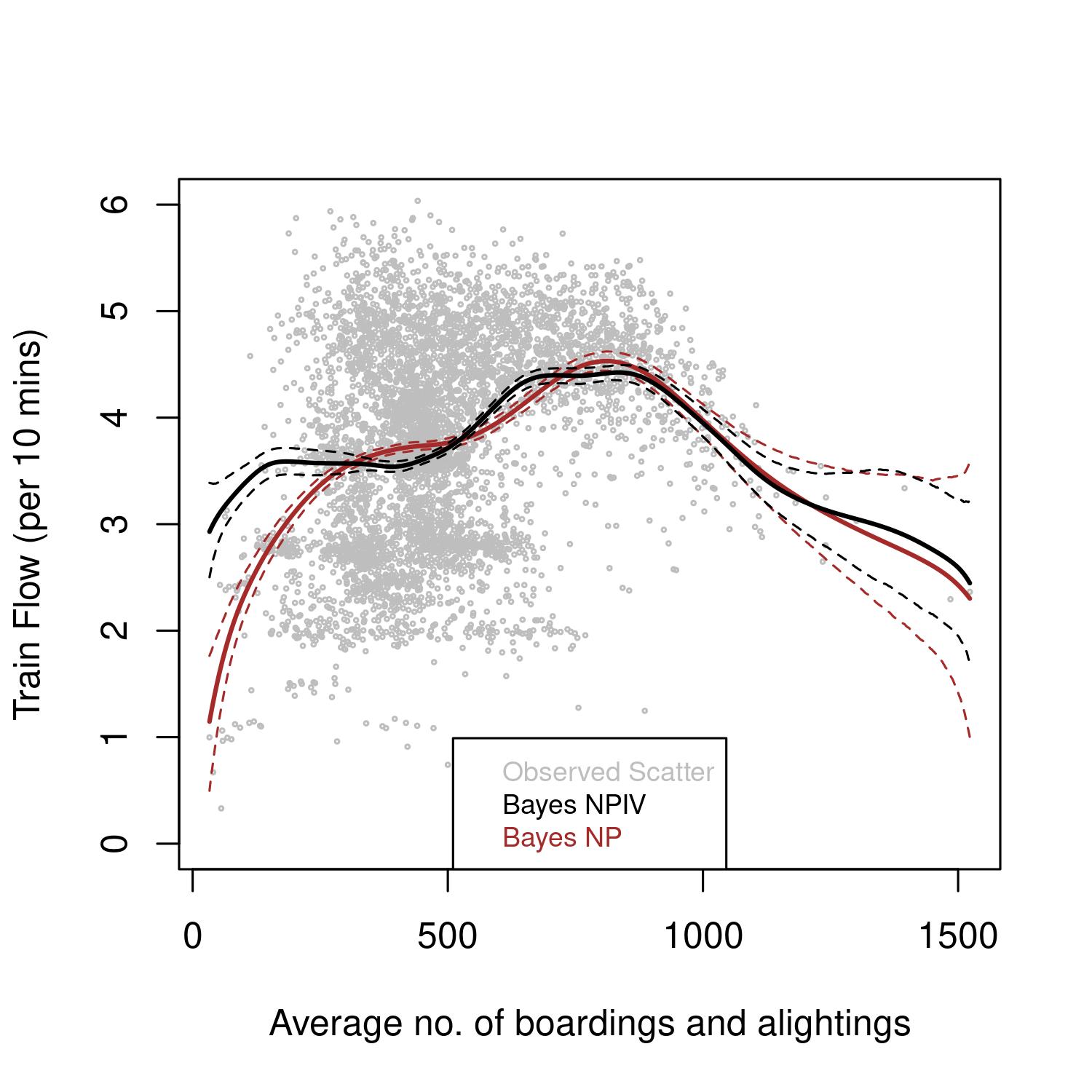}
        \subcaption{Prince Edward Station}
    \end{subfigure}
\hfill
    \begin{subfigure}{0.32\textwidth}
        \includegraphics[width=\linewidth]{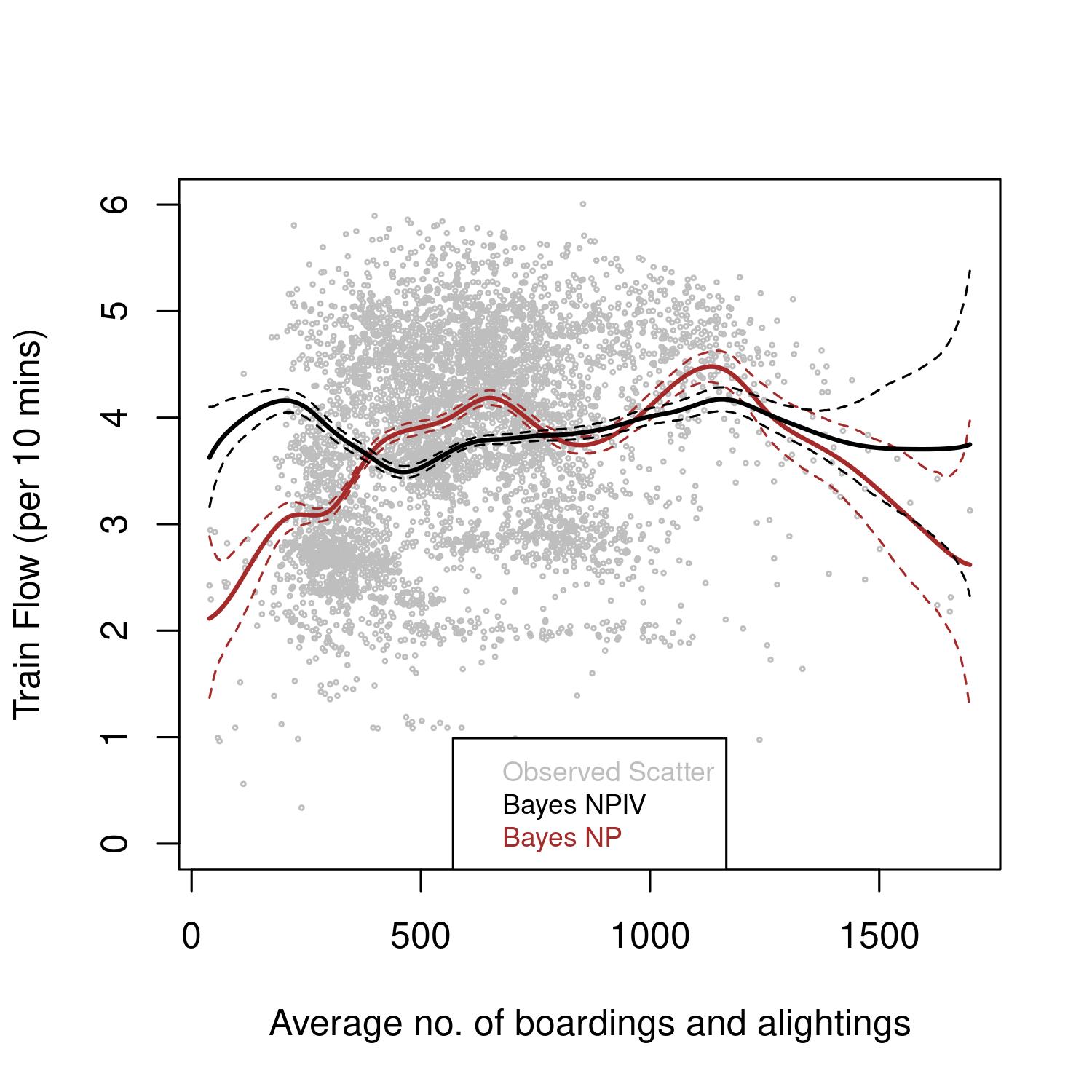}
        \subcaption{Mong Kok Station}
    \end{subfigure}
\hfill
    \begin{subfigure}{0.32\textwidth}
        \includegraphics[width=\linewidth]{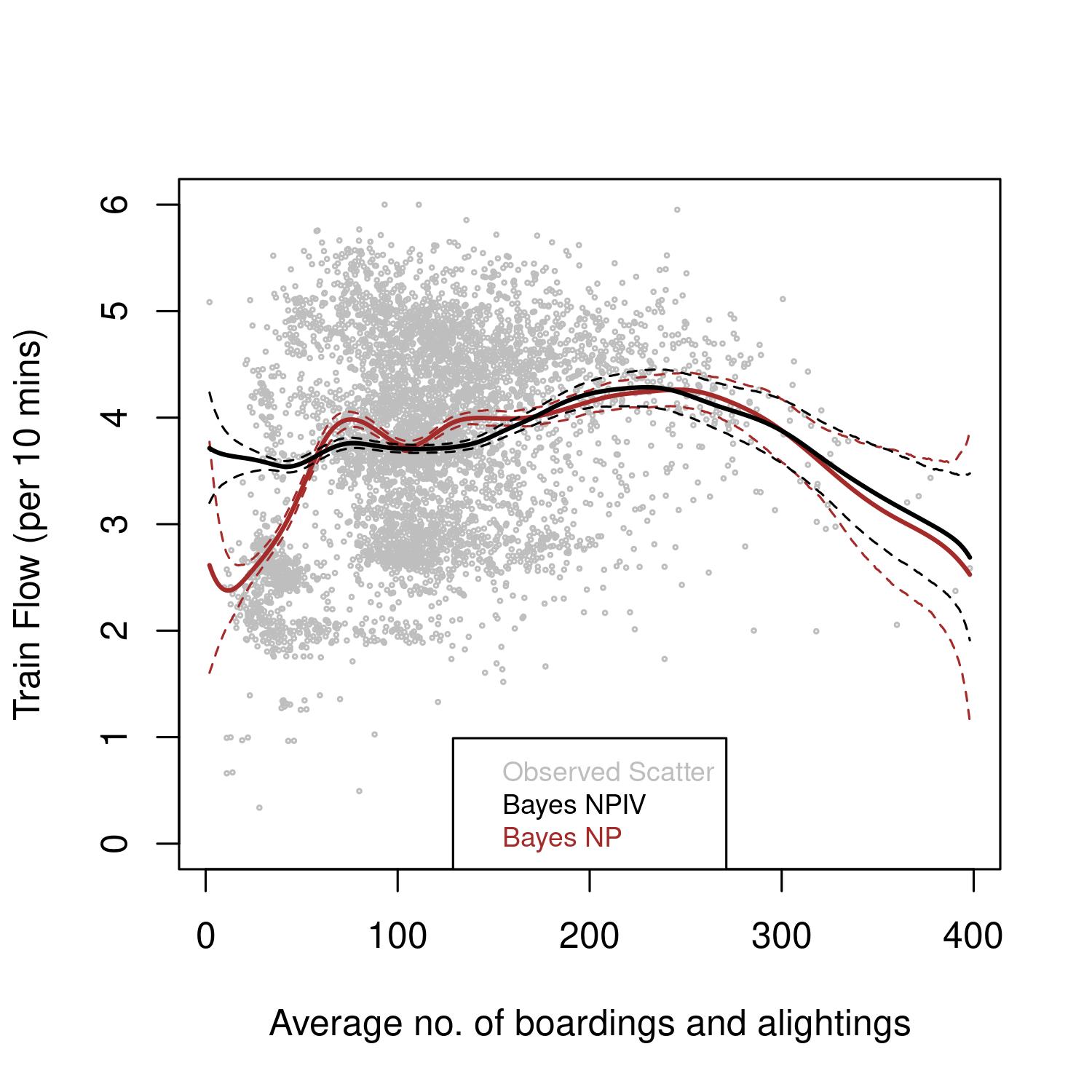}
        \subcaption{Yau Ma Tei station}
    \end{subfigure}
    \caption{Non-parametric Instrumental Variables based estimation results for train movements in the downward direction along the Kwun Tong Line (dotted lines represent 95\% credible intervals).}
    \label{fig:Res_Full_D}
\end{figure}

\begin{figure}[tp]
    \centering
    \begin{subfigure}{0.32\textwidth}
        \includegraphics[width=\linewidth]{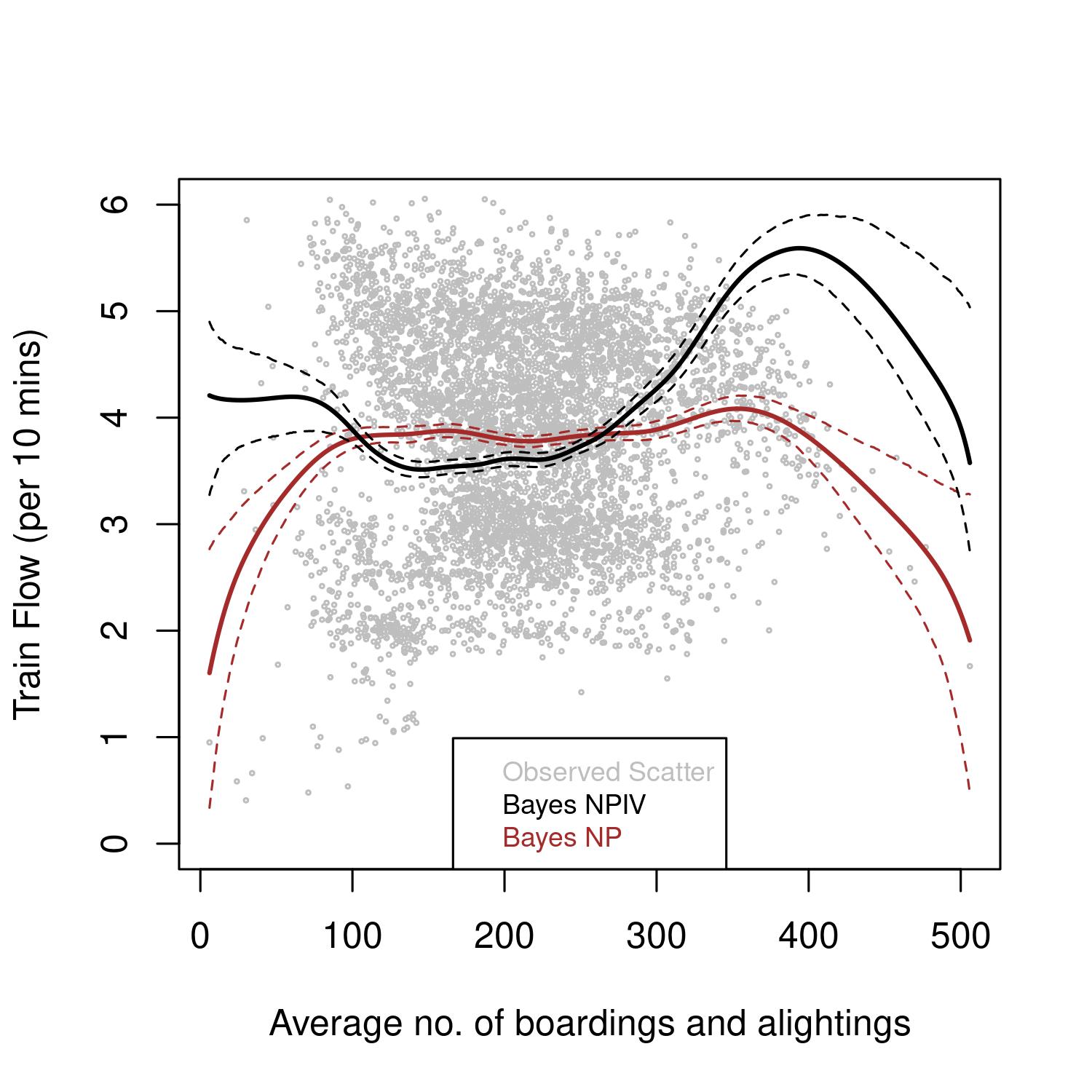}
        \subcaption{Wong Tai Sin Station}
    \end{subfigure}
\hfill
    \begin{subfigure}{0.32\textwidth}
        \includegraphics[width=\linewidth]{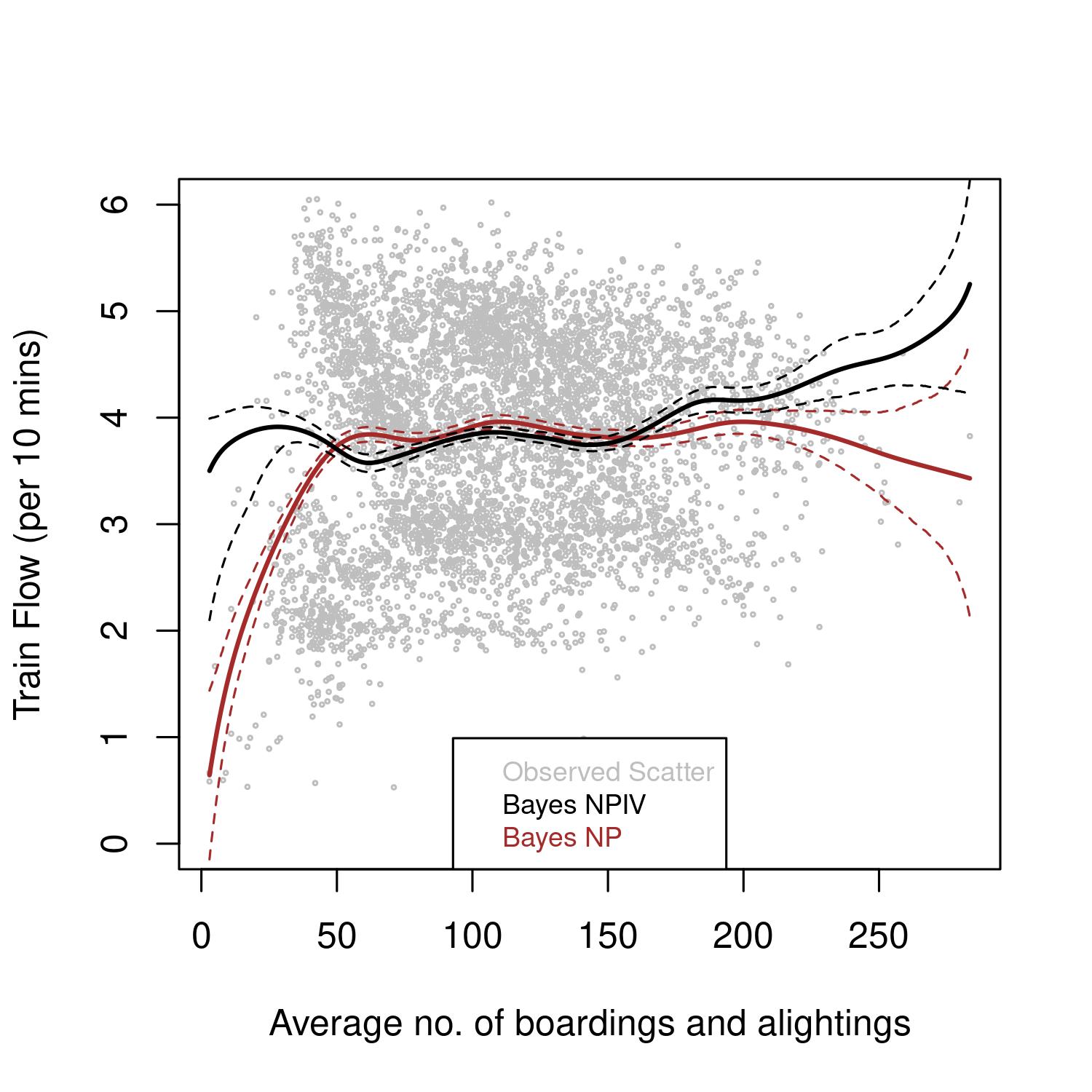}
        \subcaption{Lok Fu Station}
    \end{subfigure}
\hfill
    \begin{subfigure}{0.32\textwidth}
        \includegraphics[width=\linewidth]{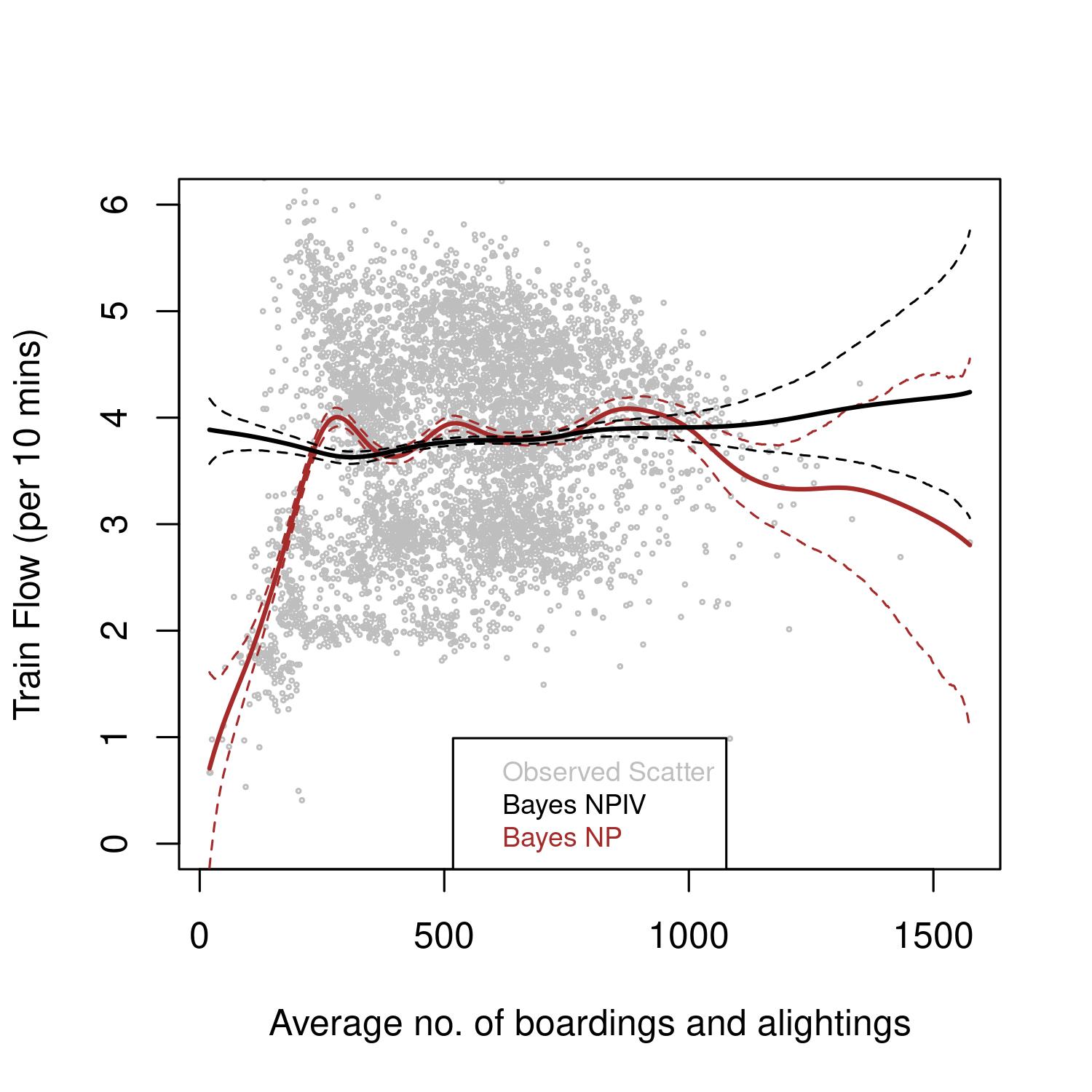}
        \subcaption{Kowloon Tong Station}
    \end{subfigure}
\hfill
    \begin{subfigure}{0.32\textwidth}
        \includegraphics[width=\linewidth]{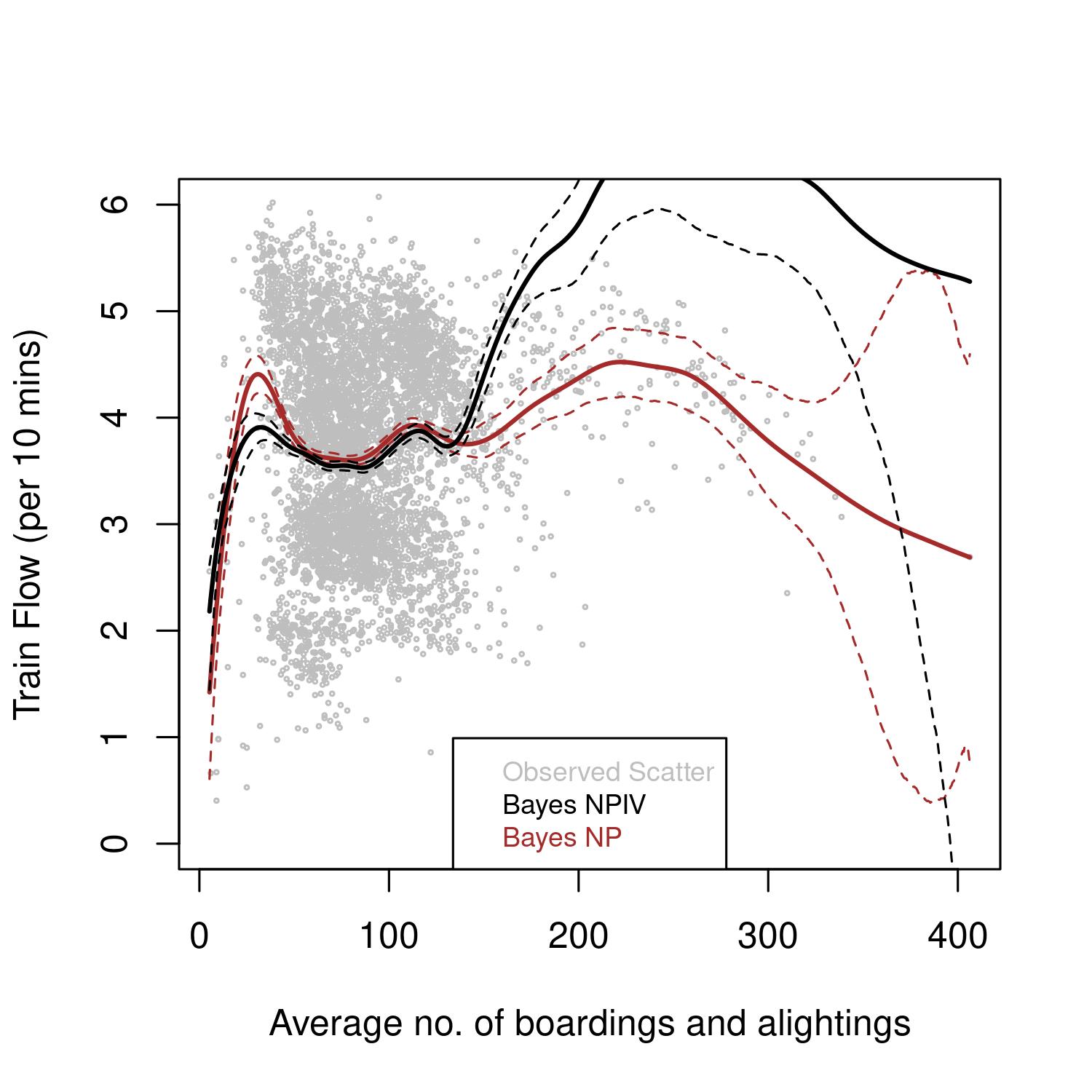}
        \subcaption{Shek Kip Mei Station}
    \end{subfigure}
\hfill
    \begin{subfigure}{0.32\textwidth}
        \includegraphics[width=\linewidth]{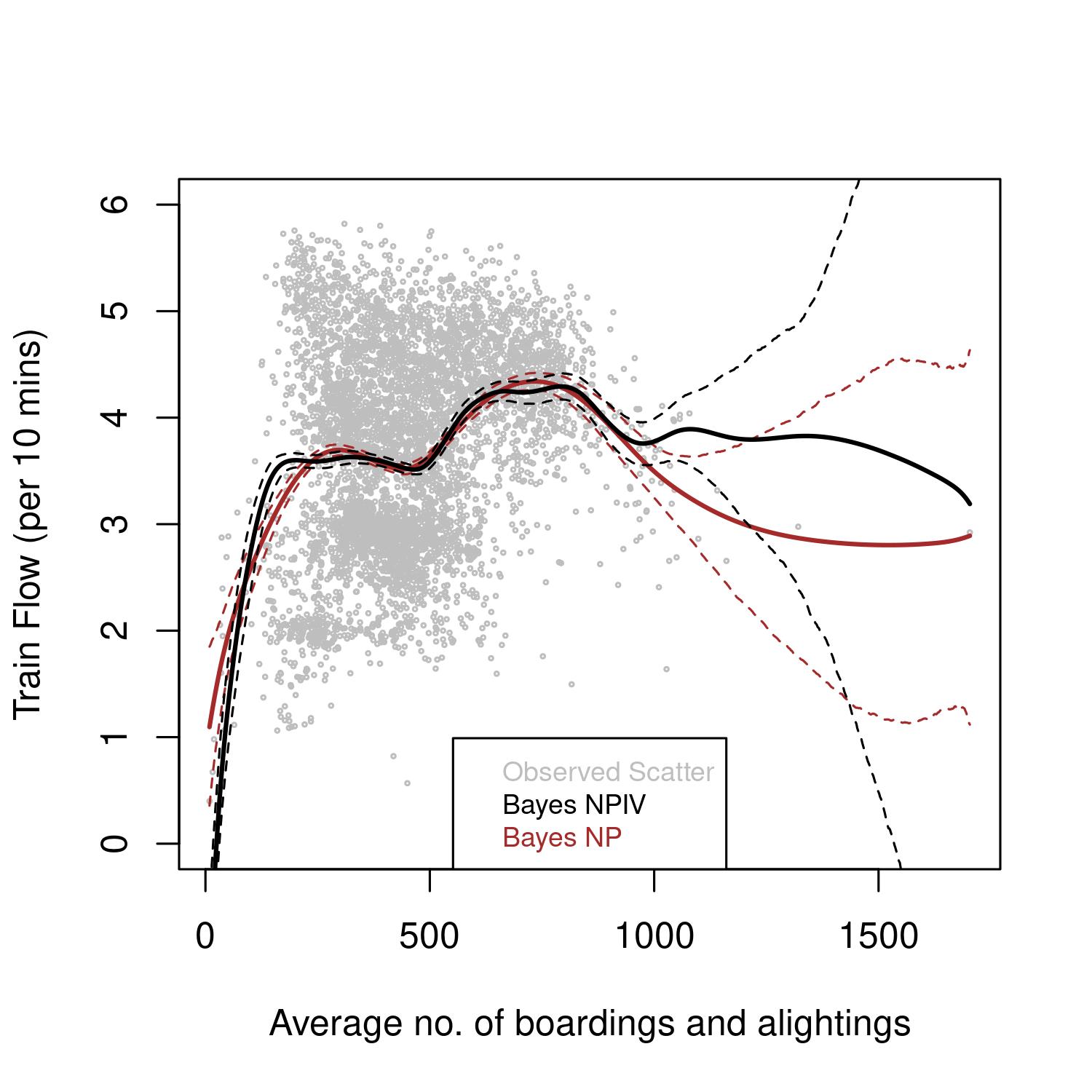}
        \subcaption{Prince Edward Station}
    \end{subfigure}
\hfill
    \begin{subfigure}{0.32\textwidth}
        \includegraphics[width=\linewidth]{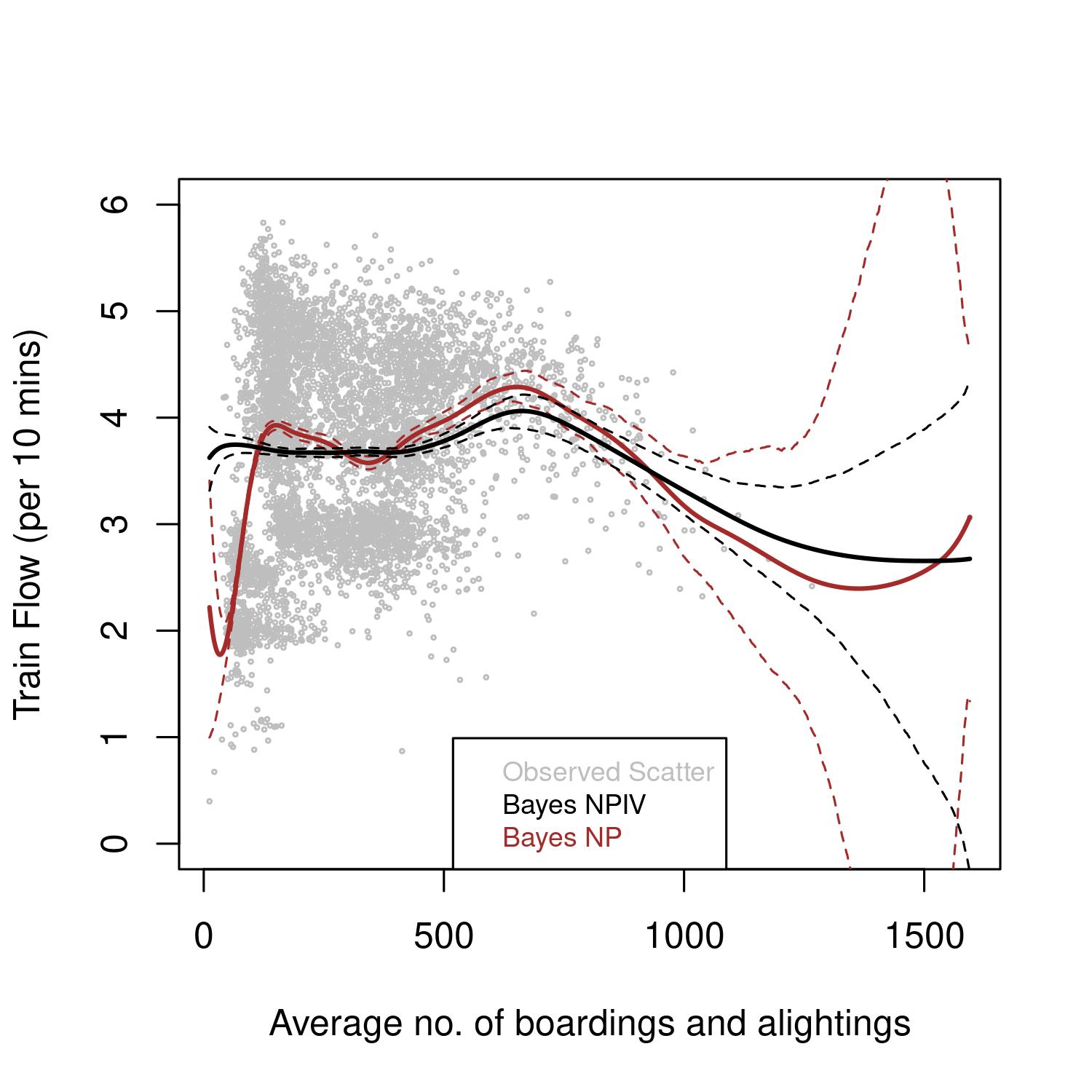}
        \subcaption{Mong Kok Station}
    \end{subfigure}
\hfill
    \begin{subfigure}{0.32\textwidth}
        \includegraphics[width=\linewidth]{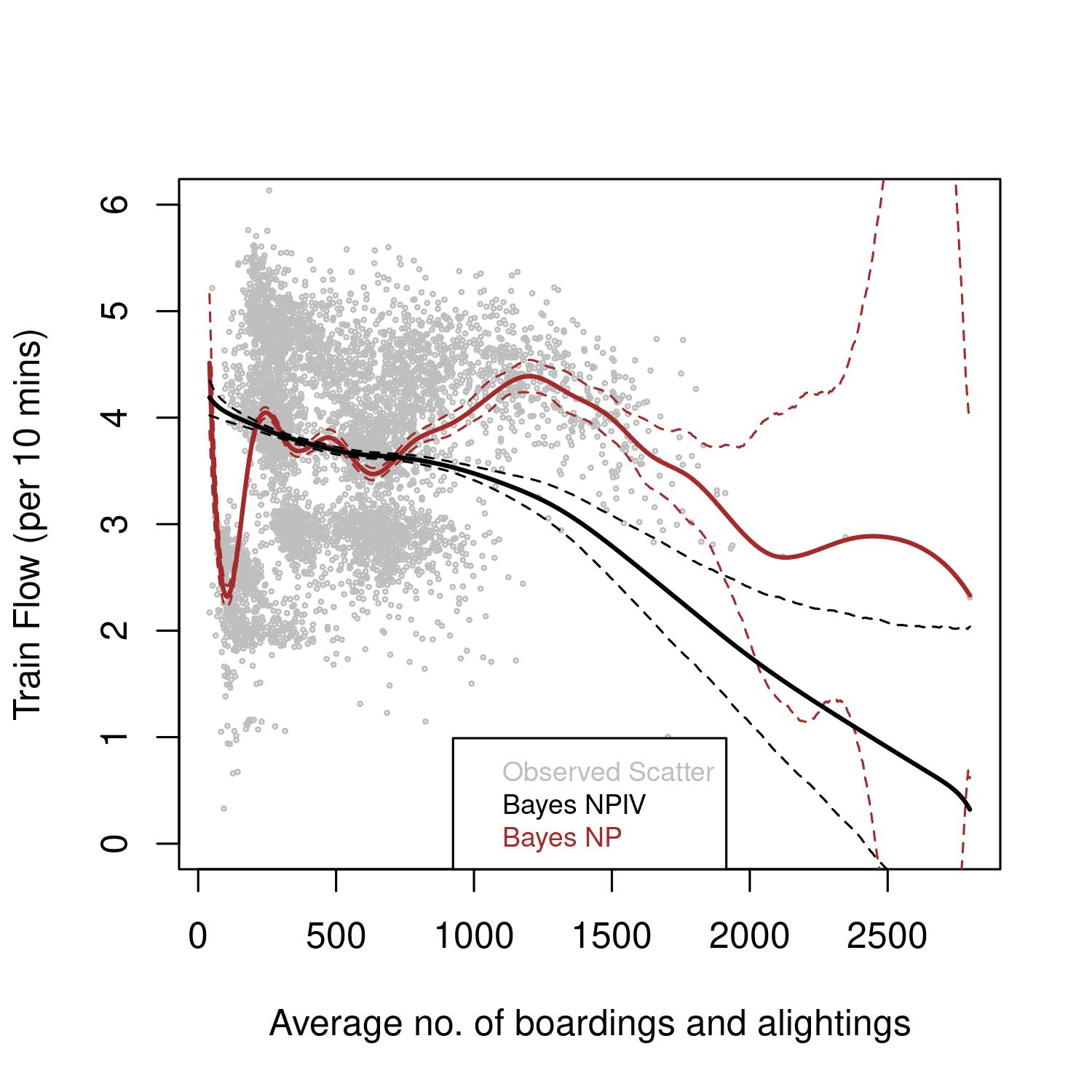}
        \subcaption{Yau Ma Tei station}
    \end{subfigure}
    \caption{Non-parametric Instrumental Variables based estimation results for train movements in the upward direction along the Kwun Tong Line (dotted lines represent 95\% credible intervals).}
    \label{fig:Res_Full_U}
\end{figure}

Figures \ref{fig:Res_Full_D} and \ref{fig:Res_Full_U} show the estimated train flow versus boarding-alighting per train curves (that is, the estimated $S(.)$ in the second stage) for all stations that are highlighted in Figure \ref{fig:map} for downward and upward directions of train movement, respectively. We observe that the \acrshort{NPIV} estimator yields nearly concave curves, however, the associated credible bands in the downward-sloping region are very wide for all stations, except for Prince Edward Station and Kowloon Tong Station in the downward direction and Prince Edward Station, Mong Kok Station  and Yau Ma Tei Station in the upward direction (see Figures \ref{fig:Res_Full_D}c, \ref{fig:Res_Full_D}e, \ref{fig:Res_Full_U}e and \ref{fig:Res_Full_U}f, \ref{fig:Res_Full_U}g). The statistical significance of the downward-sloping part is apparent in a short domain of passenger boarding-alightings at these stations. Thus, these plots provide conclusive empirical evidence to support the existence of a unique and optimal passenger volume, above which passenger boarding-alightings negatively affect the train arrival rate. The segment of the curve before this optimal passenger volume represents uncongested operations wherein the relationship between train flow and passenger boarding-alightings is either independent or is upward sloping due to the equilibrium interactions between increasing train flow (from increasing train departure frequency) and increasing passenger boarding-alightings (from increasing passenger demand). The part of the curve beyond the optimal passenger volume signifies congested operations where excessive passenger boarding-alightings at the station cause train flow to decrease.  

The stations with statistically significant downward-sloping segments act as active bottlenecks in the associated direction of train flow along the Kwun Tong Line in the \acrshort{MTR} network. In the downward direction, the optimal number of passenger boarding-alightings at Kowloon Tong Station and Prince Edward Station is around nine hundred fifty passengers and nine hundred passengers, respectively. In the upward direction, the optimal number of passenger boarding-alightings is around eight hundred passengers at Prince Edward Station,  around seven hundred passengers at Mong Kok Station, and around one thousand passengers at Yau Mai Tei Station. The corresponding maximum train inflow at all the bottleneck stations is around 4.2 trains per ten minutes, that is, the estimated minimum headway between trains is around two and a half minutes. The estimated minimum headway value in both directions is close to the planned minimum peak headway of 2.1 minutes \citep{MTR2019}. Table \ref{tab:res} summarises these results. Thus, the application of the NPIV approach allows us to adjust for any confounding bias and recover the scheduled peak headway.

The large credible intervals in the downward-sloping part of the estimated $S(.)$  at all other stations imply that there may be only a handful of instances of passenger congestion at these stations due to lower demand, in-place control measures and relatively lower operating frequency of the \acrshort{MTR} services. Thus, the estimated relationship between train flow and passenger boarding-alightings and optimal/critical passenger boarding-alightings are not universal, rather they depend upon the characteristics of the metro station such as metro demand, frequency, and spacing between consecutive stations and station design, among many other factors.  

\begin{table}[H]
\center
\caption{Summary of results.} \label{tab:res}
\resizebox{0.8\columnwidth}{!}{
\begin{tabular}{lccc}
  \hline
  Identified Bottleneck & Direction & Optimum passenger &  Associated \\
  Station & of Flow & movements &  headway \\
  \hline
  Kowloon Tong & downward & $\sim$ 950 pax per train & $\sim$ 2.4 minutes \\ 
  Prince Edward & downward & $\sim$ 900 pax per train & $\sim$ 2.4 minutes \\
  Prince Edward & upward & $\sim$ 800 pax per train & $\sim$ 2.4 minutes \\ 
  Mong Kok & upward & $\sim$ 700 pax per train & $\sim$ 2.4 minutes \\ 
  Yau Ma Tei & upward & $\sim$ 1000 pax per train & $\sim$ 2.5 minutes \\ 
  \hline
\end{tabular}}
\end{table}

\section{Discussion}

This paper presents the first step towards estimation and application of station-level \acrfull{FDs}; an analogous to traffic \acrfull{FDs} that are a key powerhouse for traffic control modelling; for metro systems. The \acrshort{FDs} are derived using large-scale \acrfull{AFC} and \acrfull{AVL} data. The estimated \acrfull{FDs} characterise the passenger-congestion phenomenon at metro stations via reproducible supply side relationships between passenger boarding-alightings and train flow. These \acrshort{FDs} could inform data-driven optimal station-level control strategies to avoid recurrent congestion delays in metro networks.

The short-term prediction of subway passenger demand has received significant attention in recent years \citep{Ding2016,Ma2018}. This study enhances the value of short-term demand prediction by estimating its causal impact on train frequencies. Understanding the dynamics of passenger boarding-alightings and train frequencies, along with estimates of optimum boarding-alighting, can help in designing strategies to control passenger boarding-alightings and minimise delays. Such control strategies may involve i) adopting platform management practices such as reducing escalator capacity, ii) deployment of staff resources to regulate the entry of passengers into bottleneck stations, and iii) pricing policies. Another strategy could be \emph{ramp metering}\footnote{A ramp meter is a basic traffic light device together with a signal controller, that are used to regulate the flow of vehicular traffic entering freeways according to current traffic conditions. Ramp metering systems have proved to be successful in decreasing traffic congestion on freeways. Similar strategies are being sought after by metro operators to regulate passenger demand in future.} of passengers entering stations to increase overall system throughput. \cite{Daganzo2005,Daganzo2007} suggest such strategies in the context of vehicular traffic control in urban networks. 

We note that metro operators around the world presently implement such strategies based on their day-to-day experience of congestion patterns at various stations. For instance, Transport for London implements different types of station control measures such as avoid train dwelling at particular stations during a specific time of the day, individual platform closures, and closures of gate lines and entrances \citep{TfL2018}. The findings of this study could assist metro operators in improving these control strategies in a data-driven manner. 

It is worth noting that the above-discussed strategies rely on controlling passenger boarding movements, but the estimated relationship includes both passenger boarding and alighting movements. This disparity does not restrict the application of the empirical results in practice because metro operators can use short-term demand prediction models to forecast the number of alighting and boarding movements at any station. Subsequently, they can adopt station-level control measures to regulate the number of boarding movements such that the instantaneous sum of boarding and alighting movements remains optimum.  

We emphasise that the objective of this study is limited to understanding the existence of traffic fundamental diagram like relationships for metro systems. We focus on presenting initial evidence on the existence of such relationships and estimating them by applying an appropriate econometric method. An application of the estimated relationships in developing practical control strategies would require a comprehensive microscopic simulation of an at-capacity metro's operations to capture the network propagation effects. Any potential long-run effects of control strategies at any station on passenger demand of other stations could be better understood in the simulation study. However, such a detailed simulation model is beyond the scope of this paper but presents an important avenue for future research. Another area of future research could be to explore the applications of fundamental diagrams to understand the level of operational service in the long run and guide improvements in the metro network.

\newpage
\appendix

\bigskip

\section{Review of congestion management strategies}
\label{S:I}

In recent years, a few studies have modelled the dynamics of metro operations while considering the physical interaction between train-congestion and passenger-congestion \citep{Seo2017, Zhang2019}. A similar literature is available for mainline railway operations \citep[][and other references therein]{Keiji2015, Wada2012}. For instance, using microscopic principles of rail operations, \cite{Seo2017} analytically derived a three-dimensional relationship between train-flow, train-density and passenger-flow along a metro line, which they refer to as the fundamental diagram of rail transit. Their relationship considers steady state operations along the corridor, that is, it assumes conditions such as same cruising speed of all trains between stations, constant headway time between successive trains and same passenger-flow arriving at each station, among others. Using Boston's subway data, \cite{Zhang2019} tried to empirically validate the proposed fundamental relationship, however, they do not find any evidence of a backward-bending relationship as originally proposed by \cite{Seo2017}. The lack of empirical evidence may be a result of assumed steady state conditions which seldom exist. Moreover, the basic idea in these studies is to derive headway-based control strategies to recover the system from knock-on delays such as keeping a moderate separation between trains with necessary adjustment in departure time from origin stations \citep{Keiji2015}, dwelling time extension at some control stations located at the upstream of the bottleneck station \citep{Wada2012} and increase of free flow speed \citep{Daganzo2009}. However, most of these  strategies address train-congestion delays without targeting the root cause of these delays -- increased passenger boarding-alightings at bottleneck stations. To summarise, although such studies attempt to understand the process of the delay propagation, they lack the empirical evidence to validate their proposed relationship. Moreover, they seldom focus on developing control strategies to directly target passenger-congestion, which is central to congestion delays in metro networks.

Another strand of the literature develops optimisation-based passenger inflow control strategies to minimise the negative impacts of recurrent congestion in metro networks during peak hours. Their underlying objectives include minimising the total waiting time experienced by passengers in the network during peak hours \citep{Shi2018,Guo2015,Yuan2020,Zhang2021}, reducing the safety risks imposed by passengers waiting on the platform \citep{Jiang2018,Zou2018}, or minimising the number of stranded passengers \citep{Wang2020,Yuan2020}. For instance, \cite{Jiang2018} developed a reinforcement learning-based method to optimise the inflow volume during a time window at each station with the aim to minimise the safety risks imposed on passengers at the metro stations. They use the number of people waiting at the platform and the frequency of passengers being stranded to measure safety. \cite{Shi2018} develop a joint optimisation model for train timetable and passenger flow control to reduce accumulation of passengers on platforms and minimise related safety risks. In another study, \cite{Zou2018} developed a static station inflow control scheme to minimise platform congestion. Their model uses demand data (\acrshort{OD} matrix) and train timetable to derive flows and capacity of each section. They further use this demand-supply information to identify bottlenecks and develop a framework to minimise platform congestion. We note that while such studies focus on minimising passenger-congestion under various objectives, they barely acknowledge the delays in the network due to passenger-congestion at stations. 

\section{The train operation model}
\label{S:II}

To simulate the movement of trains between stations under pure moving block signalling system\footnote{We consider a moving block signalling system over a fixed block signalling system as the former permits a more efficient management of queuing and delays by allowing trains to operate at lower headways \citep{Gill1994,Takeuchi2003}. Several metro lines on the London Underground, the Singapore MRT, the Hong Kong MTR and the New York City subway, among others, use moving block signalling system. Metro systems around the world are increasingly upgrading to such systems to reduce congestion-related delays in the network \citep{MTR2019b}.}, we adopt a \acrfull{CA} model. The \acrshort{CA} model was originally developed for simulation of road traffic flow \citep{NS1992}. However, owing to its ability to reproduce complex real world traffic flow phenomena in a simplistic framework \citep[for instance, see ][]{Spyropoulou2007,Meng2011}, it has also been widely used to simulate rail traffic flow \citep[refer to ][and other references therein]{Li2005,Yinping2008,Xun2013,Ning2014}.

In the \acrshort{CA} model, the rail line $i$ is divided into $L$ cells, each of length 1 metre (that is, $i \in \{1,2,...L\}$). The simulation time $T_s$ comprises of discrete time steps $t$ of 1 second each (that is, $t \in \{1,2,...T_s\}$). At each time step $t$, each cell $i$ can either be empty or occupied by the n\textsuperscript{th} train with integer velocity $v_{n,t}$ (that is, $v_{n,t} \in \{0,1...v_{max}\}$). Stations are placed at different positions along the line and corresponding dwell time is defined. Each train $n$ is indexed based on its order of entry into the system (that is, $n= \in \{1,2,...,N\}$). The position of train $n$ at time $t$ is denoted by $X_{n,t}$. The boundary conditions are open and defined as follows: (i) After each departure interval $D$, a train with velocity $v_{max}$ enters at the position $i=1$ given that the train ahead is at a safe breaking distance (given by equation \ref{eq:dn}) from the entry; (ii) At position $i=L$, trains simply exit the system. 

At each discrete time step, $t \to t+1$, the state of the system is updated according to well-defined rules, mainly governed by the following two situations:

\begin{itemize}
    \item When the (n-1)\textsuperscript{th} train is in front of the n\textsuperscript{th} train at time $t$, a comparison of the headway distance $\Delta X_{n,t} = X_{n-1,t}-X_{n,t}$ and the minimum instantaneous distance $d_{n,t}$ determines whether the n\textsuperscript{th} train will accelerate or decelerate in the next time step\footnote{The same equation also applies when the (n-1)\textsuperscript{th} train, that is in front of the n\textsuperscript{th} train, is at a station.}. The minimum instantaneous distance between successive trains operating under pure moving-block signalling is given by \citep{Yan2012}:
    \begin{equation}
    \label{eq:dn}
        d_{n,t} = \frac{v_{n,t}^2}{2b} + SM
    \end{equation}
    where, $v_{n,t}$ is the velocity of train $n$ at time $t$ and $b$ is its deceleration. The first term on the right hand side of equation \ref{eq:dn} represents the breaking distance of train $n$ at time $t$. A safety margin, $SM$, is introduced to avoid collision.
    \item When the n\textsuperscript{th} train is behind an empty station within the breaking distance, its velocity must vary such that the train can stop at the station. To obtain the updated velocity, we apply the kinematics equation:  $v_{n,t+1}^2-v_o^2=2bG_{n,t}$, where  $v_o$ is the target velocity which is zero for the train to stop at the station and $G_{n,t}$ is the distance between the station and the train $n$ at time $t$. As the \acrshort{CA} model allows only for integer values of velocity, the velocity update $v_{n,t+1}$ is given by:
    \begin{equation}
    \label{eq:vn}
        v_{n,t+1} = \textrm{int}(\sqrt{2bG_{n,t}})
    \end{equation}
\end{itemize}

Therefore, the update rules for velocity and position of a train at each time step are as follows:

\begin{enumerate}
    \item When the n\textsuperscript{th} train is behind the (n-1)\textsuperscript{th} train \\
    \textit{Step 1} Velocity update:\\
    if $\Delta X_{n,t}>d_{n,t}$, $v_{n,t+1}=\min(v_{n,t}+a,v_{max})$; \\
    elseif $\Delta X_{n,t}<d_{n,t}$, $v_{n,t+1}=\min(v_{n,t}-b,0)$; \\
    else $v_{n,t+1}=v_{n,t}$.\\
    \textit{Step 2} Position update:\\
    $X_{n,t+1}=X_{n,t}+v_{n,t+1}$.
    \item When the n\textsuperscript{th} train is behind a station
    \begin{enumerate}
        \item When the station is occupied by the (n-1)\textsuperscript{th} train \\
        The update rules are the same as case 1. 
        \item When the station is empty \\
        \textit{Step 1} Velocity update: \\
        if $G_{n,t} > d_{n,t}$, $v_{n,t+1}=\min(v_{n,t}+a,v_{max})$; \\
        elseif $G_{n,t}<d_{n,t}$, $v_{n,t+1}=\min(v_{n,t}-b,\textrm{int}(\sqrt{2bG_{n,t}}), 0)$; \\
        else $v_{n,t+1}=v_{n,t}$.\\
        \textit{Step 2} Position update:\\
        $X_{n,t+1}=X_{n,t}+v_{n,t+1}$.
    \end{enumerate}
    \item When the n\textsuperscript{th} train is at a station \\
    \textit{Step 1} Velocity update: \\
    if $t_{dwell}=T_d$ and $\Delta X_{n,t}>L_s$, $v_{n,t+1}=\min(v_n+a,v_{max})$, $t_{dwell}=0$; \\
    elseif $t_{dwell}<T_d$, $v_{n,t+1}=0$, $t_{dwell}=t_{dwell}+1$. \\
    \textit{Step 2} Position update: \\
    $X_{n,t+1}=X_{n,t}+v_{n,t+1}$. \\
    where $L_s = \frac{1}{2a} + SM$ is the safe distance to avoid any collision with the train ahead of the dwelling train, $t_{dwell}$ stores the current dwell time (that is, the time for which the train has stopped at the station until time-step $t$), and $T_d$ is the planned dwell time.
\end{enumerate}

\section{Headway-based control strategies}
\label{S:III}

\begin{figure}[h]
    \centering
        \begin{subfigure}{0.4\textwidth}
            \centering
            \includegraphics[width=1\textwidth]{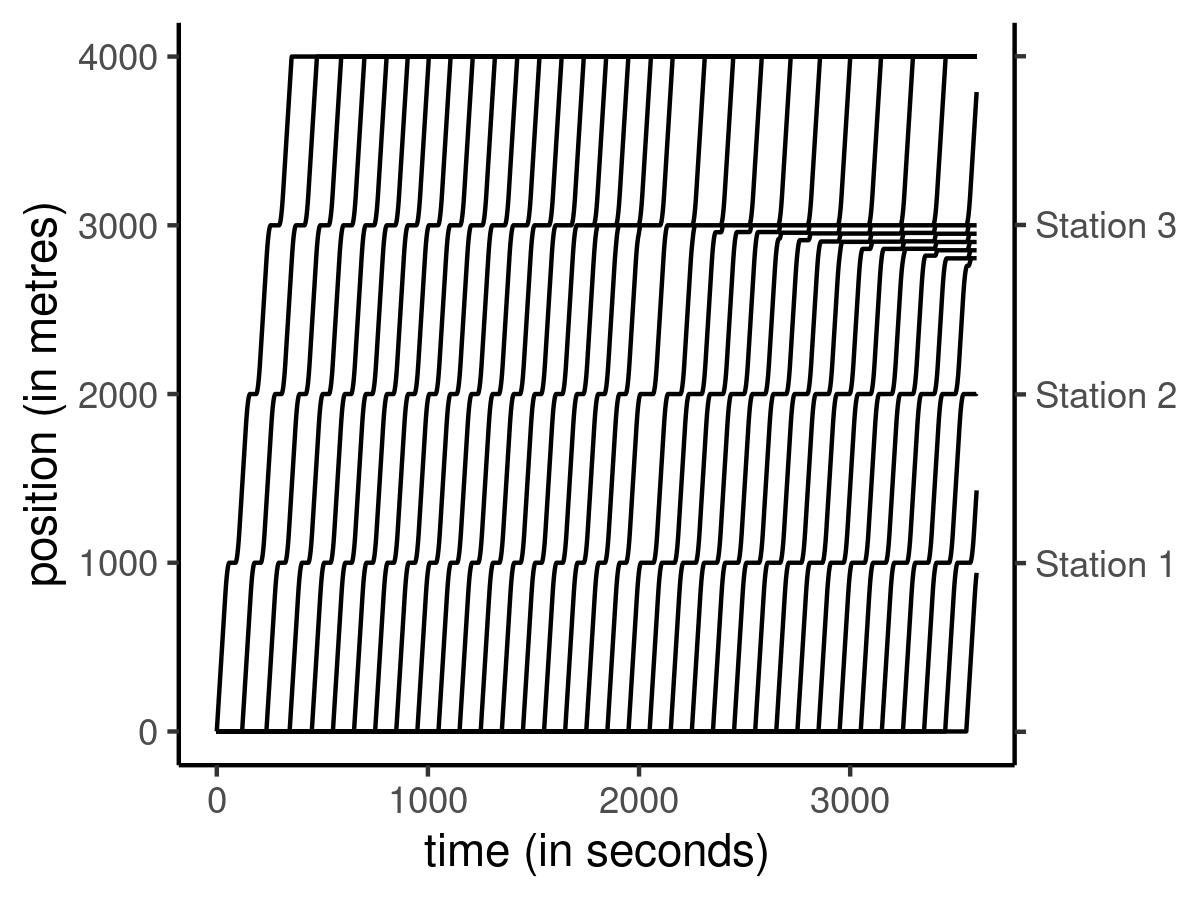}
            \caption{A headway-based control strategy.}
            \label{fig:Sim_Control_Other}
        \end{subfigure}%
        \begin{subfigure}{0.4\textwidth}
            \centering
            \includegraphics[width=1\textwidth]{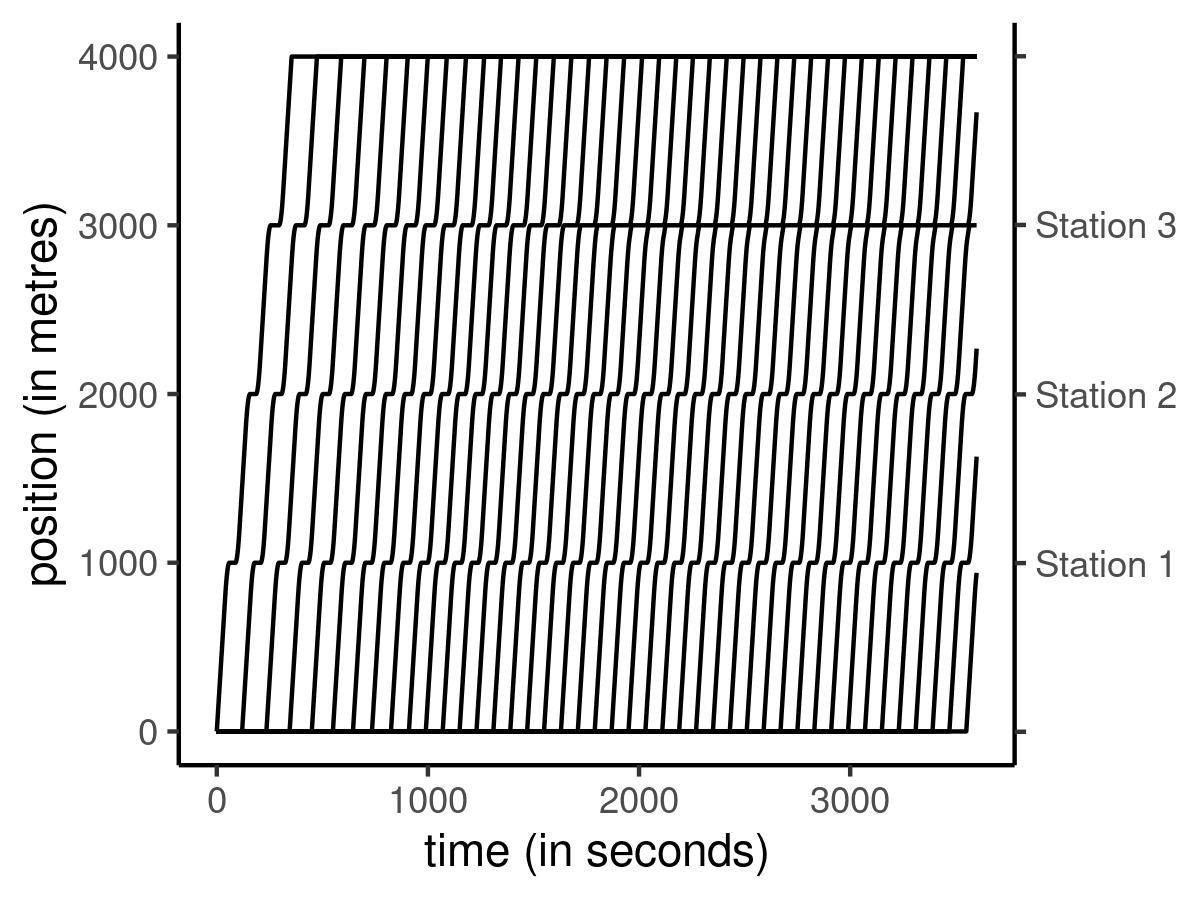}
            \caption{A combination of passenger inflow control and headway-based control strategies.}
            \label{fig:Sim_Control_Other_Mixed}
        \end{subfigure}%
    \caption{The time-space diagram representing train operations under other control strategies.}
\end{figure}

Following from the simulation exercise presented in the main paper, we compare the station-level passenger inflow control and headway-based control strategies. The headway-based strategies have been recommended in the literature \citep[for instance, see ][]{Seo2017,Keiji2015}, which enable operators to avoid train-congestion by moderating train movements per train. In our headway-based strategy, we control train movements by increasing the interval $D$ between successive trains entering into the metro system. In this controlled scenario, we set the minimum value of $D$ as 90 seconds, as opposed to 60 seconds in the \textit{no control} scenario. Moreover, to avoid queuing, we allow trains to be held longer at stations upstream of the bottleneck (that is, at stations 1 and 2) by increasing their dwell time from 30 seconds to 60 seconds. Figure \ref{fig:Sim_Control_Other} shows the time-space diagram for this headway-based control scenario. We note that train-congestion is significantly lower in the headway-based control as compared to the \textit{no control} scenario. However, the throughput of the system decreases from 39 per hour in \textit{no control} to 27 per hour in the headway-based control scenario. 

We also consider a combination of station-level passenger inflow control and headway-based strategies. We restrict the maximum passenger boarding-alighting rate at 9.75 passengers per second and we also set the minimum value of $D$ as 70 seconds, as opposed to 60 seconds in the \textit{no control} scenario. Figure \ref{fig:Sim_Control_Other_Mixed} shows the time-space diagram for this combined control scenario. The figure illustrates that queuing of trains is completely eliminated under this control strategy and the system throughput increases from 27 per hour in the headway-based control only scenario to 38 trains per hour in the combined control scenario.

This extension to our main simulation exercise demonstrates that ensuring the optimum passenger boarding-alighting per train at bottleneck stations using station-level controls can be more effective than solely headway-based strategies in reducing congestion-related delays and improving service reliability. This simple exercise also corroborates the findings from some recent studies on bus transit operations, which compare control strategies that combine limiting the number of boarding passengers at stops during peak operations and holding of buses at control stations (headway-control) with those that involve holding of buses only \citep{Delgado2009,Delgado2012}. These studies suggest that the hybrid strategy outperforms the headway-based strategy in improving service reliability by avoiding bus bunching. 

\section{Bayesian Non-parametric Instrumental Variables Approach}
\label{S:IV}

We discuss the Bayesian \acrshort{NPIV} approach \citep{Wisenfarth2014} for a model with a single confounding covariate, that is,

\begin{equation}
\label{eq:NPIV1}
q = S(n) + \epsilon_2, \quad n = h(z) + \epsilon_1
\end{equation}

Note that $\omega$ and $\xi$ are encapsulated in $\epsilon_2$, and $z$ is an instrument for the confounding regressor $n$. The relationship between $n$ and $z$  is represented by an unknown functional form $h(.)$ and $\epsilon_2$ is an idiosyncratic random error term. For the notational simplicity, we drop time-day subscripts. Bayesian \acrshort{NPIV} is a control function approach, and assumes the following standard identification restrictions: 

\begin{equation}
\label{eq:NPIV2}
E(\epsilon_1|z) = 0 \quad \textrm{and} \quad E(\epsilon_2|\epsilon_1,z) = E(\epsilon_2|\epsilon_1),
\end{equation}

\noindent which yields

\begin{equation}
\label{eq:NPIV3}
\begin{split}
E(q|n,z) & = S(n) + E(\epsilon_2|\epsilon_1,z) = S(n) + E(\epsilon_2|\epsilon_1) \\
             & = S(n) + \nu(\epsilon_1),
\end{split}
\end{equation}

\noindent where $\nu(\epsilon_1)$ is a function of the unobserved error term $\epsilon_1$. This function is known as the control function. 

To satisfy the identification restrictions presented in equation \ref{eq:NPIV2}, we need an \acrfull{IV} $z$. The \acrshort{IV} should be (i) exogenous, that is, uncorrelated with $\omega$, $\xi$, and $\epsilon_2$; (ii) relevant, that is, strongly correlated with the confounding covariate $n$. Due to the absence of suitable external instruments, we derive instruments from the available data itself. To instrument the confounding covariate (that is, average boarding-alightings per train) in the morning and afternoon hours, we use the average boarding-alightings per train in the afternoon and morning hours, respectively, from the opposite direction of train flow. For instance, we instrument average boarding-alightings observed in the ten-minute interval 8:30-8:40 hours, on workday $t$ in the upward direction of train flow, with the observation on the covariate in the ten-minute interval 18:30-18:40 hours, on workday $t$ in downward direction of train flow. We argue that due to the commuting nature of trips, the temporal variation of trips during the morning and afternoon peak and peak shoulder hours along a particular direction of flow is highly correlated with the temporal variation observed in the opposite direction of flow in afternoon and morning hours respectively. This correlation can also be inferred through the descriptive statistics summarised in Table \ref{tab:sumstats}. However, the boarding-alightings from the opposite direction of flow are exogenous because they do not directly determine the response variable $q_{it}^s$ in equation \ref{eq:spec}. To justify the relevance of the considered instrument, we present the estimated $h(.)$ in equation \ref{eq:NPIV1} in Section \ref{S:VI}. 

Conditional on the availability of a valid instrument, Bayesian \acrshort{NPIV} can  correct for confounding bias. To account for nonlinear effects of continuous covariates, both $S(.)$ and $h(.)$ (refer equation \ref{eq:NPIV1}) are specified in terms of additive predictors comprising penalised splines. Each of the functions $S()$ and $h(.)$ is approximated by a linear combination of suitable B-spline basis functions. The penalised spline approach uses a large enough number of equidistant knots in combination with a penalty to avoid over-fitting. Moreover, the joint distribution of $\epsilon_1$ and $\epsilon_2$ is specified using nonparametric Gaussian \acrfull{DPM}, which ensures robustness of the model relative to extreme observations. Efficient \acrfull{MCMC} simulation technique is employed for a fully Bayesian inference. The resulting posterior samples allow us to construct simultaneous credible bands for the non-parametric effects (i.e., $S(.)$ and $h(.)$). Thereby, the possibility of non-normal error distribution is considered and the complete variability is represented by Bayesian \acrshort{NPIV}. We now succinctly discuss specifications of the kernel error distribution in Bayesian \acrshort{NPIV}. 

To allow for a flexible distribution of error terms, the model considers a Gaussian \acrshort{DPM} with infinite mixture components, $c$, in the hierarchy illustrated in equation \ref{eq:NPIV_error}. In this equation, $\mu_c$, $\Sigma_c$ and $\pi_c$ denote the component-specific means, variances and mixing proportions. The mixture components are assumed to be independent and identically distributed with the base distribution $G_0$ of the \acrfull{DP}, where $G_0$ is given by a normal-inverse-Wishart distribution. The mixture weights are generated in a stick-breaking manner based on a Beta distribution with concentration parameter $\alpha>0$ of the \acrshort{DP}. The concentration parameter $\alpha$ determines the strength of belief in the base distribution $G_0$.

\begin{equation}
\label{eq:NPIV_error}
\begin{gathered}
(\epsilon_{1i},\epsilon_{2i}) \: \sim  \: \sum_{c=1}^\infty \pi_c \textrm{N}(\mu_c,\Sigma_c) \\
(\mu_c,\Sigma_c) \: \sim  \: G_0 = \textrm{N}(\mu|\mu_0, \tau_{\Sigma}^{-1}\Sigma) \: \textrm{IW}(\Sigma|s_\Sigma,S_\Sigma)\\
\pi_c=\upsilon_c\left(1-\sum_{j=1}^{c-1}(1-\pi_j)\right)=\upsilon_c\prod_{j=1}^{c-1}(1-\upsilon_j),\\
c=1,2,...\\
\upsilon_c  \: \sim \: \textrm{Be}(1,\alpha).\\
\end{gathered}
\end{equation}

\subsubsection*{Estimation Practicalities}

We discuss the implementation details of the Bayesian \acrshort{NPIV} method and the associated posterior analysis. Interested readers can refer to \cite{Wisenfarth2014} for details of the Gibbs sampler of the Bayesian \acrshort{NPIV} method and derivation of conditional posterior updates.

We use the \textit{BayesIV} package and \textit{DPpackage} in R to estimate the Bayesian \acrshort{NPIV}. We consider 50,000 posterior draws in the estimation, exclude the first 15,000 burn-in draws and keep every 10$^{th}$ draw from the remaining draws for the posterior analysis. The point-wise posterior mean is computed by taking the average of 3,500 posterior draws. Bayesian simultaneous credible bands are obtained using quantiles of the posterior draws. A simultaneous credible band is defined as the region $I_\delta$ such that $P_{S|data}(S \in I_\delta) = 1-\delta$, that is, the posterior probability that the entire true function $S(.)$ is inside the region given the data equals to $1-\delta$. The Bayesian simultaneous credible bands are constructed using the point-wise credible intervals derived from the $\delta/2$ and $1-\delta/2$ quantiles of the posterior samples of $S(.)$ from the \acrshort{MCMC} output such that $(1-\delta)100\%$ of the sampled curves are contained in the credible band. Similar process is used to obtain the credible intervals of $h(.)$.

\section{Summary of Data}
\label{S:V}

Table \ref{tab:sumstats} presents the summary statistics for ten-minute train flows and average number of boardings and alightings at each analysed station. We note that the four interchange stations -- Kowloon Tong, Prince Edward, Mong Kok and Yau Ma Tei -- are associated with higher level of passenger boardings and alightings in either of the two or both directions of train flow as compared to other stations. 

\section{Relevance of Instruments}
\label{S:VI}

\begin{figure}[tp]
    \centering
    \begin{subfigure}{0.32\textwidth}
        \includegraphics[width=\linewidth]{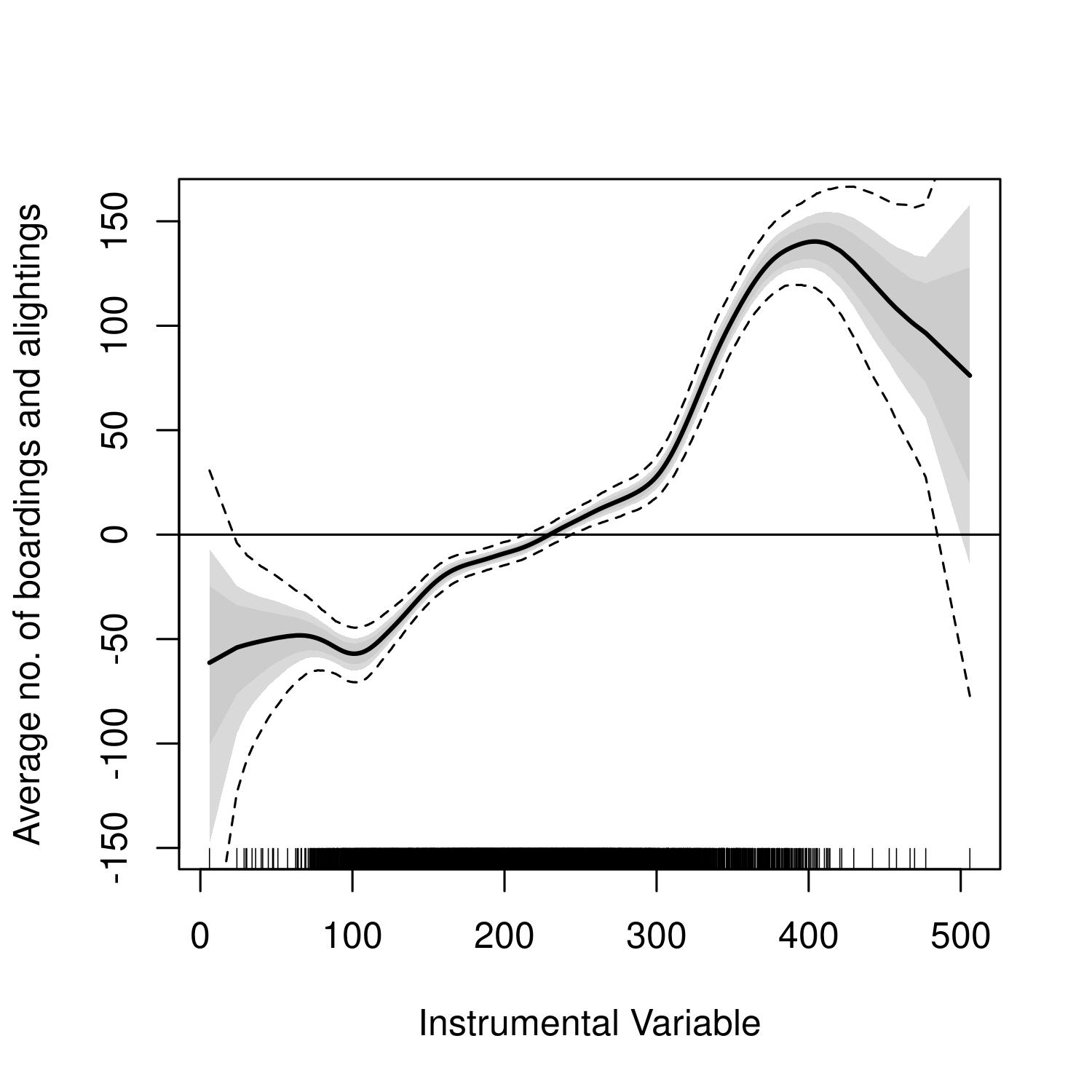}
        \subcaption{Wong Tai Sin Station}
    \end{subfigure}
\hfill
    \begin{subfigure}{0.32\textwidth}
        \includegraphics[width=\linewidth]{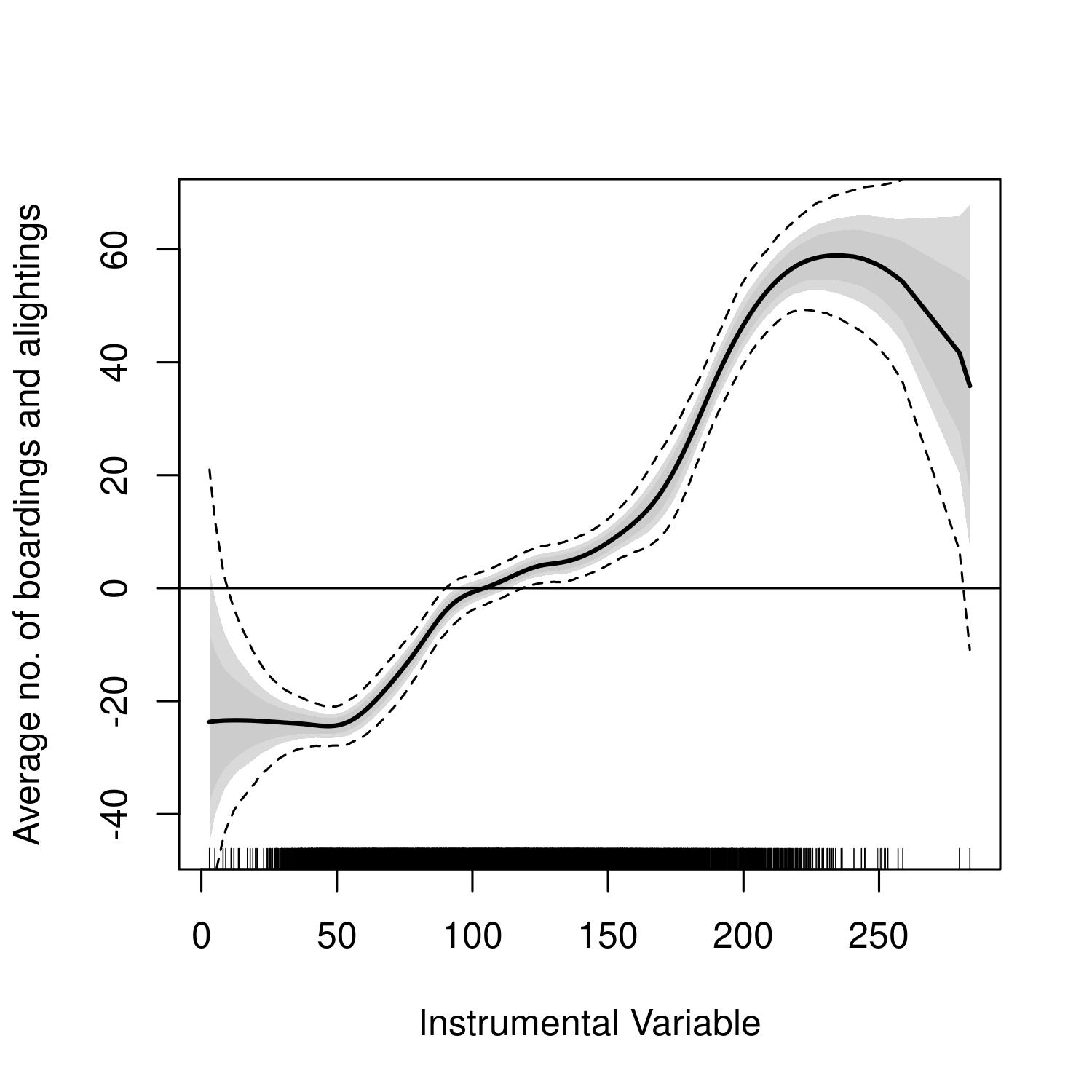}
        \subcaption{Lok Fu Station}
    \end{subfigure}
\hfill
    \begin{subfigure}{0.32\textwidth}
        \includegraphics[width=\linewidth]{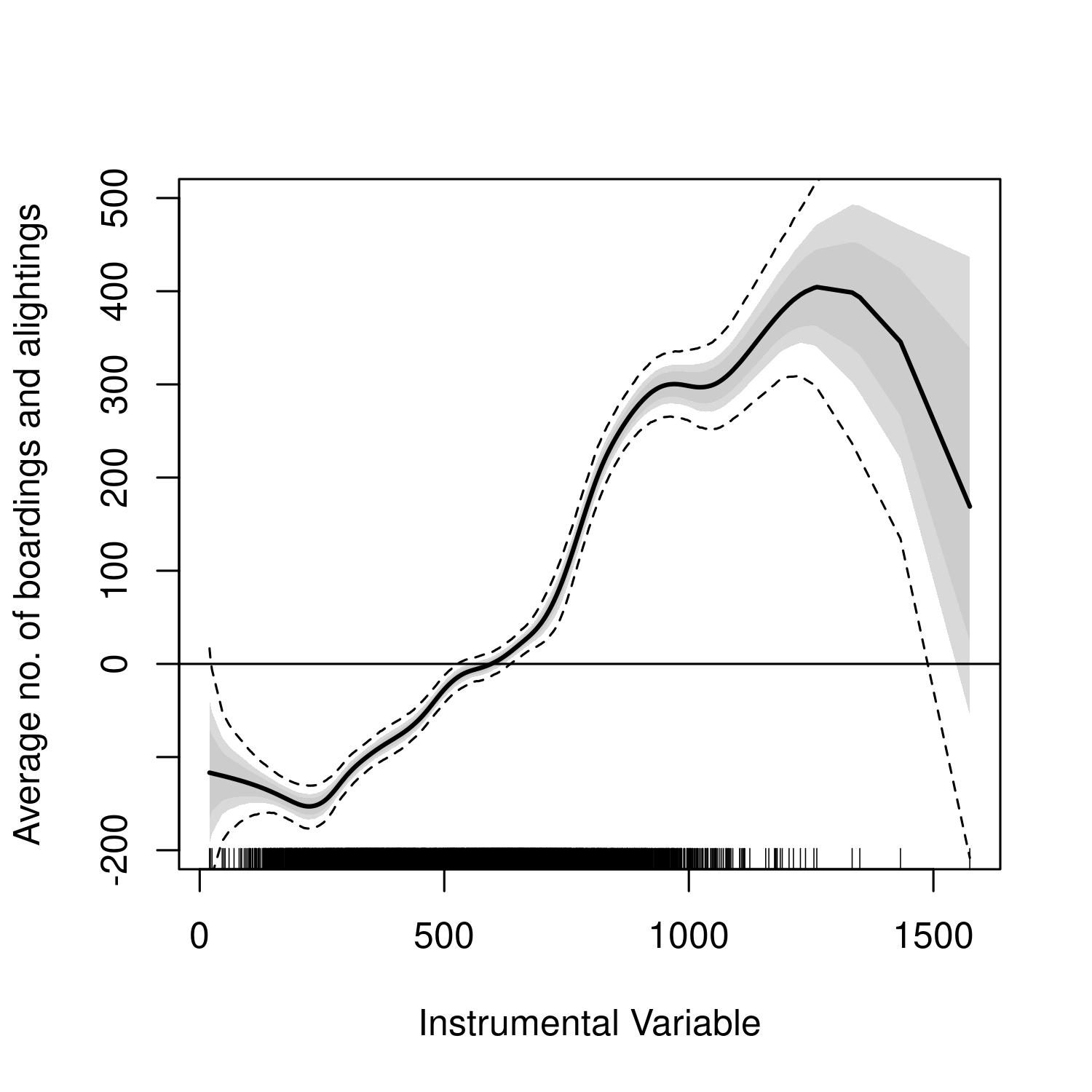}
        \subcaption{Kowloon Tong Station}
    \end{subfigure}
\hfill
    \begin{subfigure}{0.32\textwidth}
        \includegraphics[width=\linewidth]{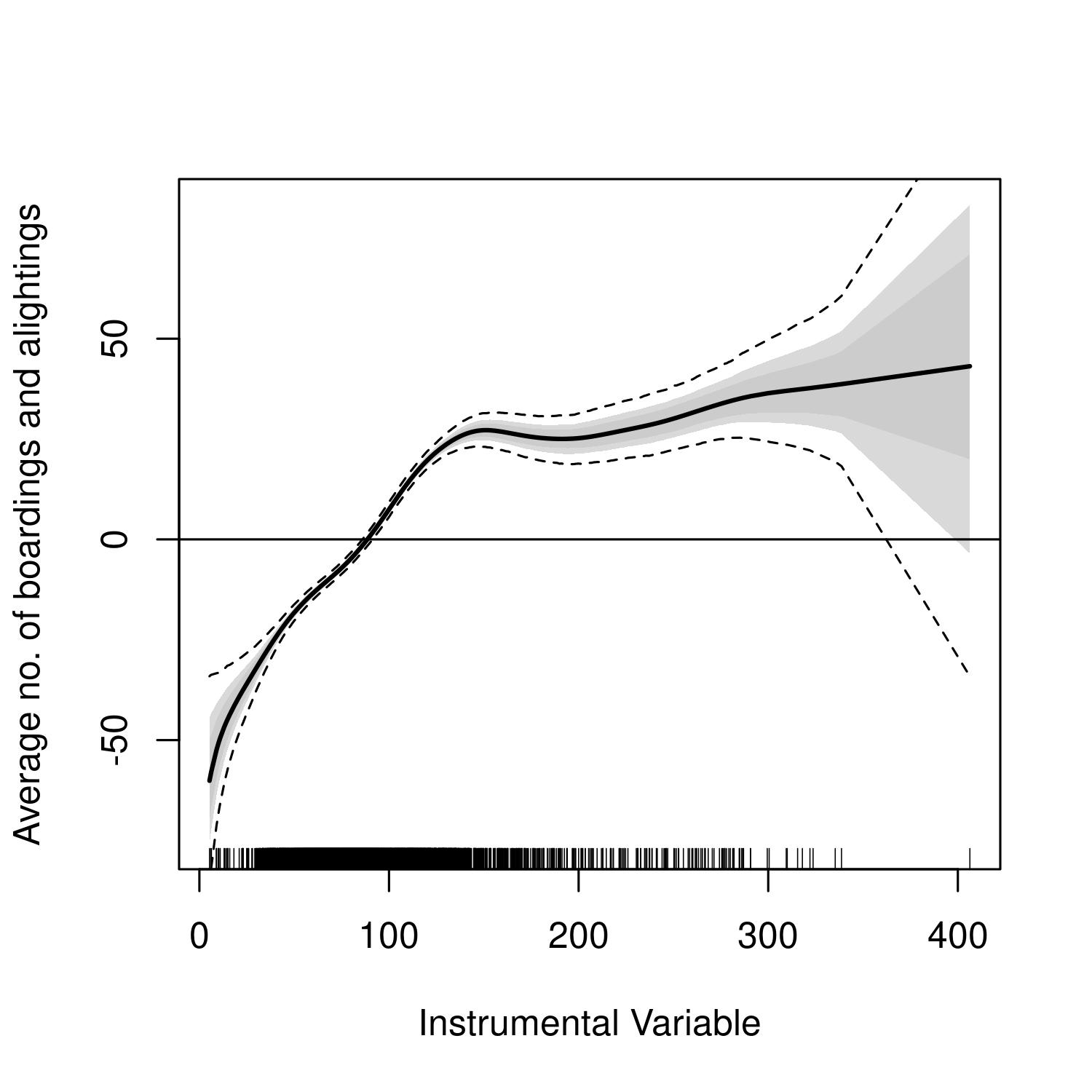}
        \subcaption{Shek Kip Mei Station}
    \end{subfigure}
\hfill
    \begin{subfigure}{0.32\textwidth}
        \includegraphics[width=\linewidth]{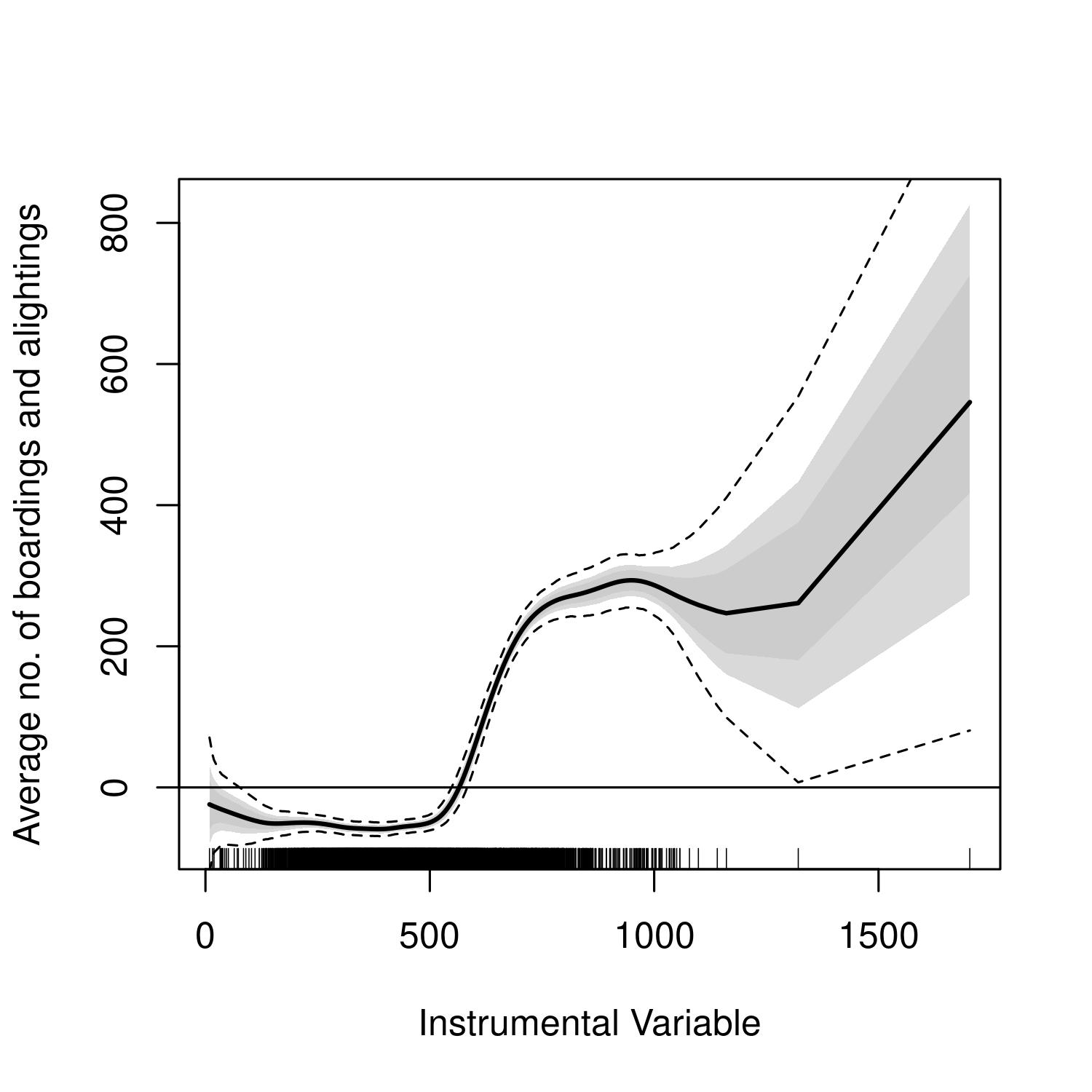}
        \subcaption{Prince Edward Station}
    \end{subfigure}
\hfill
    \begin{subfigure}{0.32\textwidth}
        \includegraphics[width=\linewidth]{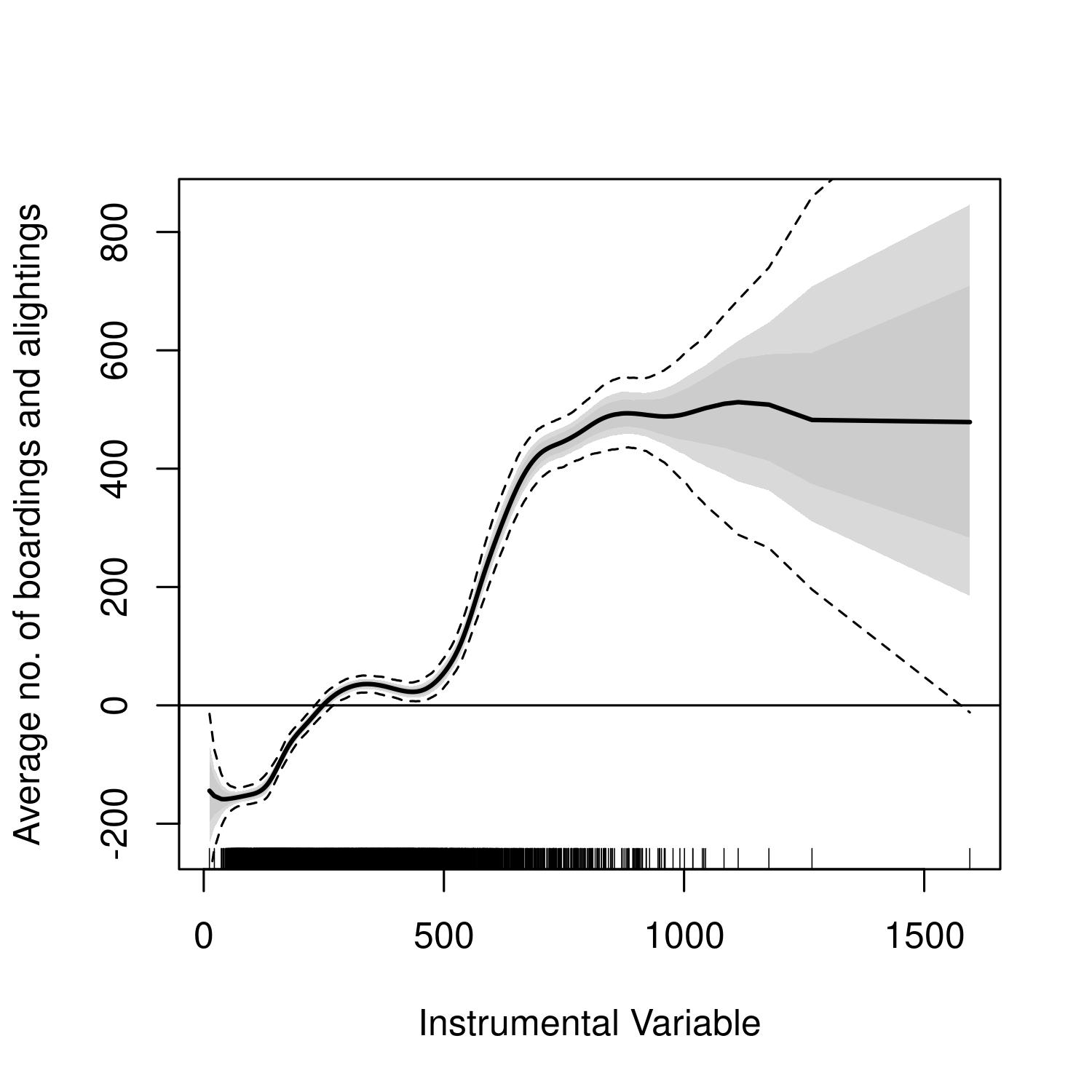}
        \subcaption{Mong Kok Station}
    \end{subfigure}
\hfill
    \begin{subfigure}{0.32\textwidth}
        \includegraphics[width=\linewidth]{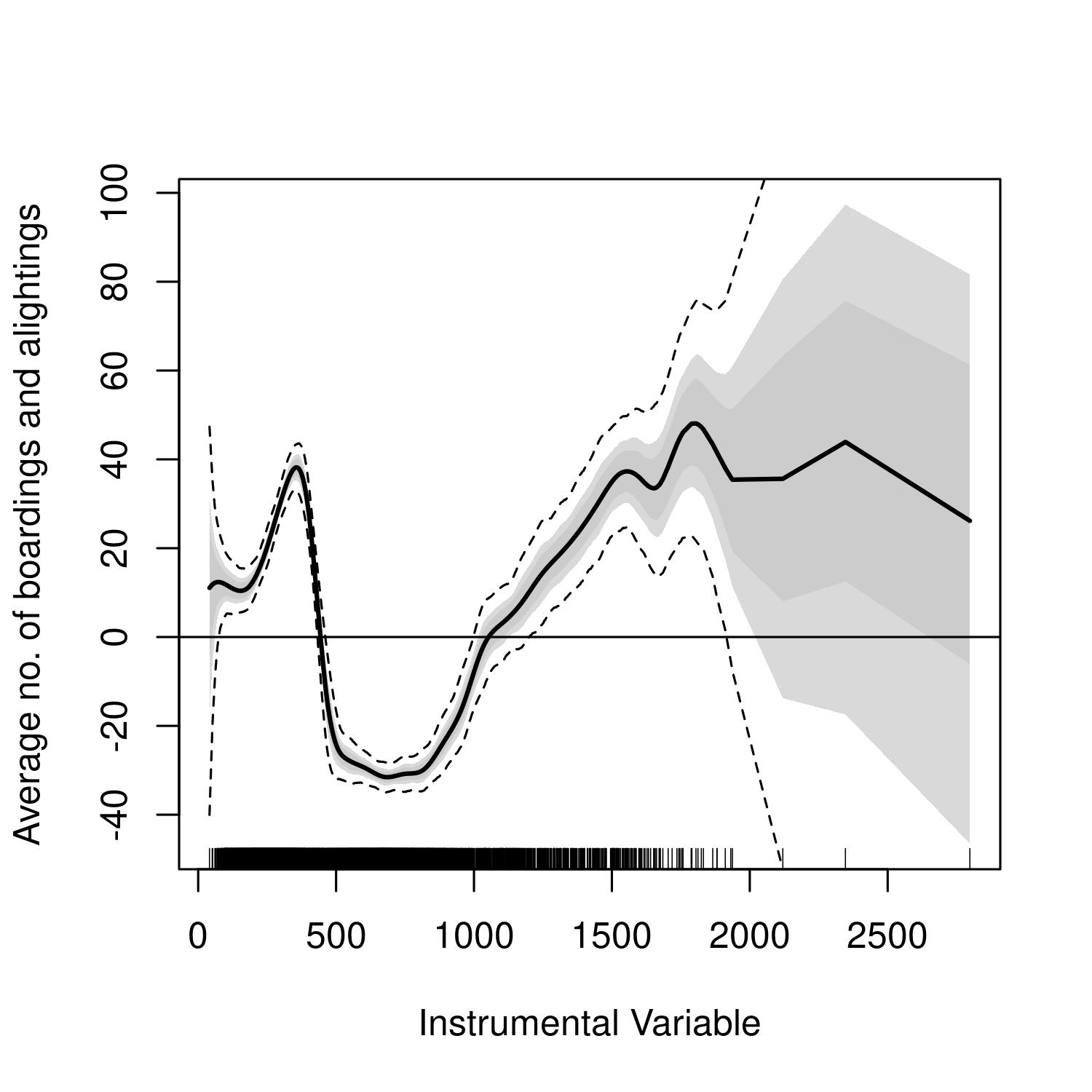}
        \subcaption{Yau Ma Tei station}
    \end{subfigure}
    \caption[]{Relevance of instruments in analysis of train movements in the downward direction along the Kwun Tong Line (dotted lines represent 95\% credible intervals).}
    \label{fig:Inst_Str_D}
\end{figure}

\begin{figure}[tp]
    \centering
    \begin{subfigure}{0.32\textwidth}
        \includegraphics[width=\linewidth]{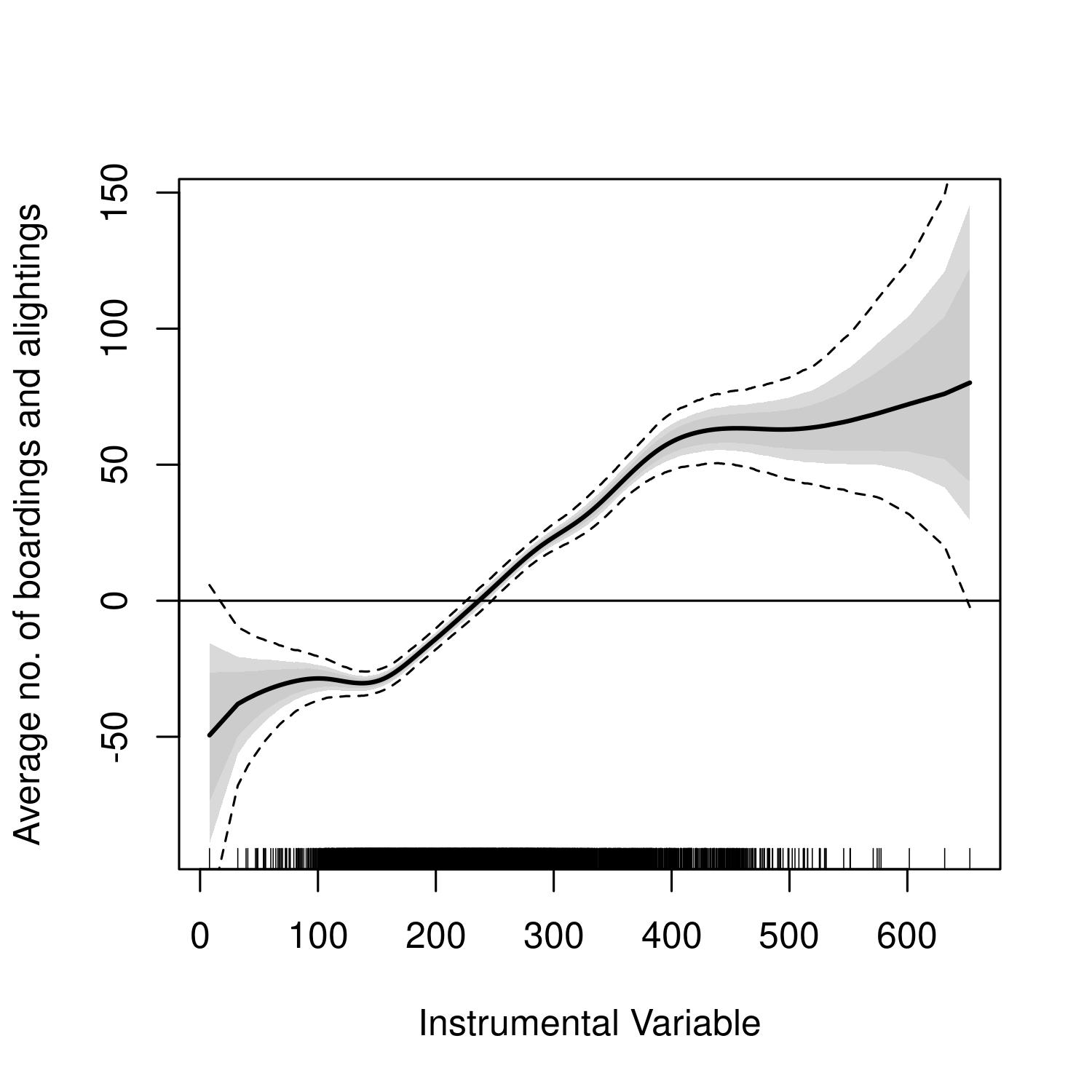}
        \subcaption{Wong Tai Sin Station}
    \end{subfigure}
\hfill
    \begin{subfigure}{0.32\textwidth}
        \includegraphics[width=\linewidth]{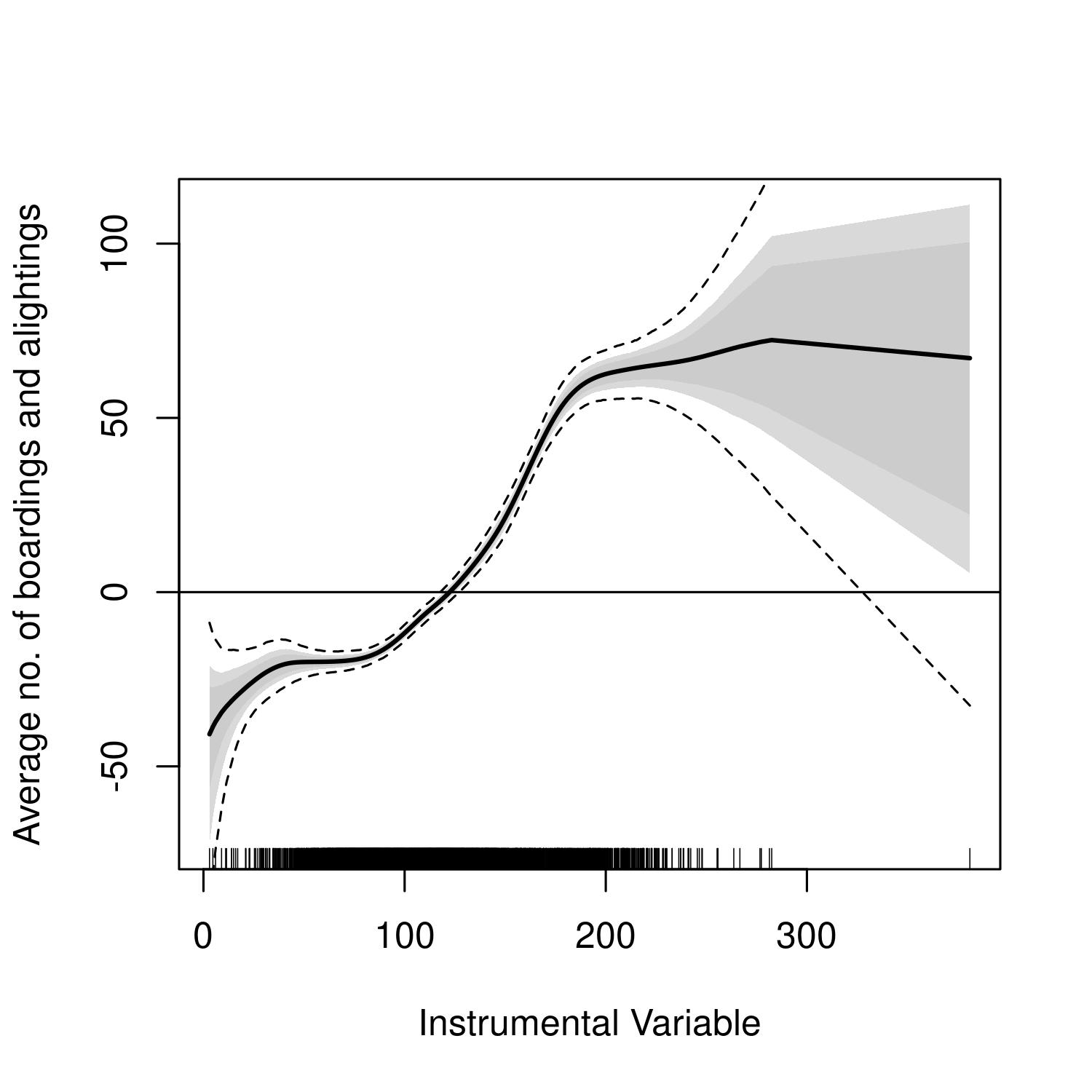}
        \subcaption{Lok Fu Station}
    \end{subfigure}
\hfill
    \begin{subfigure}{0.32\textwidth}
        \includegraphics[width=\linewidth]{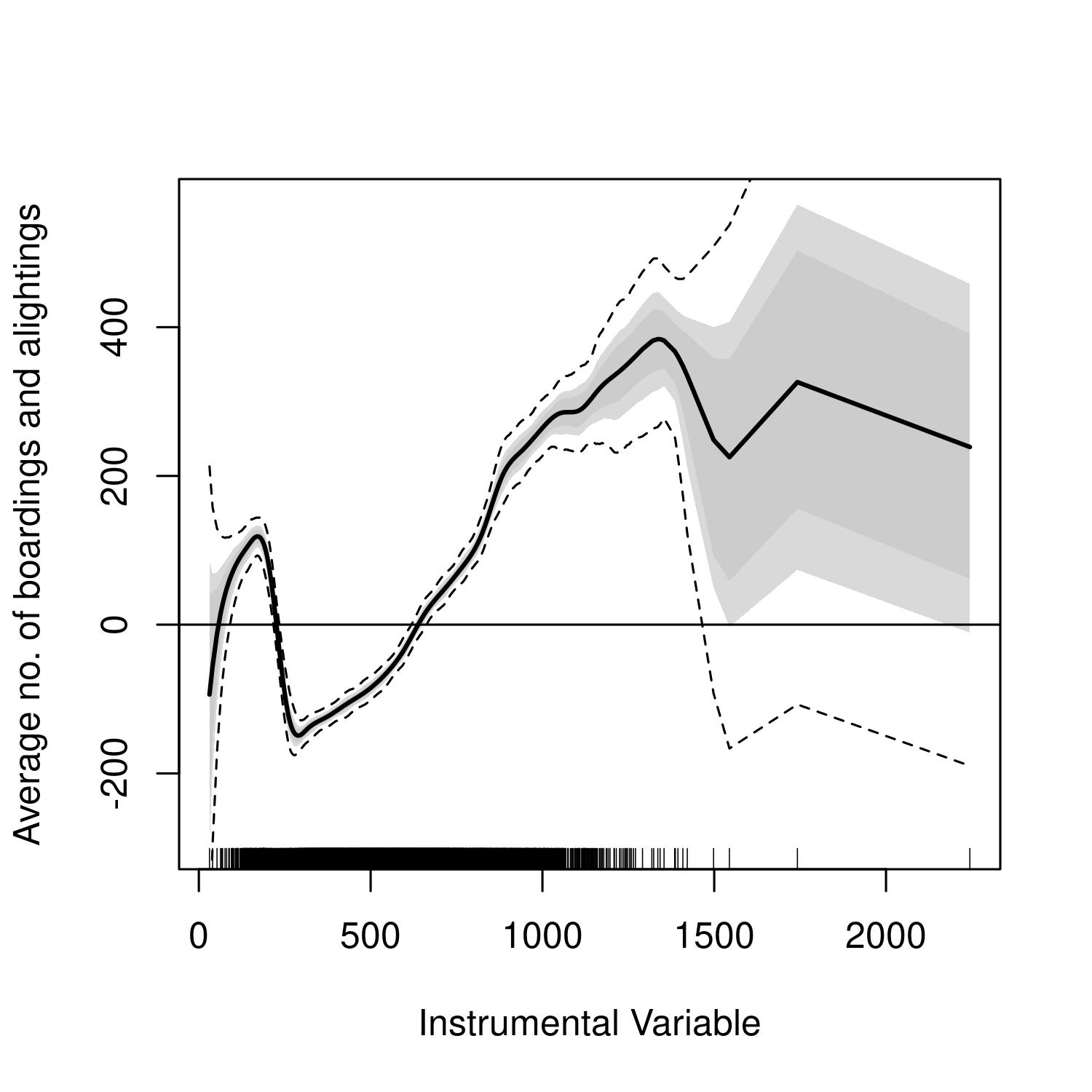}
        \subcaption{Kowloon Tong Station}
    \end{subfigure}
\hfill
    \begin{subfigure}{0.32\textwidth}
        \includegraphics[width=\linewidth]{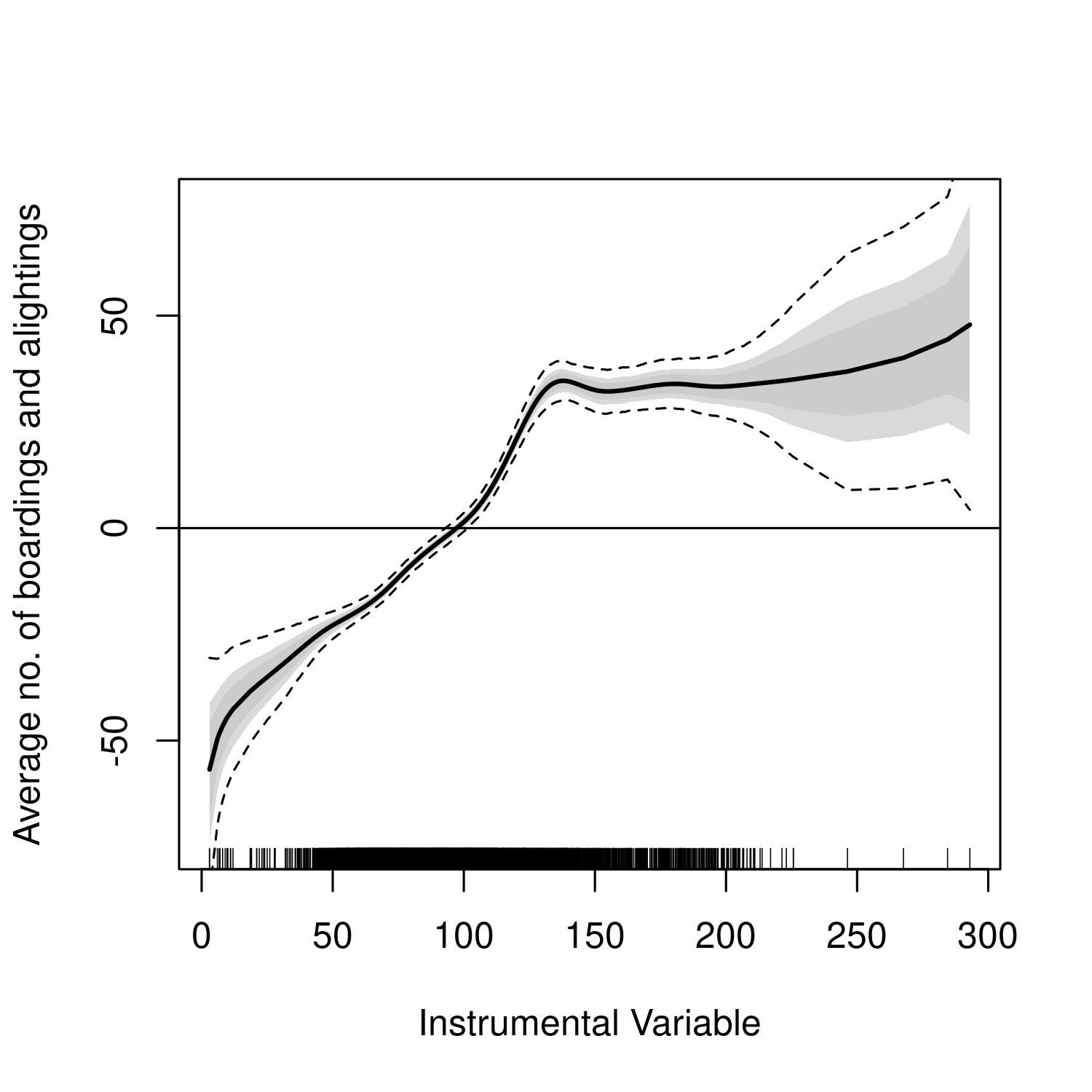}
        \subcaption{Shek Kip Mei Station}
    \end{subfigure}
\hfill
    \begin{subfigure}{0.32\textwidth}
        \includegraphics[width=\linewidth]{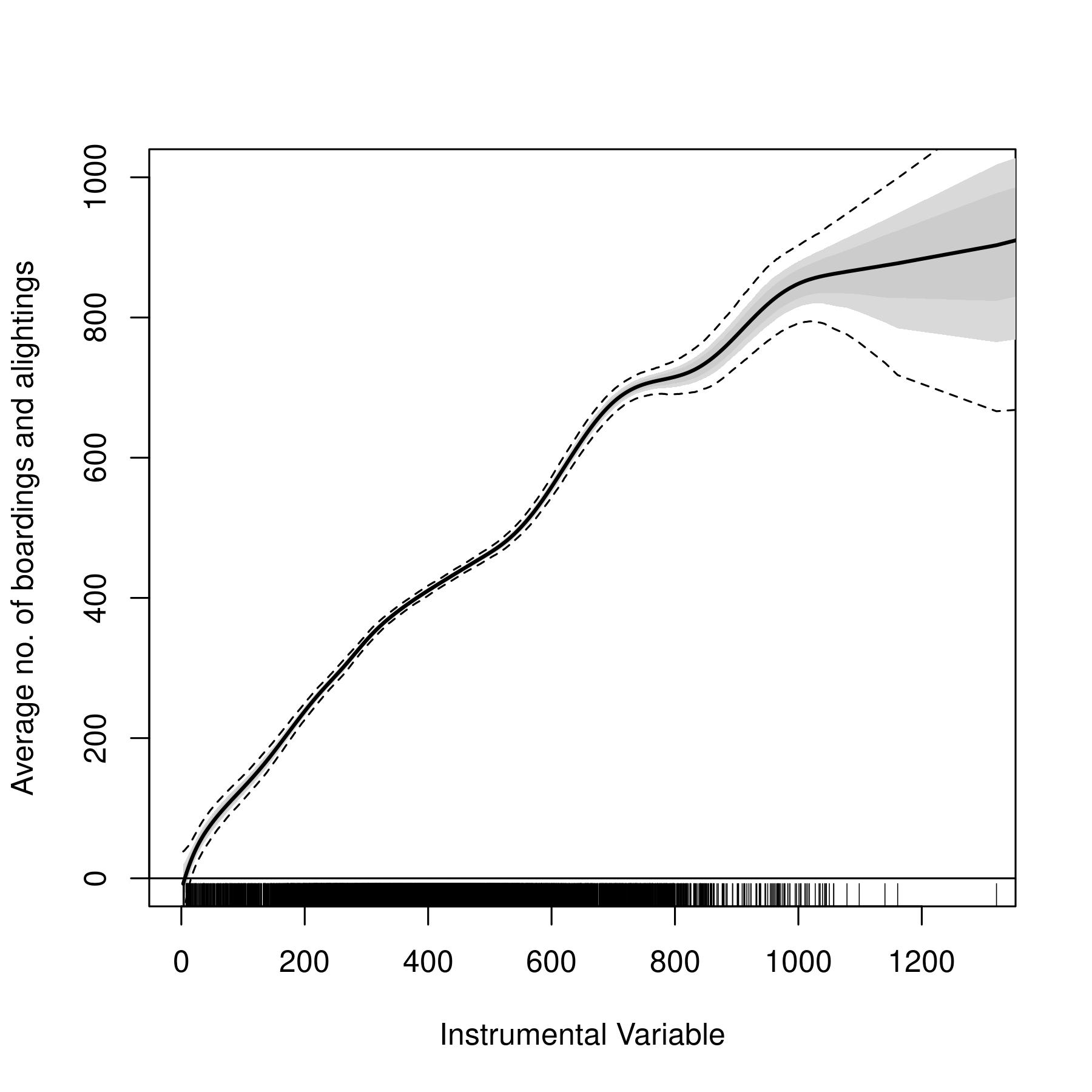}
        \subcaption{Prince Edward Station}
    \end{subfigure}
\hfill
    \begin{subfigure}{0.32\textwidth}
        \includegraphics[width=\linewidth]{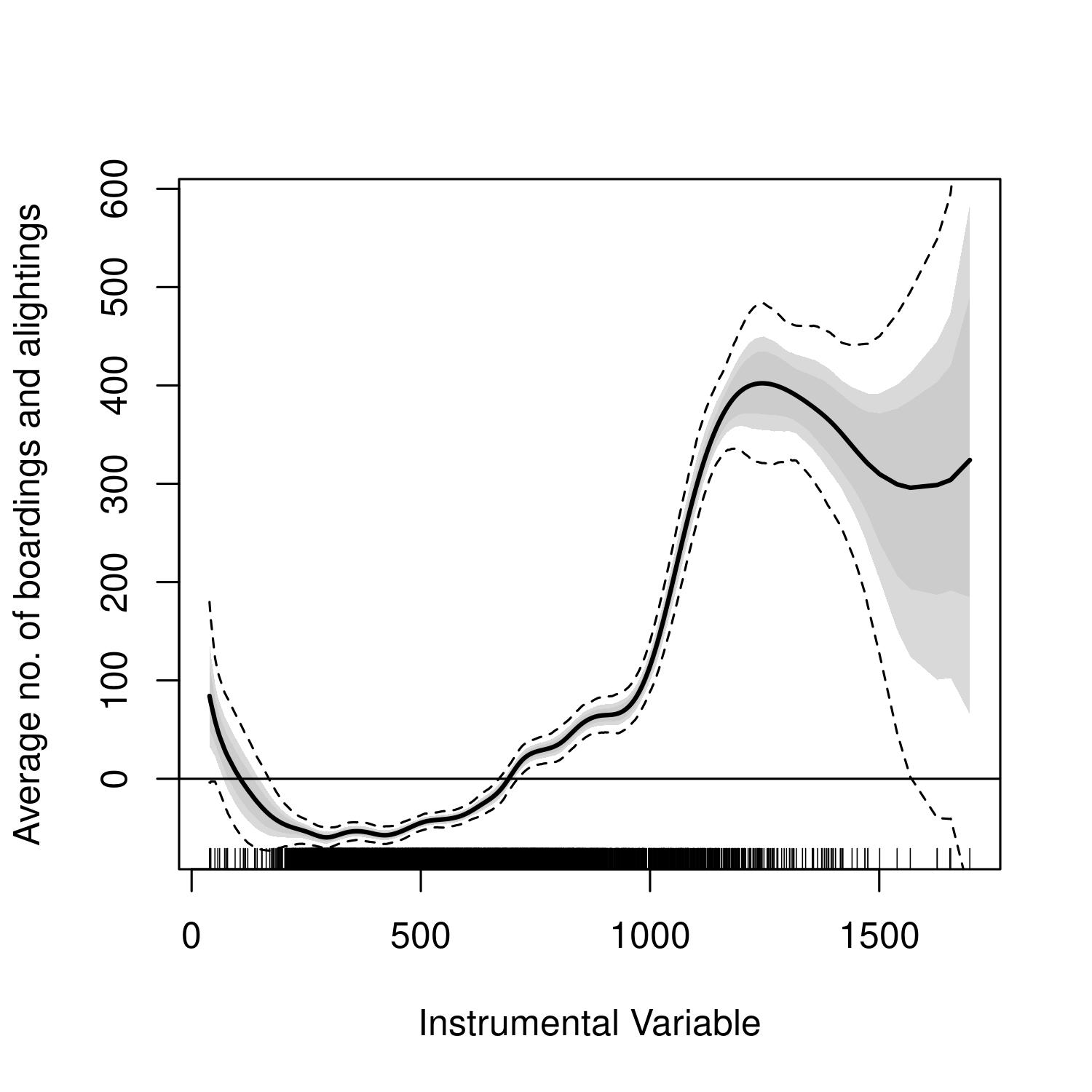}
        \subcaption{Mong Kok Station}
    \end{subfigure}
\hfill
    \begin{subfigure}{0.32\textwidth}
        \includegraphics[width=\linewidth]{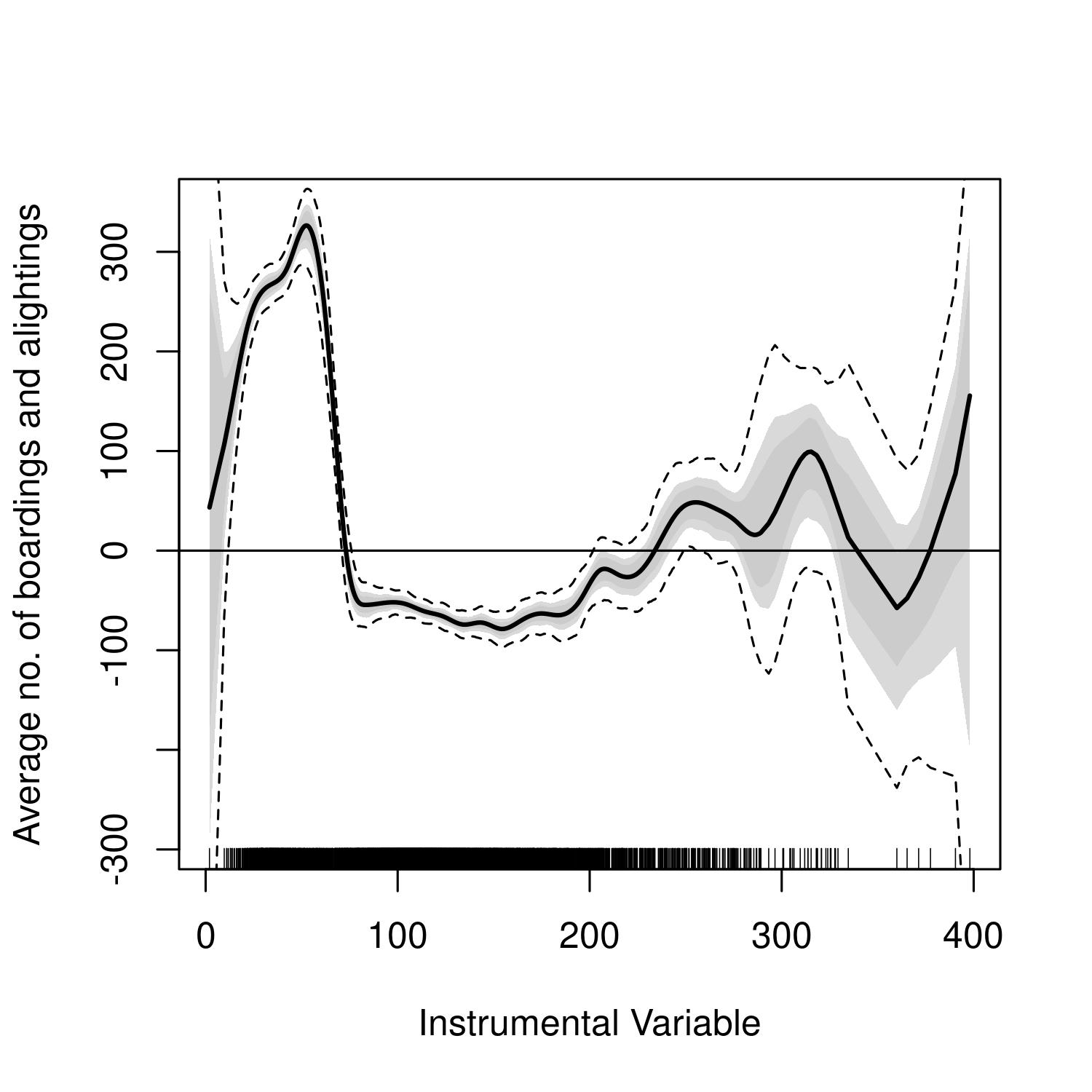}
        \subcaption{Yau Ma Tei station}
    \end{subfigure}
    \caption[]{Relevance of instruments in analysis of train movements in the upward direction along the Kwun Tong Line (dotted lines represent 95\% credible intervals).}
    \label{fig:Inst_Str_U}
\end{figure}

Figures \ref{fig:Inst_Str_D} and \ref{fig:Inst_Str_U} illustrate the results (that is, the estimated $h(.)$) from regression of the confounding covariate over the chosen instrument for the two directions of train flows at all analysed stations. 

From these figures, we notice a strong correlation between the instrument and the confounding covariate. These figures provide supporting evidence that the selected instruments satisfy the relevance condition.

\section{Distribution of Errors}
\label{S:VII}

\begin{figure}[tp]
    \centering
    \begin{subfigure}{0.32\textwidth}
        \includegraphics[width=\linewidth]{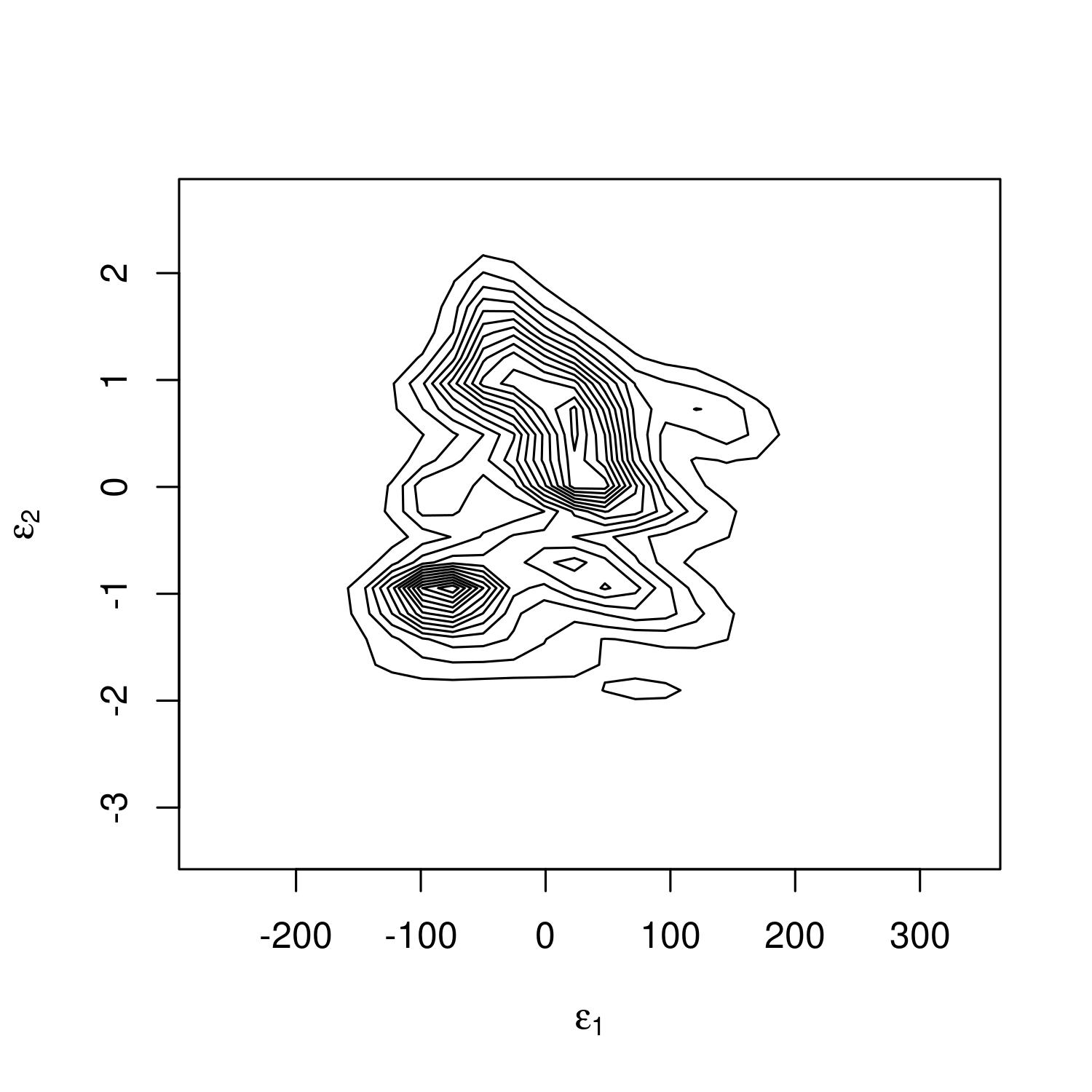}
        \subcaption{Wong Tai Sin Station}
    \end{subfigure}
\hfill
    \begin{subfigure}{0.32\textwidth}
        \includegraphics[width=\linewidth]{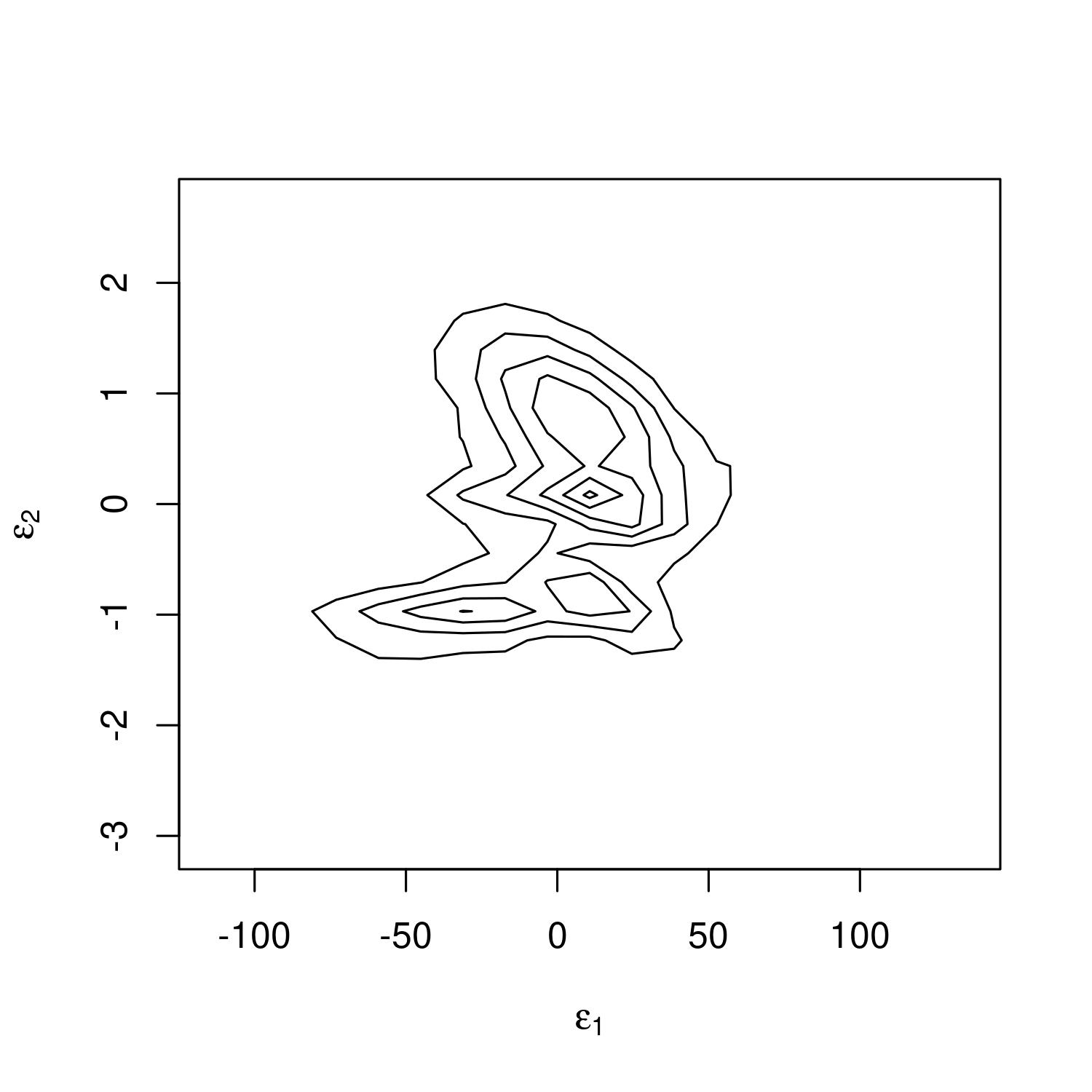}
        \subcaption{Lok Fu Station}
    \end{subfigure}
\hfill
    \begin{subfigure}{0.32\textwidth}
        \includegraphics[width=\linewidth]{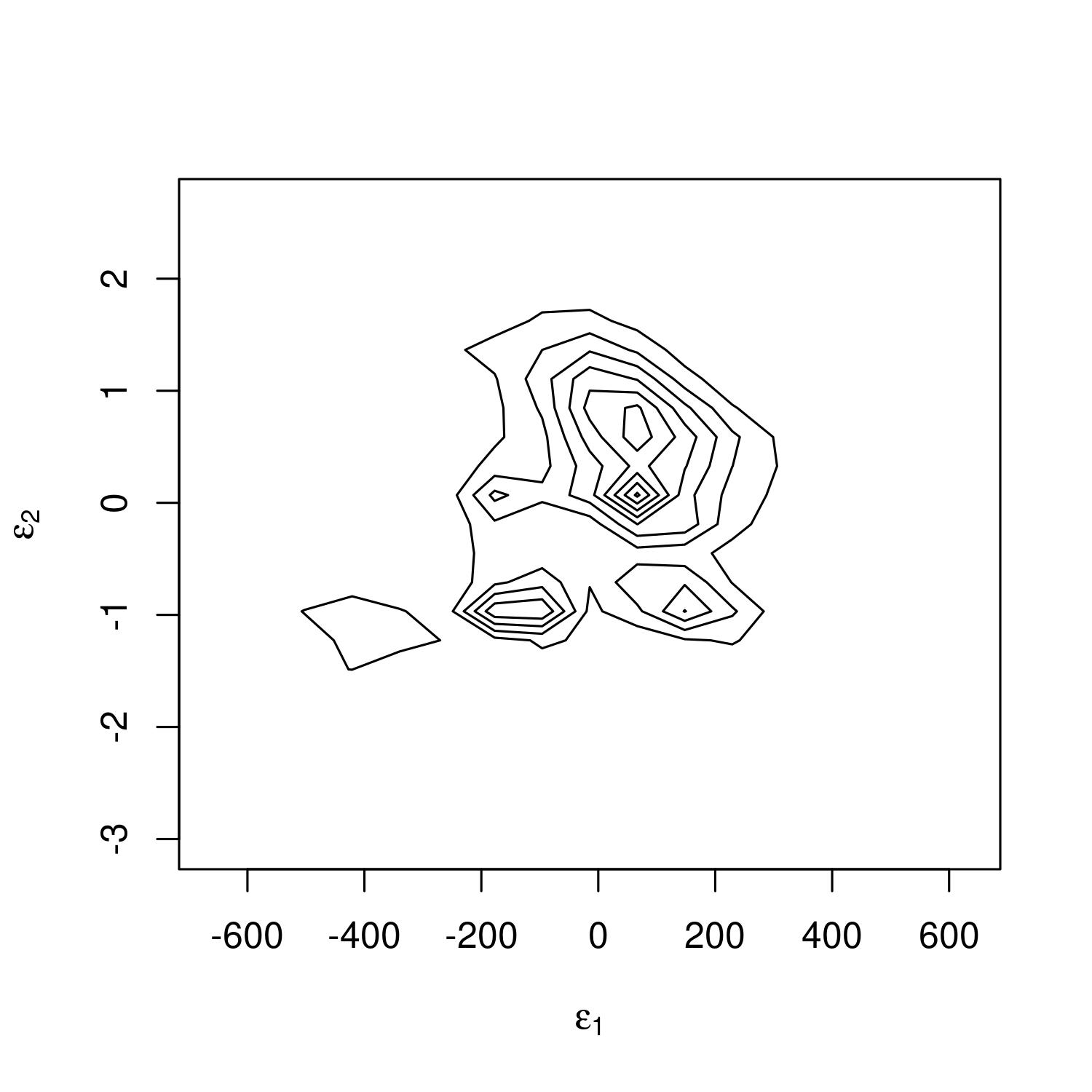}
        \subcaption{Kowloon Tong Station}
    \end{subfigure}
\hfill
    \begin{subfigure}{0.32\textwidth}
        \includegraphics[width=\linewidth]{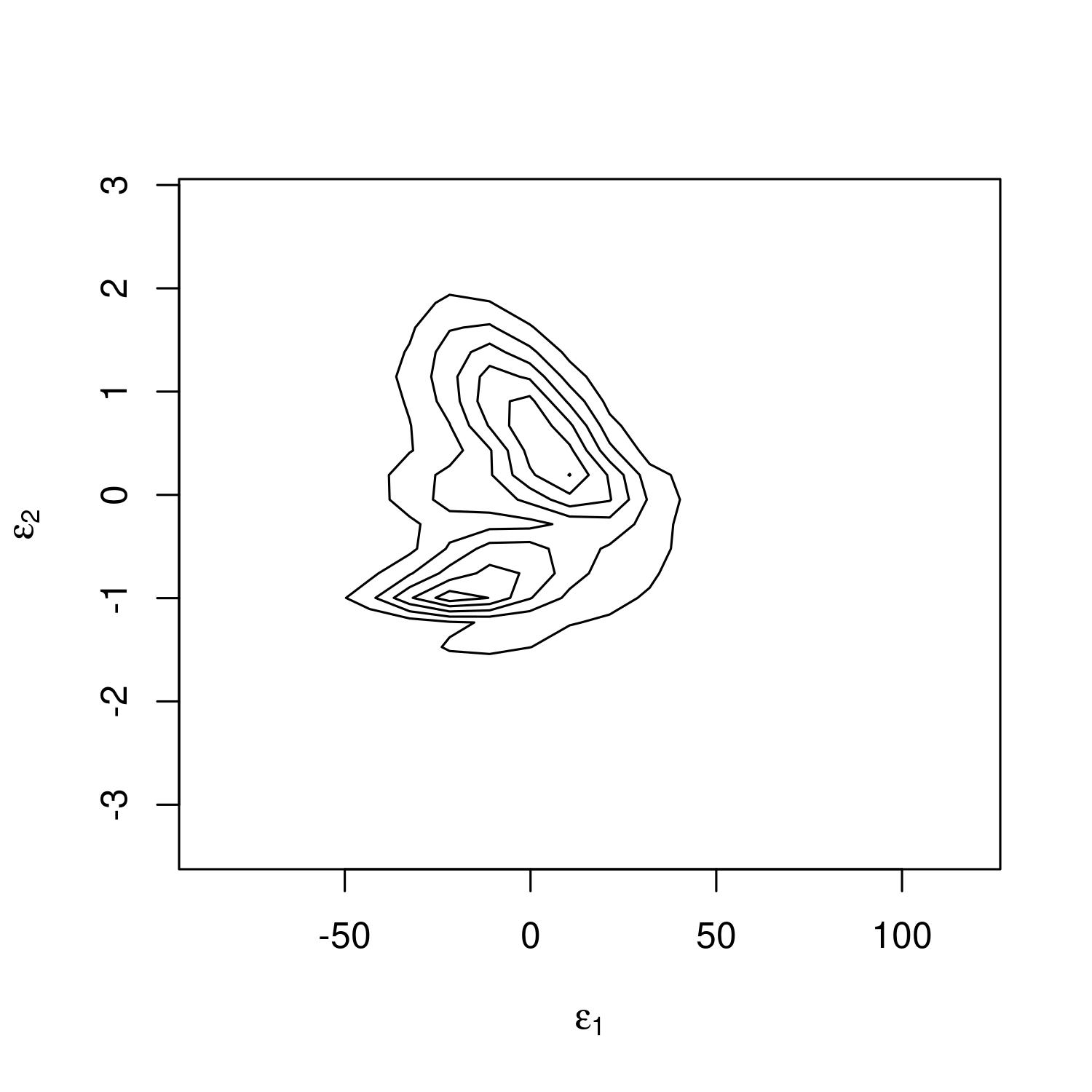}
        \subcaption{Shek Kip Mei Station}
    \end{subfigure}
\hfill
    \begin{subfigure}{0.32\textwidth}
        \includegraphics[width=\linewidth]{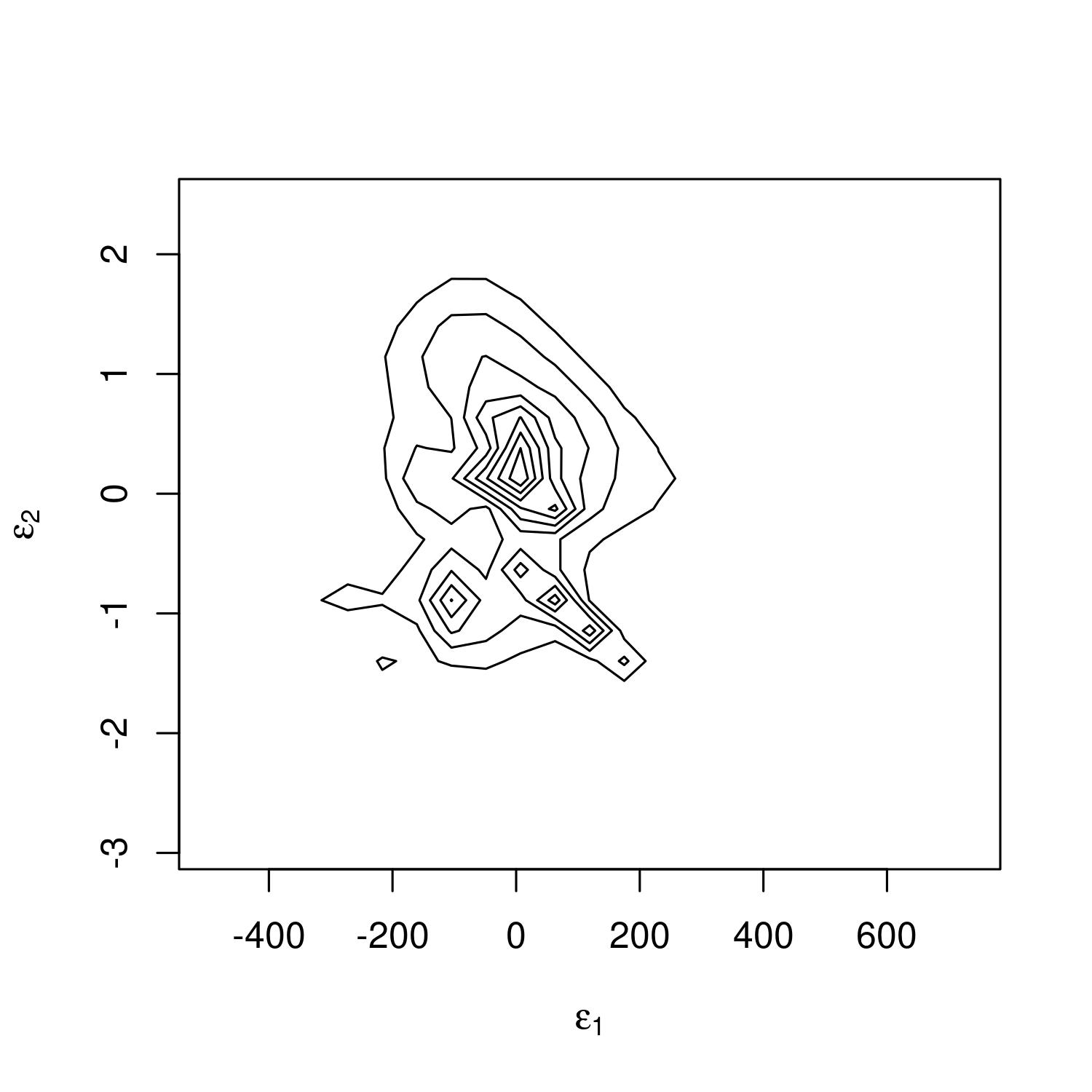}
        \subcaption{Prince Edward Station}
    \end{subfigure}
\hfill
    \begin{subfigure}{0.32\textwidth}
        \includegraphics[width=\linewidth]{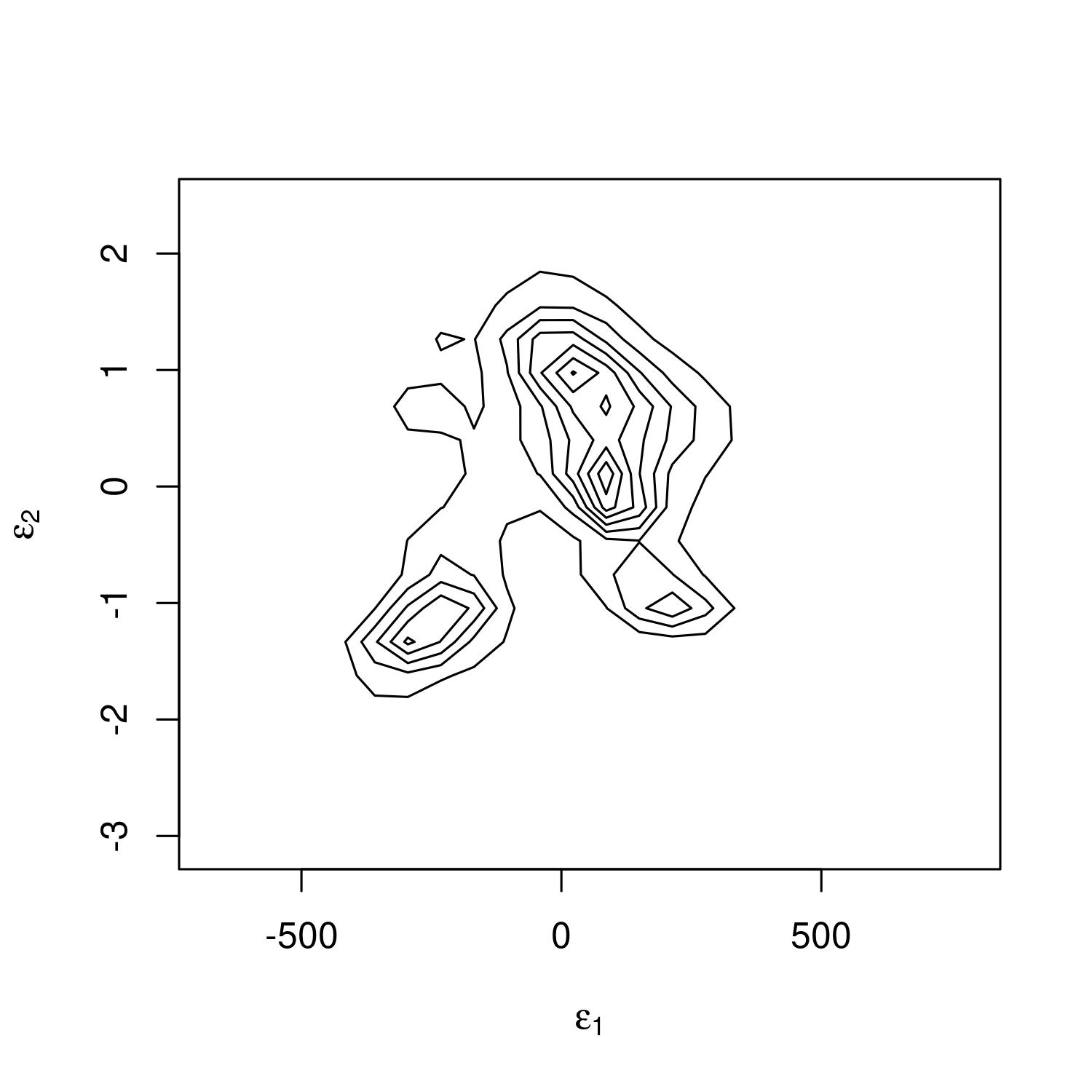}
        \subcaption{Mong Kok Station}
    \end{subfigure}
\hfill
    \begin{subfigure}{0.32\textwidth}
        \includegraphics[width=\linewidth]{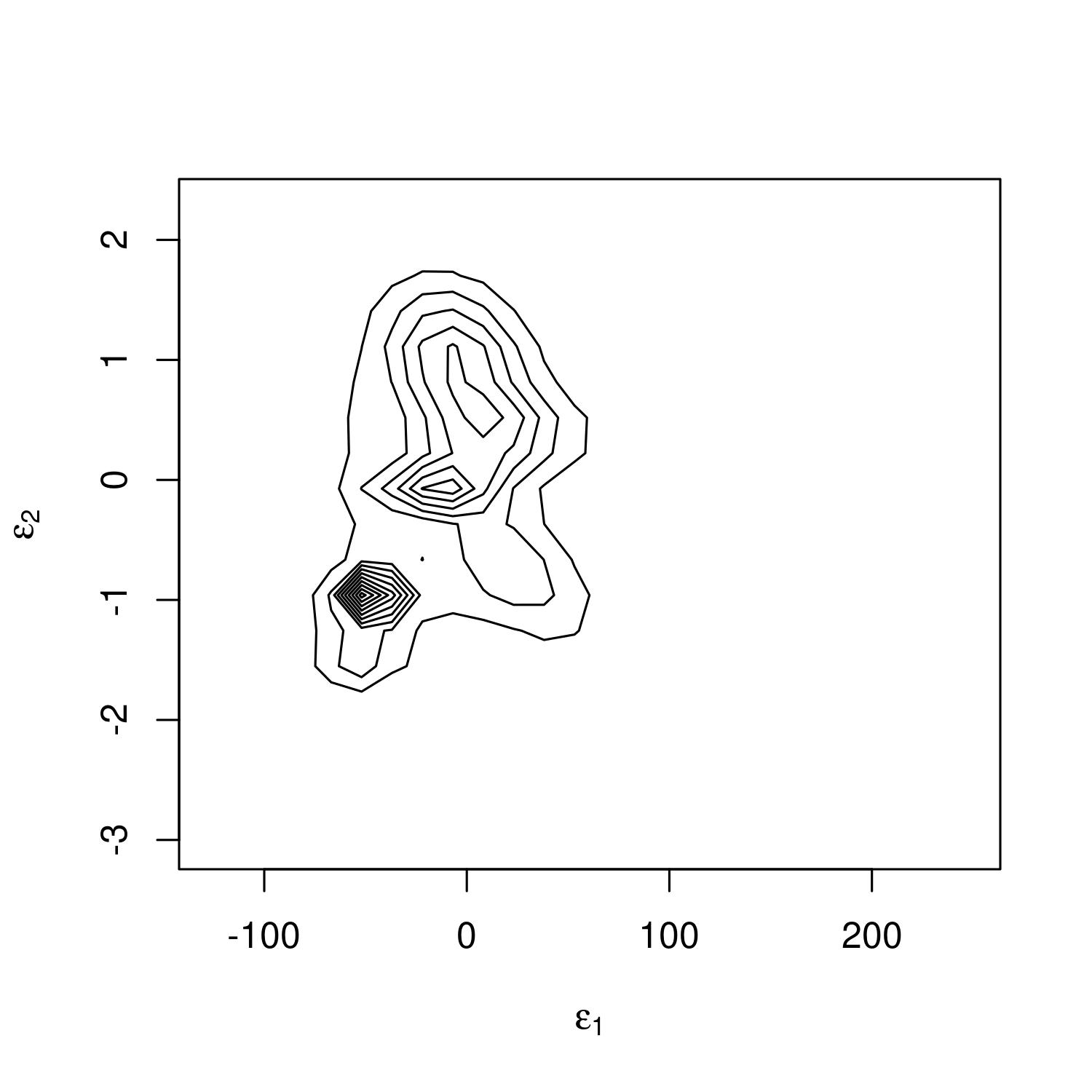}
        \subcaption{Yau Ma Tei station}
    \end{subfigure}
    \caption[]{Distribution of errors in analysis of train movements in the downward direction along the Kwun Tong Line (dotted lines represent 95\% credible intervals).}
    \label{fig:Err_Dist_D}
\end{figure}

\begin{figure}[tp]
    \centering
    \begin{subfigure}{0.32\textwidth}
        \includegraphics[width=\linewidth]{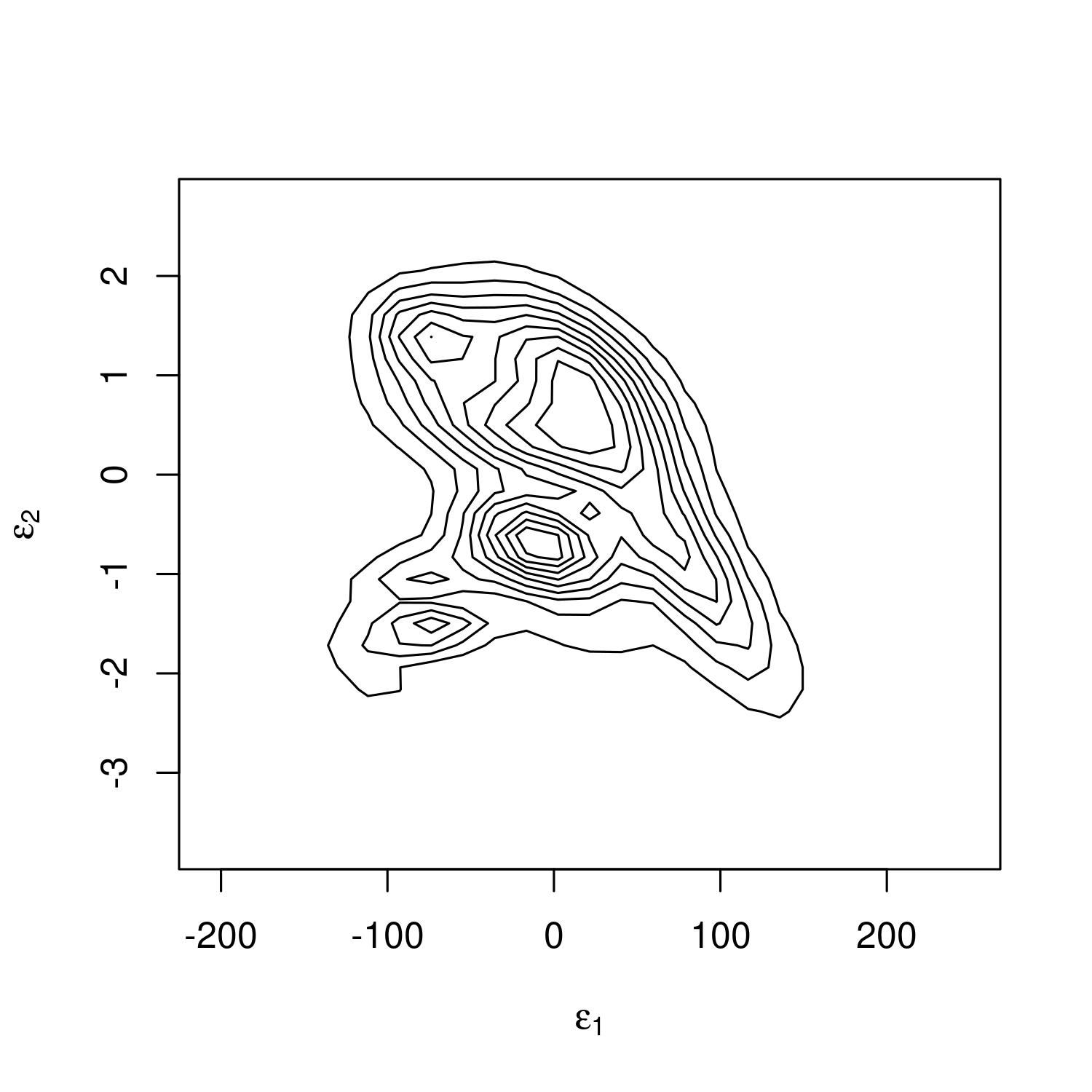}
        \subcaption{Wong Tai Sin Station}
    \end{subfigure}
\hfill
    \begin{subfigure}{0.32\textwidth}
        \includegraphics[width=\linewidth]{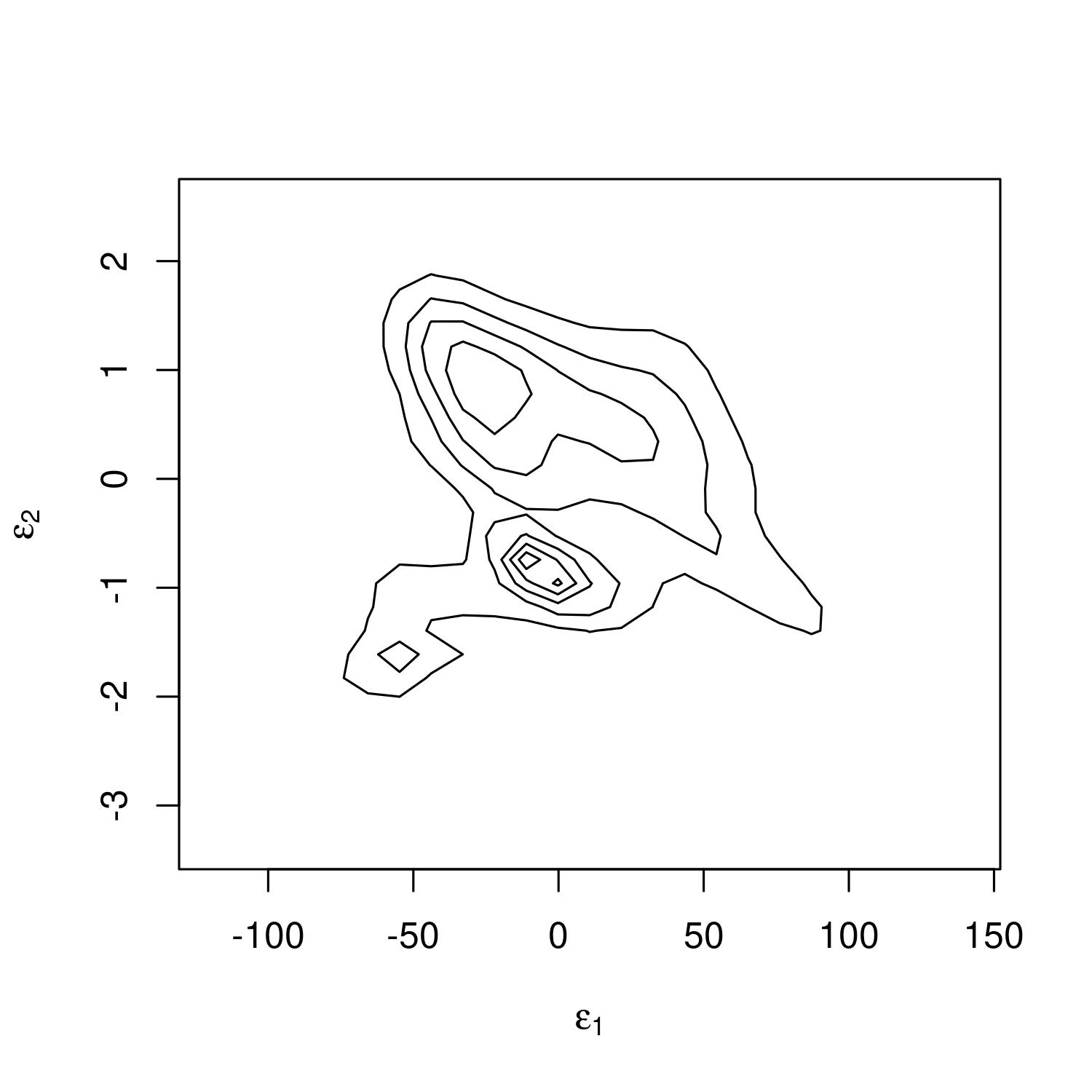}
        \subcaption{Lok Fu Station}
    \end{subfigure}
\hfill
    \begin{subfigure}{0.32\textwidth}
        \includegraphics[width=\linewidth]{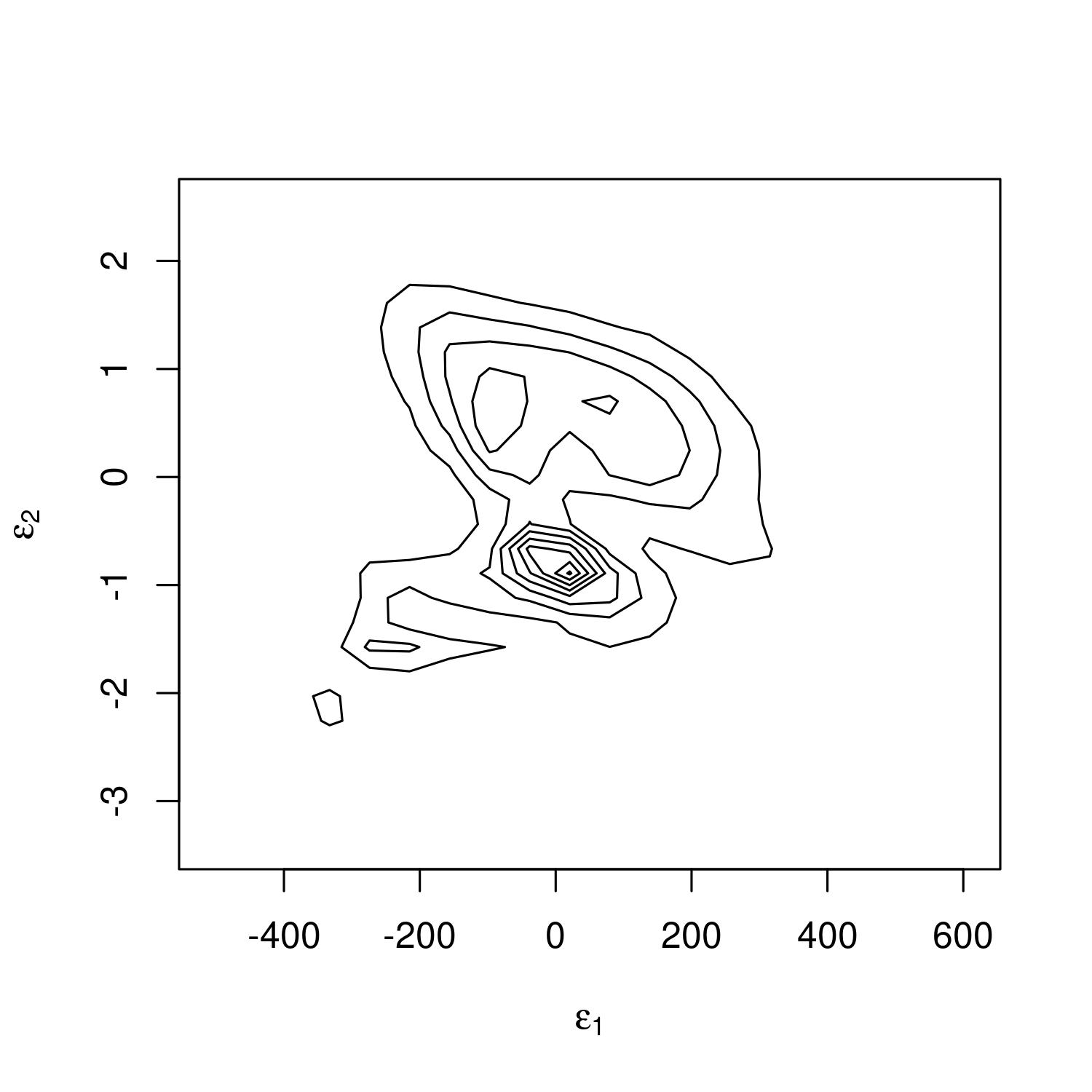}
        \subcaption{Kowloon Tong Station}
    \end{subfigure}
\hfill
    \begin{subfigure}{0.32\textwidth}
        \includegraphics[width=\linewidth]{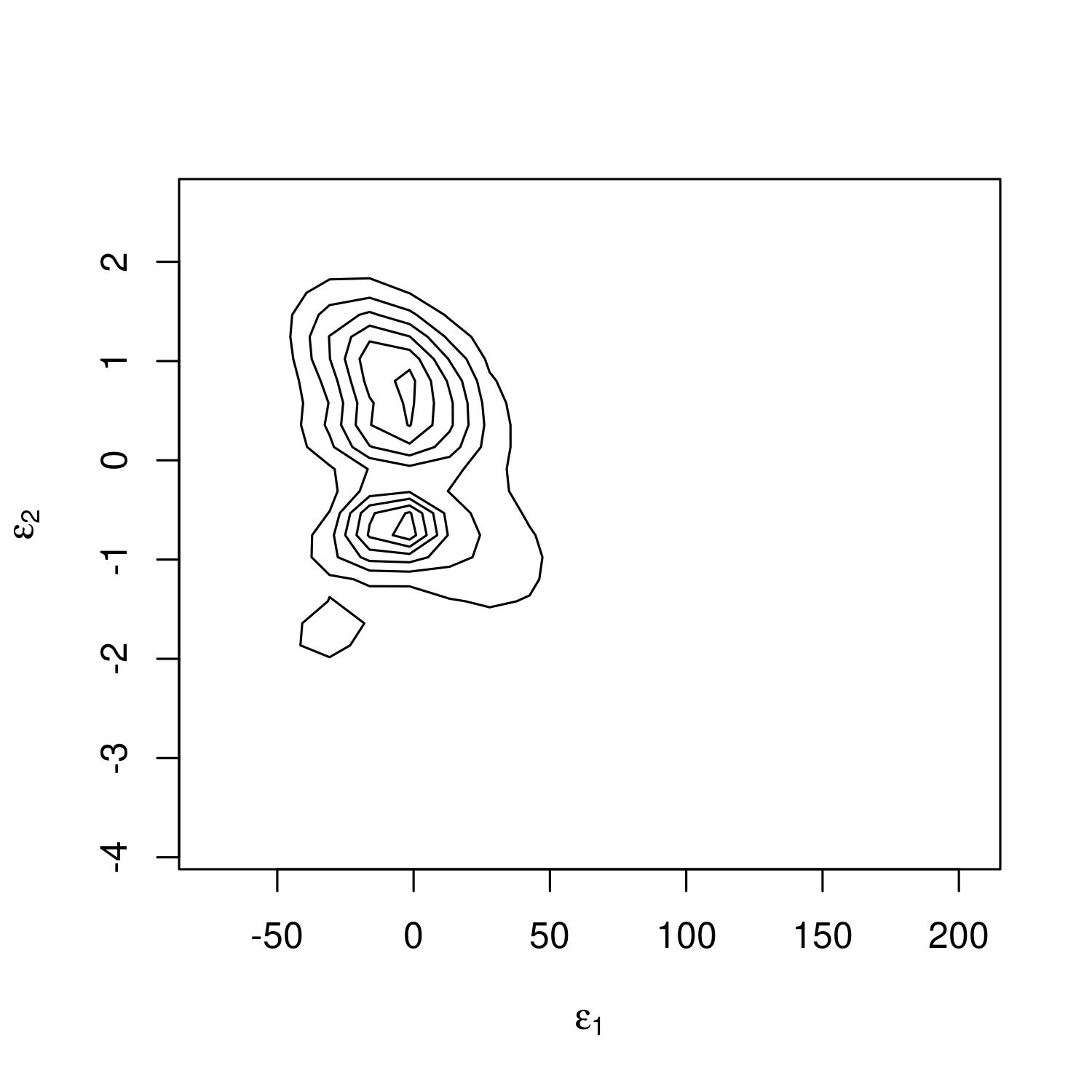}
        \subcaption{Shek Kip Mei Station}
    \end{subfigure}
\hfill
    \begin{subfigure}{0.32\textwidth}
        \includegraphics[width=\linewidth]{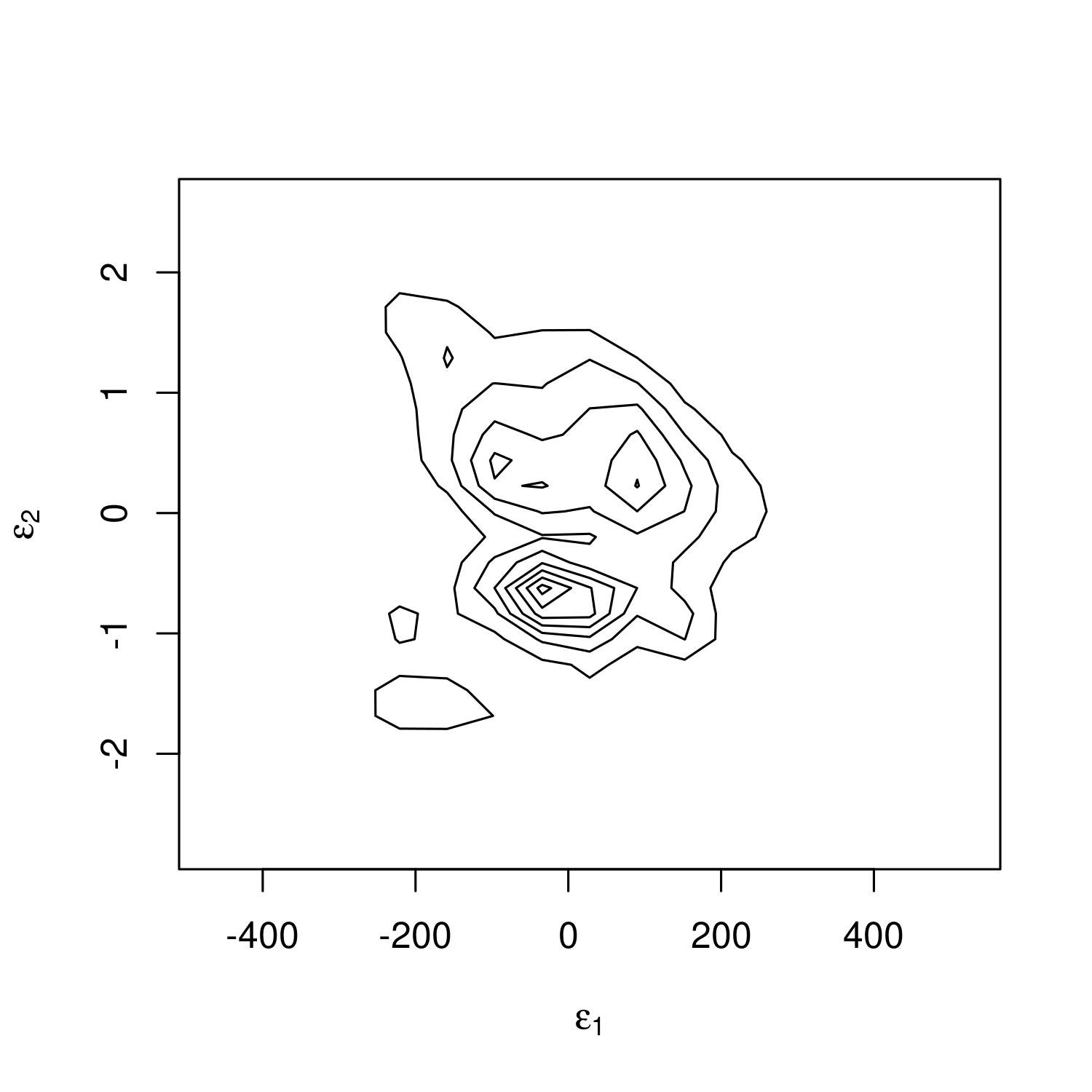}
        \subcaption{Prince Edward Station}
    \end{subfigure}
\hfill
    \begin{subfigure}{0.32\textwidth}
        \includegraphics[width=\linewidth]{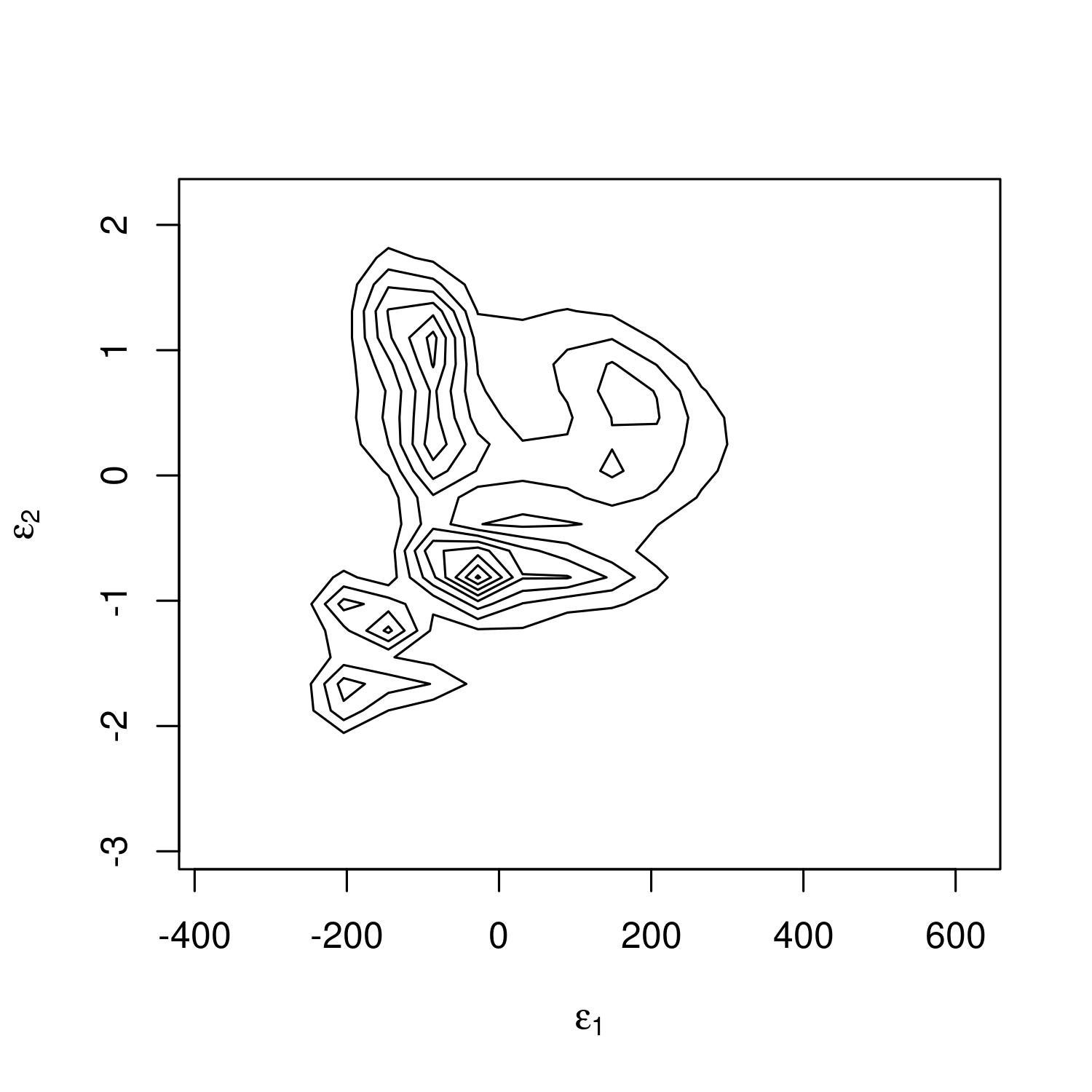}
        \subcaption{Mong Kok Station}
    \end{subfigure}
\hfill
    \begin{subfigure}{0.32\textwidth}
        \includegraphics[width=\linewidth]{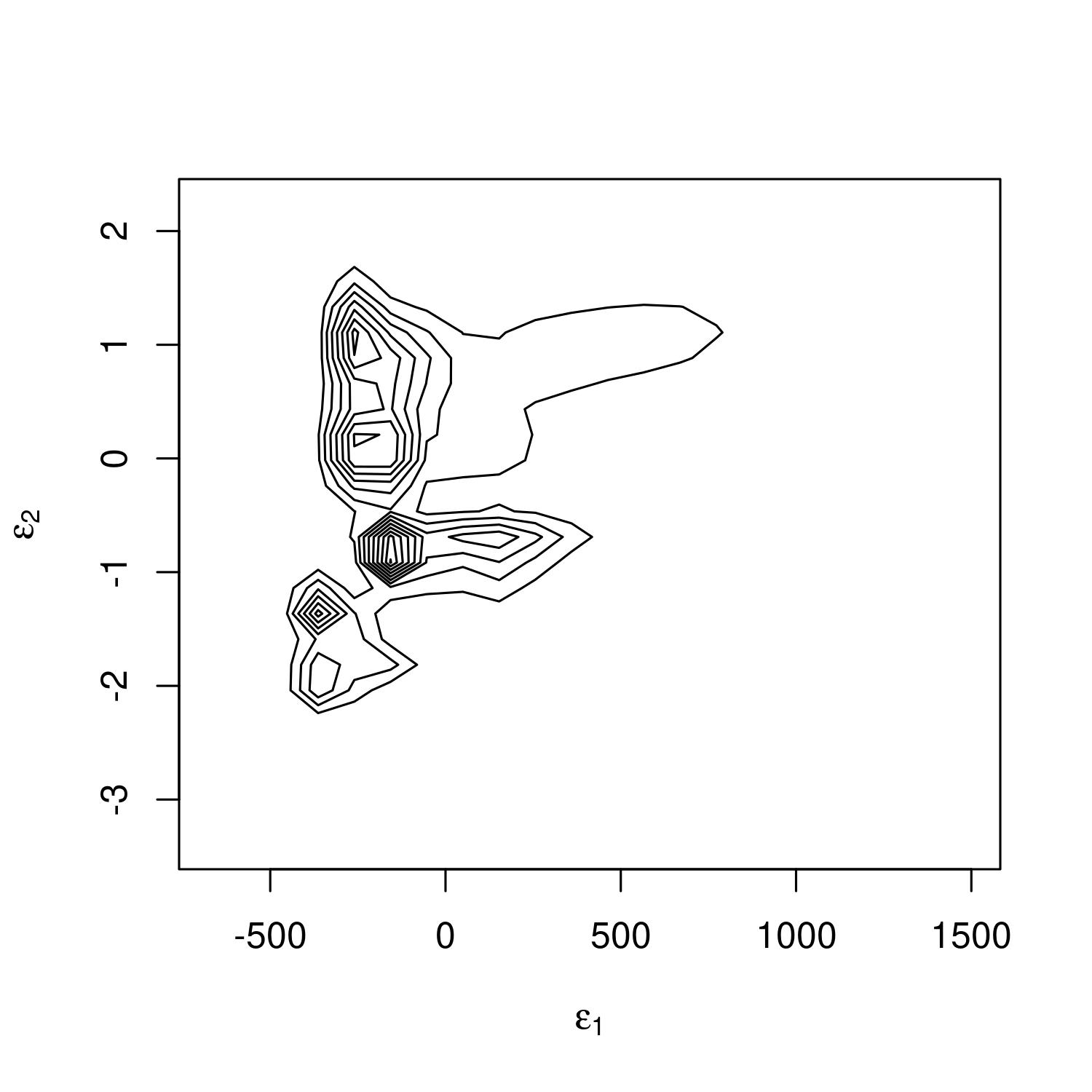}
        \subcaption{Yau Ma Tei station}
    \end{subfigure}
    \caption[]{Distribution of errors in analysis of train movements in the upward direction along the Kwun Tong Line (dotted lines represent 95\% credible intervals).}
    \label{fig:Err_Dist_U}
\end{figure}

Figures \ref{fig:Err_Dist_D} and \ref{fig:Err_Dist_U} show the contour plot of the joint distribution of errors from the second stage ($\epsilon_2$) and the first stage ($\epsilon_1$) for both directions of train flows at all analysed stations. From these figures, we observe the joint error distribution is bi-modal. 

The results suggest that the estimates of $S(.)$ from traditional econometric methods could have poor finite sample properties because they generally assume uni-modal symmetric and thin-tailed Gaussian error distributions. The adopted Bayesian \acrshort{NPIV} method addresses all these potential challenges by allowing for a flexible distribution of errors, instead of assuming a restrictive parametric error distribution.

\bigskip

\section*{Acknowledgement}
Authors are grateful to Hong Kong \acrshort{MTR} for providing data used in the empirical study. They are also thankful to Manuel Wiesenfarth for helpful email conversations and support in figuring out the implementation details of the adopted method. The authors are also grateful to Alexander Barron from the Transport Strategy Centre at Imperial College London for his insightful suggestions on the manuscript.

\printglossary[type=\acronymtype]

\begin{table}[tp]
\caption{\vspace{-.5cm}Summary statistics for variables used in the analysis.} \label{tab:sumstats}
\begin{center}
\resizebox{\textwidth}{!}{
\begin{tabular}{llllrrrrr}
  \hline
  Station & Direction & Time-of-day & Variable &  Obs. & Min & Max & Mean & Std.Dev \\ 
  \hline
  Wong Tai & downward & morning & train flow (tr/10min) & 2652 & 0.40 & 6.08 & 3.64 & 1.07 \\
  & & & boarding-alightings & 2652 & 32.00 & 653.00 & 268.24 & 92.52 \\
  & & afternoon & train flow (tr/10min) & 2943 & 0.47 & 6.61 & 3.72 & 0.83 \\
  & & & boarding-alightings & 2943 & 8.00 & 529.75 & 202.99 & 57.56 \\
  & upward & morning & train flow (tr/10min) & 2428 & 0.41 & 6.05 & 3.71 & 1.16 \\
  & & & boarding-alightings & 2428 & 6.00 & 506.0 & 178.67 & 62.23 \\
  & & afternoon & train flow (tr/10min) & 2940 & 0.48 & 6.05 & 3.85 & 0.84 \\
  & & & boarding-alightings & 2940 & 40.00 & 477.00 & 247.23 & 59.54 \\
  Lok Fu & downward & morning & train flow (tr/10min) & 2640 & 0.40 & 7.30 & 3.68 & 1.08 \\
  & & & boarding-alightings & 2640 & 3.00 & 381.00 & 120.92 & 47.42 \\
  & & afternoon & train flow (tr/10min) & 2448 & 0.53 & 6.05 & 3.69 & 1.15 \\
  & & & boarding-alightings & 2448 & 3.00 & 283.50 & 78.28 & 31.56 \\
  & upward & morning & train flow (tr/10min) & 2942 & 0.60 & 6.34 & 3.74 & 0.84 \\
  & & & boarding-alightings & 2942 & 4.67 & 214.00 & 104.43 & 27.82 \\
   & & afternoon & train flow (tr/10min) & 2943 & 0.53 & 5.97 & 3.83 & 0.84 \\
  & & & boarding-alightings & 2943 & 5.00 & 279.67 & 139.35 & 36.44 \\
  Kowloon Tong & downward & morning & train flow (tr/10min) & 2649 & 0.40 & 7.22 & 3.66 & 1.07 \\
  & & & boarding-alightings & 2649 & 31.00 & 2244.00 & 636.39 & 260.37 \\
  & & afternoon & train flow (tr/10min) & 2942 & 0.94 & 6.26 & 3.71 & 0.82 \\
  & & & boarding-alightings & 2942 & 170.00 & 1241.67 & 536.97 & 154.68 \\
  & upward & morning & train flow (tr/10min) & 2490 & 0.41 & 6.25 & 3.64 & 1.14 \\
  & & & boarding-alightings & 2490 & 20.00 & 1574.50 & 397.68 & 170.82 \\
  & & afternoon & train flow (tr/10min) & 2943 & 0.50 & 5.81 & 3.80 & 0.81 \\
  & & & boarding-alightings & 2943 & 202.00 & 1432.67 & 664.96 & 148.36 \\
  Shek Kip Mei & downward & morning & train flow (tr/10min) & 2584 & 0.40 & 6.57 & 3.69 & 1.08 \\
  & & & boarding-alightings & 2584 & 3.00 & 293.00 & 101.15 & 36.31 \\
  & & afternoon & train flow (tr/10min) & 2940 & 0.50 & 6.05 & 3.72 & 0.83 \\
  & & & boarding-alightings & 2940 & 3.00 & 217.50 & 91.49 & 31.36 \\
  & upward & morning & train flow (tr/10min) & 2501 & 0.40 & 6.02 & 3.64 & 1.14 \\
  & & & boarding-alightings & 2501 & 5.33 & 406.33 & 85.16 & 50.15 \\
  & & afternoon & train flow (tr/10min) & 2941 & 0.86 & 6.07 & 3.79 & 0.81 \\
  & & & boarding-alightings & 2941 & 6.33 & 173.00 & 94.00 & 23.05 \\
  Prince Edward & downward & morning & train flow (tr/10min) & 2585 & 0.33 & 7.00 & 3.69 & 1.09 \\
  & & & boarding-alightings & 2585 & 33.00 & 1523.00 & 477.85 & 187.98 \\
  & & afternoon & train flow (tr/10min) & 2943 & 1.08 & 6.04 & 3.74 & 0.82 \\
  & & & boarding-alightings & 2943 & 112.67 & 1485.00 & 495.24 & 174.99 \\
  & upward & morning & train flow (tr/10min) & 2500 & 0.40 & 5.08 & 3.63 & 1.10\\
  & & & boarding-alightings & 2500 & 9.00 & 1703.50 & 407.71 & 192.08\\
  & & afternoon & train flow (tr/10min) & 2943 & 0.57 & 5.82 & 3.73 & 0.79 \\
  & & & boarding-alightings & 2943 & 47.00 & 1078.67 & 494.01 & 127.55 \\
  Mong Kok & downward & morning & train flow (tr/10min) & 2583 & 0.34 & 8.04 & 3.70 & 1.10 \\
  & & & boarding-alightings & 2583 & 39.00 & 1697.68 & 679.95 & 273.58 \\
  & & afternoon & train flow (tr/10min) & 2944 & 0.56 & 6.32 & 3.74 & 0.82 \\
  & & & boarding-alightings & 2944 & 96.00 & 1654.00 & 505.47 & 171.92 \\
  & upward & morning & train flow (tr/10min) & 2503 & 0.40 & 5.83 & 3.62 & 1.09 \\
  & & & boarding-alightings & 2503 & 12.00 & 547.00 & 160.19 & 64.86 \\
  & & afternoon & train flow (tr/10min) & 2943 & 0.87 & 5.71 & 3.71 & 0.77 \\
  & & & boarding-alightings & 2943 & 131.00 & 1595.00 & 429.57 & 140.30 \\
  Yau Ma Tei & downward & morning & train flow (tr/10min) & 2569 & 0.34 & 8.25 & 3.69 & 1.09 \\
  & & & boarding-alightings & 2569 & 2.00 & 245.60 & 87.03 & 39.20 \\
  & & afternoon & train flow (tr/10min) & 2942 & 0.49 & 6.00 & 3.75 & 0.80 \\
  & & & boarding-alightings & 2942 & 43.00 & 398.00 & 147.13 & 47.62 \\
  & upward & morning & train flow (tr/10min) & 2504 & 0.33 & 6.13 & 3.61 & 1.08 \\
  & & & boarding-alightings & 2504 & 41.00 & 991.50 & 276.35 & 110.63 \\
  & & afternoon & train flow (tr/10min) & 2935 & 1.00 & 5.60 & 3.68 & 0.78 \\
  & & & boarding-alightings & 2935 & 129.67 & 2798.00 & 774.91 & 275.43 \\
  \hline
  \multicolumn{8}{l} {*Obs.: Number of observations, Std. Dev.: Standard Deviation,tr/10min: trains per ten minutes} \\ 
  \multicolumn{8}{l} {**boarding-alightings represent average no. of boardings and alightings per train} \\ 
\end{tabular}}
\end{center}
\end{table}

\newpage
\bibliographystyle{agsm}
\bibliography{TransitFD}

@article{TfL2020,
  title={Transport for {L}ondon travel in {London} performance report},
  author={{TfL}},
  journal={Transport for London},
  year={2020},
  url = {\url{https://content.tfl.gov.uk/travel-in-london-report-13.pdf}}
}

@inproceedings{Seo2017,
	author={Toru Seo and Kentaro Wada and Daisuke Fukuda},
	year={2017},
	title={A macroscopic and dynamic model of urban rail transit with delay and congestion},
	booktitle={96th Annual Meeting of the Transportation Research Board}
}

@article{Horcher2017,
	author={Daniel Hörcher and Daniel J. Graham and Richard J. Anderson},
	year={2017},
	title={Crowding cost estimation with large scale smart card and vehicle location data},
	journal={Transportation Research Part B: Methodological},
	volume={95},
	pages={105-125}
}

@article{Wisenfarth2014,
	author={Manuel Wiesenfarth and Carlos Matías Hisgen and Thomas Kneib and Carmen Cadarso-Suarez},
	year={2014},
	title={Bayesian nonparametric instrumental variables regression based on penalized splines and dirichlet process mixtures},
	journal={Journal of Business and Economic Statistics},
	volume={32},
	number={3},
	pages={468-482}
}

@article{Tirachini2013,
  title={Crowding in public transport systems: effects on users, operation and implications for the estimation of demand},
  author={Tirachini, Alejandro and Hensher, David A and Rose, John M},
  journal={Transportation research part A: policy and practice},
  volume={53},
  pages={36-52},
  year={2013},
  publisher={Elsevier}
}

@article{Keiji2015,
  title={Simulation analysis of train operation to recover knock-on delay under high-frequency intervals},
  author={Keiji, Kariyazaki and Naohiko, Hibino and Shigeru, Morichi},
  journal={Case Studies on Transport Policy},
  volume={3},
  number={1},
  pages={92-98},
  year={2015},
  publisher={Elsevier}
}

@inproceedings{Zhang2019,
  title={Fundamental diagram of urban rail transit: An empirical investigation by {B}oston’s subway data},
  author={Zhang, Jiahua and Wada, Kentaro},
  year={2019},
  booktitle={8th Symposium of the European Association for Research in Transportation}
}

@article{Yan2012,
  title={Modeling and simulation for urban rail traffic problem based on cellular automata},
  author={Yan, Xu and Cheng-Xun, Cao and Ming-Hua, Li and Jin-Long, Luo},
  journal={Communications in Theoretical Physics},
  volume={58},
  number={6},
  pages={847},
  year={2012},
  publisher={IOP Publishing}
}

@article{Daganzo2009,
  title={A headway-based approach to eliminate bus bunching: {S}ystematic analysis and comparisons},
  author={Daganzo, Carlos F},
  journal={Transportation Research Part B: Methodological},
  volume={43},
  number={10},
  pages={913-921},
  year={2009},
  publisher={Elsevier}
}

@article{Carey1994,
  title={Stochastic approximation to the effects of headways on knock-on delays of trains},
  author={Carey, Malachy and Kwieci{\'n}ski, Andrzej},
  journal={Transportation Research Part B: Methodological},
  volume={28},
  number={4},
  pages={251-267},
  year={1994},
  publisher={Elsevier}
}

@inproceedings{Wada2012,
  title={A control strategy to prevent propagating delays in high-frequency railway systems},
  author={Wada, Kentaro and Akamatsu, Takashi and Osawa, Minoru},
  booktitle={The 1st European Symposium on Quantitative Methods in Transportation Systems},
  year={2012}
}

@article{LA2019,
  title={Delays on the {L}ondon {U}nderground caused by overcrowding},
  author={{London Assembly}},
  journal={Questions to the Mayor, Mayor of London},
  year={2019},
  url = {\url{http://tiny.cc/4mvlsz}}
}

@article{Ind2017,
  title={Tube passengers wasted 400,000 hours in 2016 because of overcrowding delays},
  author={{Independent}},
  journal={Independent, UK},
  year={2017},
  url = {\url{http://tiny.cc/0mvlsz}}
}

@article{TfL2018,
  title={Transport for {L}ondon customer service and operational performance report},
  author={{TfL}},
  journal={Transport for London},
  year={2018},
  url = {\url{http://tiny.cc/8ywlsz}}
}

@article{Daganzo2005,
  title={Improving city mobility through gridlock control: an approach and some ideas},
  author={Daganzo, Carlos F},
  year={2005}
}

@article{Daganzo2007,
  title={Urban gridlock: {M}acroscopic modeling and mitigation approaches},
  author={Daganzo, Carlos F},
  journal={Transportation Research Part B: Methodological},
  volume={41},
  number={1},
  pages={49--62},
  year={2007},
  publisher={Elsevier}
}

@article{NS1992,
  title={A cellular automaton model for freeway traffic},
  author={Nagel, Kai and Schreckenberg, Michael},
  journal={Journal de physique I},
  volume={2},
  number={12},
  pages={2221--2229},
  year={1992},
  publisher={EDP Sciences}
}

@article{Li2005,
  title={Cellular automaton model for railway traffic},
  author={Ke Ping Li and Zi You Gao and Bin Ning},
  year={2005},
  journal={Journal of Computational Physics},
  volume={209},
  number={1},
  pages={179-192},
  publisher={Elsevier}
}

@article{Ning2014,
  title={An integrated control model for headway regulation and energy saving in urban rail transit},
  author={Ning, Bin and Xun, Jing and Gao, Shigen and Zhang, Lingying},
  journal={IEEE Transactions on Intelligent Transportation Systems},
  volume={16},
  number={3},
  pages={1469--1478},
  year={2014},
  publisher={IEEE}
}

@article{Yinping2008,
  title={Modeling study for tracking operation of subway trains based on cellular automata},
  author={Yinping, FU and Ziyou, GAO and Keping, Li},
  journal={Journal of Transportation Systems Engineering and Information Technology},
  volume={8},
  number={4},
  pages={89--95},
  year={2008},
  publisher={Elsevier}
}

@article{Xun2013,
  title={The impact of end-to-end communication delay on railway traffic flow using cellular automata model},
  author={Xun, Jing and Ning, Bin and Li, Ke-ping and Zhang, Wei-bin},
  journal={Transportation Research Part C: Emerging Technologies},
  volume={35},
  pages={127--140},
  year={2013},
  publisher={Elsevier}
}

@article{Spyropoulou2007,
  title={Modelling a signal controlled traffic stream using cellular automata},
  author={Spyropoulou, Ioanna},
  journal={Transportation Research Part C: Emerging Technologies},
  volume={15},
  number={3},
  pages={175--190},
  year={2007},
  publisher={Elsevier}
}

@article{Meng2011,
  title={An improved cellular automata model for heterogeneous work zone traffic},
  author={Meng, Qiang and Weng, Jinxian},
  journal={Transportation research part C: emerging technologies},
  volume={19},
  number={6},
  pages={1263--1275},
  year={2011},
  publisher={Elsevier}
}

@article{MTR2019,
  title={{Kwun {T}ong {L}ine, {H}ong {K}ong {MTR}}},
  author={{Travel China Guide}},
  year={2019},
  url = {\url{https://tinyurl.com/y45467z4}}
}

@article{Ding2016,
  title={Predicting short-term subway ridership and prioritizing its influential factors using gradient boosting decision trees},
  author={Ding, Chuan and Wang, Donggen and Ma, Xiaolei and Li, Haiying},
  journal={Sustainability},
  volume={8},
  number={11},
  pages={1100},
  year={2016},
  publisher={Multidisciplinary Digital Publishing Institute}
}

@article{Ma2018,
  title={Parallel architecture of convolutional bi-directional {LSTM} neural networks for network-wide metro ridership prediction},
  author={Ma, Xiaolei and Zhang, Jiyu and Du, Bowen and Ding, Chuan and Sun, Leilei},
  journal={IEEE Transactions on Intelligent Transportation Systems},
  volume={20},
  number={6},
  pages={2278--2288},
  year={2018},
  publisher={IEEE}
}

@misc{Gill1994,
  title={Railway signalling system},
  author={Gill, David C},
  year={1994},
  month=nov # "~22",
  publisher={Google Patents},
  note={US Patent 5,366,183}
}

@article{Takeuchi2003,
  title={Moving block signalling dynamics: performance measures and re-starting queued electric trains},
  author={Takeuchi, H and Goodman, CJ and Sone, S},
  journal={IEE Proceedings-electric power applications},
  volume={150},
  number={4},
  pages={483--492},
  year={2003},
  publisher={IET}
}

@article{MTR2019b,
  title={{Investing for the future}},
  author={{Hong Kong MTR}},
  year={2019},
  url = {\url{https://tinyurl.com/y459kjnx}}
}

@article{Delgado2012,
  title={How much can holding and/or limiting boarding improve transit performance?},
  author={Delgado, Felipe and Munoz, Juan Carlos and Giesen, Ricardo},
  journal={Transportation Research Part B: Methodological},
  volume={46},
  number={9},
  pages={1202--1217},
  year={2012},
  publisher={Elsevier}
}

@article{Guo2015,
  title={Cooperative passenger inflow control in urban mass transit network with constraint on capacity of station},
  author={Guo, Jianyuan and Jia, Limin and Qin, Yong and Zhou, Huijuan},
  journal={Discrete Dynamics in Nature and Society},
  volume={2015},
  year={2015},
  publisher={Hindawi}
}

@article{Zou2018,
  title={Managing recurrent congestion of subway network in peak hours with station inflow control},
  author={Zou, Qingru and Yao, Xiangming and Zhao, Peng and Dou, Fei and Yang, Taoyuan},
  journal={Journal of Advanced Transportation},
  volume={2018},
  year={2018},
  publisher={Hindawi}
}

@article{Jiang2018,
  title={Reinforcement learning approach for coordinated passenger inflow control of urban rail transit in peak hours},
  author={Jiang, Zhibin and Fan, Wei and Liu, Wei and Zhu, Bingqin and Gu, Jinjing},
  journal={Transportation Research Part C: Emerging Technologies},
  volume={88},
  pages={1--16},
  year={2018},
  publisher={Elsevier}
}

@article{Shi2018,
  title={Service-oriented train timetabling with collaborative passenger flow control on an oversaturated metro line: An integer linear optimization approach},
  author={Shi, Jungang and Yang, Lixing and Yang, Jing and Gao, Ziyou},
  journal={Transportation Research Part B: Methodological},
  volume={110},
  pages={26--59},
  year={2018},
  publisher={Elsevier}
}

@article{Delgado2009,
  title={Real-time control of buses in a transit corridor based on vehicle holding and boarding limits},
  author={Delgado, Felipe and Munoz, Juan Carlos and Giesen, Ricardo and Cipriano, Aldo},
  journal={Transportation Research Record},
  volume={2090},
  number={1},
  pages={59--67},
  year={2009},
  publisher={SAGE Publications Sage CA: Los Angeles, CA}
}

@book{Daganzo1997,
	author={Carlos F. Daganzo},
	year={1997},
	title={Fundamentals of transportation and traffic operations},
	publisher={Pergamon Oxford},
	volume={30}
}

@article{Bansal2020,
  title={A Dynamic Choice Model with Heterogeneous Decision Rules: {A}pplication in Estimating the User Cost of Rail Crowding},
  author={Bansal, Prateek and H{\"o}rcher, Daniel and Graham, Daniel J},
  journal={arXiv preprint arXiv:2007.03682},
  year={2020}
}

@article{Yuan2020,
  title={Passenger flow control strategies for urban rail transit networks},
  author={Yuan, Fuya and Sun, Huijun and Kang, Liujiang and Wu, Jianjun},
  journal={Applied Mathematical Modelling},
  volume={82},
  pages={168--188},
  year={2020},
  publisher={Elsevier}
}

@article{Wang2020,
  title={Multistation coordinated and dynamic passenger inflow control for a metro line},
  author={Wang, Xingrong and Wu, Jianjun and Yang, Xin and Guo, Xin and Yin, Haodong and Sun, Huijun},
  journal={IET Intelligent Transport Systems},
  volume={14},
  number={9},
  pages={1068--1078},
  year={2020},
  publisher={IET}
}

@article{Loder2019,
  title={Understanding traffic capacity of urban networks},
  author={Loder, Allister and Amb{\"u}hl, Lukas and Menendez, Monica and Axhausen, Kay W},
  journal={Scientific reports},
  volume={9},
  number={1},
  pages={1--10},
  year={2019},
  publisher={Nature Publishing Group}
}

@article{Geroliminis2008,
  title={Existence of urban-scale macroscopic fundamental diagrams: Some experimental findings},
  author={Geroliminis, Nikolas and Daganzo, Carlos F},
  journal={Transportation Research Part B: Methodological},
  volume={42},
  number={9},
  pages={759--770},
  year={2008},
  publisher={Elsevier}
}

@article{Papageorgiou2003,
	author={Markos Papageorgiou and Christina Diakaki and Vaya Dinopoulou and Apostolos Kotsialos and Yibing Wang},
	year={2003},
	title={Review of road traffic control strategies},
	journal={Proceedings of the IEEE},
	volume={91},
	number={12},
	pages={2043-2067}
}

@book{Small2007,
	author={Kenneth A. Small and Erik T. Verhoef},
	year={2007},
	title={The economics of urban transportation},
	publisher={Routledge}
}

@article{Horowitz2011,
	author={Joel L. Horowitz},
	year={2011},
	title={Applied Nonparametric Instrumental Variables Estimation},
	journal={Econometrica},
	volume={79},
	number={2},
	pages={347-394}
}

@article{Newey2003,
	author={Whitney K. Newey and James L. Powell},
	year={2003},
	title={Instrumental Variables Estimation of Nonparametric Models},
	journal={Econometrica},
	volume={71},
	number={5},
	pages={1565-1578}
}

@article{Chetverikov2017,
	author={Denis Chetverikov and Daniel Wilhelm},
	year={2017},
	title={Nonparametric Instrumental Variables Estimation under Monotonicity},
	journal={Econometrica},
	volume={85},
	number={4},
	pages={1303-1320}
}

@article{Newey2013,
	author={Whitney K. Newey},
	year={2013},
	title={Nonparametric Instrumental Variables Estimation},
	journal={American Economic Review},
	volume={103},
	number={3},
	pages={550-556}
}

@article{Amirgholy2017,
  title={Optimal design of sustainable transit systems in congested urban networks: A macroscopic approach},
  author={Amirgholy, Mahyar and Shahabi, Mehrdad and Gao, H Oliver},
  journal={Transportation Research Part E: Logistics and Transportation Review},
  volume={103},
  pages={261--285},
  year={2017},
  publisher={Elsevier}
}

@article{Zhang2021,
  title={Model and algorithm of coordinated flow controlling with station-based constraints in a metro system},
  author={Zhang, Ping and Sun, Huijun and Qu, Yunchao and Yin, Haodong and Jin, Jian Gang and Wu, Jianjun},
  journal={Transportation Research Part E: Logistics and Transportation Review},
  volume={148},
  pages={102274},
  year={2021},
  publisher={Elsevier}
}

@article{Silva2015,
  title={Predicting traffic volumes and estimating the effects of shocks in massive transportation systems},
  author={Silva, Ricardo and Kang, Soong Moon and Airoldi, Edoardo M},
  journal={Proceedings of the National Academy of Sciences},
  volume={112},
  number={18},
  pages={5643--5648},
  year={2015},
  publisher={National Acad Sciences}
}

\end{document}